\documentclass[extra]{gji}
\usepackage[pass]{geometry}
\usepackage{graphicx}
\usepackage{epstopdf}

\usepackage{amssymb}
\usepackage{amsmath}
\usepackage{psfrag}
\usepackage{pstool}
\title[Reactive and shear-driven instabilities]  
  {Magmatic channelisation by reactive and shear-driven instabilities at mid-ocean ridges: a combined analysis} 
    \author[D. W. Rees Jones, H. Zhang and R. F. Katz]
  {D. W. Rees Jones$^{1,2,}$\thanks{david.reesjones@st-andrews.ac.uk}, H. Zhang$^{2,3}$ and R. F. Katz$^2$\\
  $^1$ University of St Andrews, School of Mathematics and Statistics, \\ Mathematical Institute, North Haugh, St Andrews, KY16 9SS, United Kingdom. \\
  $^2$ University of Oxford, Department of Earth Sciences, \\ South Parks Road, Oxford, OX1 3AN, United Kingdom. \\
  $^3$ Peking University, Department of Mechanics and Engineering Science, \\ Beijing, 100871, China.
  }

\usepackage{lineno}
\newcommand*\patchAmsMathEnvironmentForLineno[1]{%
  \expandafter\let\csname old#1\expandafter\endcsname\csname #1\endcsname
  \expandafter\let\csname oldend#1\expandafter\endcsname\csname end#1\endcsname
  \renewenvironment{#1}%
     {\linenomath\csname old#1\endcsname}%
     {\csname oldend#1\endcsname\endlinenomath}}%
\newcommand*\patchBothAmsMathEnvironmentsForLineno[1]{%
  \patchAmsMathEnvironmentForLineno{#1}%
  \patchAmsMathEnvironmentForLineno{#1*}}%
\AtBeginDocument{%
\patchBothAmsMathEnvironmentsForLineno{equation}%
\patchBothAmsMathEnvironmentsForLineno{align}%
\patchBothAmsMathEnvironmentsForLineno{flalign}%
\patchBothAmsMathEnvironmentsForLineno{alignat}%
\patchBothAmsMathEnvironmentsForLineno{gather}%
\patchBothAmsMathEnvironmentsForLineno{multline}%
}

\begin{document}

\maketitle
\begin{summary}
It is generally accepted that melt extraction from the mantle at mid-ocean ridges is concentrated in narrow regions of elevated melt fraction called channels. Two feedback mechanisms have been proposed to explain why these channels grow by linear instability: shear flow of partially molten mantle and reactive flow of the ascending magma. These two mechanisms have been studied extensively, in isolation from each other, through theory and laboratory experiments as well as field and geophysical observations. Here, we develop a consistent theory that accounts for both proposed mechanisms and allows us to weigh their relative contributions. We show that interaction of the two feedback mechanisms is insignificant and that the total linear growth rate of channels is well-approximated by summing their independent growth rates. Furthermore, we explain how their competition is governed by the orientation of channels with respect to gravity and mantle shear. By itself, analysis of the reaction-infiltration instability predicts the formation of tube-shaped channels. We show that with the addition of even a small amount of extension in the horizontal, the combined instability favours tabular channels, consistent with the observed morphology of dunite bodies in ophiolites. We apply the new theory to mid-ocean ridges by calculating the accumulated growth and rotation of channels along streamlines of the solid flow. We show that reactive flow is the dominant instability mechanism deep beneath the ridge axis, where the most unstable orientation of high-porosity channels is sub-vertical.  Channels are then rotated by the solid flow away from the vertical. The contribution of the shear-driven instability is confined to the margins of the melting region. Within the limitations of our study, the shear-driven feedback does not appear to be responsible for significant melt focusing or for the shallowly dipping seismic anisotropy that has been obtained by seismic inversions.
\end{summary}


\section{Introduction} \label{sec:intro}
At mid-ocean ridges, plate spreading induces upwelling of the mantle, causing decompression melting.
This melting occurs inside a volume (the melting region) that extends to a depth and distance of order 100~km from the ridge axis.  The rock within the volume is partially molten, consisting of a crystalline solid with liquid melt along the boundaries of solid grains. 
The melt resides in an interconnected, permeable network of pores, such that it can segregate from the residue and migrate over large distances, driven by buoyancy and pressure gradients. 
Several lines of evidence suggest that the migration is not spatially uniform, but is rather localised in channels of elevated melt fraction (porosity).

A key line of evidence for channelised transport comes from geological observations of tabular bodies of nearly pure olivine (dunite) in ophiolites, which are otherwise dominantly of olivine+pyroxene lithology (harzburgite). These have been interpreted as the relics of former channels in which focused melt flow has dissolved all pyroxene and replaced it with olivine \citep{quick1982,kelemen90,kelemen95,kelemen97,kelemen00,braun02}. Similar features are observed in laboratory experiments in which Si-undersaturated melts are forced to traverse a porous, olivine+pyroxene matrix \citep{pec15, pec17}. 
Furthermore, there is well-documented chemical disequilibrium between erupted lavas and the harzburgitic uppermost mantle. This surprising observation can be reconciled by channelized flow; transport through dunite channels isolates rising melts from the surrounding mantle harzburgite \citep{kelemen95}. Moreover, young lavas are typically found to contain isotopes of the uranium-series decay chain with disequilibrium activity ratios \citep{sims02}.  This has been interpreted to indicate rapid melt transport from depth in the mantle, which would be a consequence of the hypothetically channelised flow \citep{jull02, Elliot2014}.

Physical models and laboratory experiments point to two possible types of fluid-mechanical instability that can cause localisation of magmatic flow. 
First, chemical reactions during magmatic ascent can dissolve the host rock due to a solubility gradient, driving a reaction-infiltration instability \citep{kelemen90,aharonov95,kelemen95b}.
Second, shear of the solid rock coupled with the fact that its viscous strength decreases with melt fraction cause a shear-driven banding instability \citep{stevenson89,holtzman03a}. 
These mechanisms have been extensively studied in isolation, and their outcomes have been invoked separately to explain natural observations.

The key objective of the present paper is to assess the relative contributions of these two instabilities to melt transport within the mid-ocean ridge melting region. In particular, it is to discover the dynamic and parametric conditions under which the shear-driven instability can contribute significantly to the overall pattern of channelised melt transport. It has been widely assumed that the shear-driven instability makes an important contribution to melt flow and mantle dynamics \citep[e.g.,][]{kohlstedt09, kawakatsu09, Holtzman2010}, but no quantitative assessment that also includes reactive flow has been published. In the context of our simplified analysis, we test and ultimately challenge this assumption. Instead we argue that the reaction-infiltration instability is dominant and that shear-driven instability may be insignificant beneath mid-ocean ridges.

We reviewed research into the reaction-infiltration instability in \citet{reesjones2018-jfm} and augment that review briefly here.
Recent laboratory experiments at high temperature and pressure showed that highly permeable, cylindrical conduits form due to the reactive flow of Si-undersaturated melt \citep{pec15,pec17,pec20}. 
Similar cylindrical features also arise from reactive instabilities in a class of porous media called mushy layers \citep{Tait92,Worster97}.
The cylindrical geometry of channels in mushy layers is noteworthy because it contrasts with the tabular nature of dunite bodies in ophiolites. 
Theoretical work has focused on linear stability analysis and two-dimensional numerical calculations \citep{aharonov95, spiegelman01,hewitt10,liang10, hesse11,Schiemenz2011,Szymczak13,Szymczak14,Jordan2015}.
Numerical calculations of reactive flow were extended by \citet{Baltzell2015} to include the effect of mantle shear. 
However, that study used a constant shear viscosity rather than one that is porosity-weakening.
Thus it considered the effect of shear in stretching and rotating high-porosity features caused by the reactive instability, but excluded the potential for the shear flow to drive a melt-localising instability.

The shear-driven instability was reviewed by \citet{kohlstedt09}.
The instability was predicted theoretically by \citet{stevenson89} before being confirmed experimentally by \citet{holtzman03a}. 
It has since been the subject of extensive laboratory experiments \citep{holtzman07,king11a,qi15} and theoretical study \citep{spiegelman03b, katz06,butler09,alisic16}. 
Recent consensus is that it is controlled by viscous anisotropy \citep{takei09a,takei09b,takei09c,butler12,takei13,katz13,qi15}, although the details remain incompletely understood. 
The emergent patterns, including the associated development of a distinct mode of olivine lattice preferred orientation \citep{holtzman03b} and hence seismic anisotropy \citep{Holtzman2010}, have been invoked to explain proxy measurements.

A set of laboratory experiments reported by \cite{king11b} considered the combined role of reaction and shear in generating melt bands.  However, the experimental conditions are so far from the natural system that it is difficult to draw general conclusions from that study.

The natural system of primary interest here is the mid-ocean ridge (MOR). MORs are a fundamental component of plate tectonics, the predominant locus of present terrestrial magmatism, and a context in which melt channelization is inferred from observations, as discussed above. 
The stability analysis of \citet{butler09} suggested that shear-driven porosity bands would form here. However, that study neglected the role of buoyancy and ongoing melting.
\citet{Vestrum2020} considered the effect of both buoyancy and a uniform background melting rate (but excluded reactive melting). 
They showed that the consequences of uniform melting depend on details of the rheological model. 
Furthermore, they showed that buoyancy does not affect the magnitude of the shear bands; rather, it causes bands to travel as porosity waves.  These studies cannot, however, address the relative importance of reactive flow to melt localisation.
The relative importance was tested in terms of its geochemical consequences by \citet{liu19}, who considered models with various distributions of both reactive channels and shear-induced bands.  However, the distribution and character of localised porosity features in \citet{liu19} were prescribed, rather than emerging dynamically.

The broad goal of this paper is to develop a theoretical understanding of the combined dynamics of reactive and shear-driven instabilities in the partially molten upper mantle. This allows us to assess their relevance for magmatic flow localisation at MORs, at length- and time-scales that are necessarily very different from those of the laboratory. Insights gained here may have wider implications for magmatism.

In section~\ref{sec:methods}, we develop a theoretical method to determine the linear growth rate of perturbations in the form of alternating bands of higher and lower porosity.  The theory allows for the perturbations to evolve by both reactive and shear mechanisms simultaneously.  We show that this theory can reproduce previous estimates of their growth rate in both the reaction-only and shear-only limits.

In section~\ref{sec:results_local}, we apply this theory to the idealized scenario of an infinite, partially molten material with a uniform background magmatic flow, a linear solubility gradient aligned with gravity, and a linear shear of the solid matrix. 
We identify a parameter that describes the relative importance of reaction and shear.
In this idealised context we show that there are two distinguished directions that control the orientation of high-porosity pathways: one in the direction of gravity (the vertical) over which there is a chemical solubility gradient, and the other in the direction of the maximum tensile deviatoric stress. 
We calculate the growth of the instability and show that the optimal orientation for growth of porosity perturbations depends on these directions and also the ratio of shear-to-reactive growth rates.
Indeed, within a plane normal to gravity, it is shear that breaks the horizontal isotropy. We show that the most unstable, fastest growing features are tabular bodies extending horizontally in the direction in which there is no component of the shear flow.
This is important because dunites are observed to be tabular features, rather than the cylindrical conduits favoured by the pure reaction-infiltration instability.  
In this section we also discuss the role of compaction, chemical advection and diffusion, all of which play a role in determining the wavelength-dependence of the growth rate. 

In section~\ref{sec:methods_MOR} we present a methodology and in section~\ref{sec:results_MOR} we present results from the first combined assessment of both reactive and shear-driven instability at mid-ocean ridges.
We build on the approach of \citet{Gebhardt2016}, 
allowing the amplitude and orientation of perturbations to evolve along streamlines of the solid flow, based on their local growth rate.  But whereas \citet{Gebhardt2016} considered only shear-driven growth, in our case, the growth rate is a consequence of both reaction and shear. 
We show that their contributions are not spatially uniform and predict how they vary along corner-flow streamlines.
Generally, we find that the initial growth of porosity bands is dominated by reaction. 
Within the space of the melting region and the plausible ranges of control parameters, our results indicate that melt channels are sub-vertical features that undergo some rotation by the shear flow. 
Shear-driven instability is predicted to contribute at the margins of the melting region. 

In section~\ref{sec:discussion}, we review the implications of these findings for understanding the relative importance of reaction and shear in driving channelized melt extraction from the mantle. We discuss implications for interpreting the origins of tabular dunites embedded in a harzburgitic upper mantle. 
We also discuss melt focusing towards the ridge axis, as well as seismic anisotropy that has been attributed to aligned, melt-rich bands.

\section{Methods: local analysis of combined instability} \label{sec:methods}
\begin{table}
 \caption{State variables and selected formulae}
 \begin{tabular}{ll} \label{tab:state}
  Symbol & Property \\
    $ (x,y,z)$ & Cartesian co-ordinates     \\
  $ t $ &    Time \\
  $ \phi $ & Porosity     \\
  $ \Gamma $ & Volumetric melt rate     \\
  $ \boldsymbol{v}_s $ &  Solid velocity:   $ \boldsymbol{v}_s = \boldsymbol{u} +\nabla U$  \\
  $ \boldsymbol{u} $ &  Divergence-free part of solid velocity   \\
   $\nabla U $ &  Curl-free part of solid velocity   \\
   $ \boldsymbol{v}_l $ &  Liquid velocity \\
   $ \boldsymbol{v}_D $ &  Darcy velocity: $ \phi ( \boldsymbol{v}_l - \boldsymbol{v}_s  )$ \\
   $\mathcal{C}$ & Compaction rate: $\mathcal{C} = \nabla \cdot \boldsymbol{v}_s $ \\
   $\mathrm{\textbf{D}}_s $ & Deviatoric strain rate:  $\mathrm{\textbf{D}}_s = \frac{1}{2} \left[ \nabla \boldsymbol{v}_s +\nabla  \boldsymbol{v}_s ^T - \frac{2}{3}\mathcal{C}\mathrm{\textbf{I}} \right]$ \\
 $P _l $ &  Liquid pressure \\ 
 $ c_l $ & Liquid concentration \\
 $X $ & Liquid undersaturation \\   
 \end{tabular} \hfill\
 \vspace{0.5cm}
%
 \caption{Physical properties and selected formulae}
 \begin{tabular}{ll} \label{tab:properties}
  Symbol & Property \\
$\rho_s$ & Density of solid   \\
  $\rho_l$ & Density of liquid   \\
  $\Delta \rho $ & Density difference: $\Delta \rho =\rho_s-\rho_l $  \\
  $\overline{\rho} $ & Bulk density: $\overline{\rho} = \phi \rho_l +(1-\phi) \rho_s $ \\
  $\boldsymbol{g}$ & Gravity vector \\
  $g$ & Gravitational acceleration: $g=\vert \boldsymbol{g} \vert$ \\
  $K$ & Fluid mobility (permeability/fluid viscosity) \\
  $\eta$ & Shear viscosity \\ 
  $\zeta$ & Bulk viscosity \\
  $ R $ &  Reaction rate constant   \\
  $ \alpha  $ & Inverse reactivity   \\
  $ \beta $ & Equilibrium concentration gradient  \\ 
  $D_X $ & Chemical diffusivity in liquid phase   \\
 \end{tabular} \hfill\
\end{table}

\subsection{Equations governing two-phase flow} \label{sec:2-phase-flow}
The partially molten upper mantle can be modelled as a region of two-phase flow.
The continuum model that we use, developed by \citet{mckenzie84}, is based on averaging all the quantities of interest across a control volume containing both solid and liquid phase.
Table~\ref{tab:state} lists the state variables and table~\ref{tab:properties} lists the physical properties that we define in this section.
Our formulation makes a Boussinesq approximation: the solid density $\rho_s$ and liquid density $\rho_l$ are taken to be constants, and their difference $\Delta \rho = \rho_s - \rho_l$ is neglected everywhere, except in so far as it drives segregation of the liquid by buoyancy.

Mass conservation in the liquid phase is given by
\begin{equation}
    \frac{\partial \phi}{\partial t} + \nabla \cdot (\phi \boldsymbol{v}_l ) = \Gamma, \label{eq:mass_l}
\end{equation}
where $t$ is time, $\phi$ is the porosity, $\boldsymbol{v}_l$ is the liquid velocity and $\Gamma$ is the volumetric melting rate.
Similarly, mass conservation in the solid phase is given by
\begin{equation}
    \frac{\partial (1-\phi)}{\partial t} + \nabla \cdot ((1-\phi) \boldsymbol{v}_s  ) = -\Gamma, \label{eq:mass_s}
\end{equation}
where $\boldsymbol{v}_s$ is the solid velocity. 
The compaction rate is defined as $\mathcal{C}=\nabla \cdot \boldsymbol{v}_s$.
It is convenient to sum equations~(\ref{eq:mass_l}) and (\ref{eq:mass_s}) to form an equation for the compaction rate:
\begin{equation}
    \mathcal{C} + \nabla \cdot \boldsymbol{v}_D   = 0, \label{eq:compaction1}
\end{equation}
where the Darcy (melt segregation) velocity is defined as
 \begin{equation}
    \boldsymbol{v}_D \equiv \phi ( \boldsymbol{v}_l - \boldsymbol{v}_s  ). \label{eq:Darcy_def}
\end{equation}
Then equation~(\ref{eq:mass_s}) can be rewritten
\begin{equation}
    \frac{\partial \phi}{\partial t} +  \boldsymbol{v}_s \cdot \nabla \phi = \Gamma +(1-\phi) \mathcal{C}. \label{eq:mass_s_2}
\end{equation}
In this formulation, porosity, moving with the solid phase, evolves due to melting rate $\Gamma$ and compaction $\mathcal{C}$. 
Note that $\mathcal{C}>0$ represents a decompaction of the mantle matrix, i.e., an increase in the porosity.
Equations~(\ref{eq:compaction1}) and (\ref{eq:mass_s_2}) define two-phase mass conservation in our system.

Next, two-phase momentum conservation can be written using a `Stokes--Darcy' formulation. 
The Darcy velocity 
 \begin{equation}
    \boldsymbol{v}_D =  -K\left(\nabla P_l - \rho_l \boldsymbol{g} \right). \label{eq:Darcy}
\end{equation}
is driven by gradients in the liquid pressure $P_l$ and buoyancy ($\boldsymbol{g}$ is gravity).
$K$ is the liquid mobility, which is defined as the permeability divided by the liquid viscosity.
We will often refer to $K$ as the permeability, since we assume the liquid viscosity is a constant so variation in $K$ comes from variation in permeability. 
The Stokes part of the system can be written
\begin{equation}
    \nabla P_l = \nabla \cdot \left[2\eta \mathrm{\textbf{D}}_s \right] + \nabla  (\zeta\mathcal{C}) + \overline{\rho}\boldsymbol{g},\label{eq:Momentum}
\end{equation}
where $\eta$ is the shear viscosity (Newtonian), $\zeta$ is the bulk viscosity, $\overline{\rho} = \phi \rho_l +(1-\phi) \rho_s =\rho_l +(1-\phi)\Delta \rho $ is the bulk density and
\begin{equation}
    \mathrm{\textbf{D}}_s = \frac{1}{2} \left[ \nabla \boldsymbol{v}_s +\nabla  \boldsymbol{v}_s ^T - \frac{2}{3}\mathcal{C}\mathrm{\textbf{I}} \right]
\end{equation}
is the deviatoric strain rate tensor ($\mathrm{\textbf{I}}$ is the identity tensor, $^T$ is the transpose operator).

We discuss constitutive laws and the dependence of material properties on porosity in the next section. 
For now, we note that the shear-driven mode of instability relies on the fact that $\eta$ decreases with $\phi$. 
This motivates expanding out the derivatives involving $\eta$. 
We also apply a Helmholtz decomposition of the solid velocity into a shearing (incompressible) part and a compacting (compressible, but irrotational) part
\begin{equation}
    \boldsymbol{v}_s = \boldsymbol{u} + \nabla U, \quad \nabla \cdot  \boldsymbol{u} = 0,
\end{equation}
where $U$ is a scalar potential that can be related to $\mathcal{C}$ by the relationship \mbox{$\mathcal{C}=\nabla^2 U$}. 
Then equation~(\ref{eq:Momentum}) becomes:
\begin{equation}
    \nabla P_l =  2\mathrm{\textbf{D}}_s  \cdot \nabla \eta + \eta \nabla^2 \boldsymbol{u}+ \frac{4}{3} \eta \nabla \mathcal{C} + \nabla  (\zeta\mathcal{C}) + \overline{\rho}\boldsymbol{g}.\label{eq:Momentum2}
\end{equation}
The terms involving pressure, $\eta \nabla^2 \boldsymbol{u}$ and gravity can be recognised as the usual, single-phase form of Stokes law with a constant Newtonian viscosity.
We can eliminate various terms by taking the curl of equation~\eqref{eq:Momentum2} to form a type of vorticity equation
\begin{equation}
    0 =  \nabla \times \left[ 2\mathrm{\textbf{D}}_s  \cdot \nabla \eta + \eta \nabla^2 \boldsymbol{u}+ \frac{4}{3} \eta \nabla \mathcal{C} - \phi \Delta \rho \boldsymbol{g} \right].\label{eq:Momentum3}
\end{equation}
We can also substitute equation~\eqref{eq:Momentum2} into equation~(\ref{eq:Darcy}) to obtain
\begin{equation}
    \boldsymbol{v}_D =  -K\left[ 2 \mathrm{\textbf{D}}_s  \cdot \nabla \eta + \eta \nabla^2 \boldsymbol{u} + \frac{4}{3} \eta \nabla \mathcal{C} + \nabla  (\zeta\mathcal{C}) + (1-\phi) \Delta \rho \boldsymbol{g}  \right]. \label{eq:Darcy2}
\end{equation}

The melting rate $\Gamma$ is determined by chemical disequilibrium. 
We describe the model that we and others have used in more detail in \citet{reesjones2018-jfm}; the original model is due to \citet{aharonov95}.
The crucial ingredients are that the melting rate is linearly proportional to the chemical disequilibrium 
\begin{equation}
    \Gamma = R X, \label{eq:melt-rate}
\end{equation}
where $R$ is the reaction rate constant and $X$ is the chemical undersaturation. 
That $R$ is constant is a reasonable assumption during the initial growth of the instability; later $R$ will decrease as the concentration of the soluble component in the solid phase decreases.
Equation~\eqref{eq:melt-rate} is a first order, linear, kinetic reaction rate equation.
Then $X$ is governed by an advection-diffusion-reaction equation, which can be written in the form
\begin{equation}
    -\phi \frac{\partial X}{\partial t} + \phi \boldsymbol{v}_l \cdot \nabla \left( \beta z  - X \right) = \alpha \Gamma -  \nabla \cdot( \phi D_X \nabla X), \label{eq:chemistry1}
\end{equation}
where $\beta$ is the constant gradient of the equilibrium chemical concentration, sometimes called the solubility gradient, which we assume is orientated in the vertical direction. Under these assumptions, the liquid concentration can be written in terms of the undersaturation as \mbox{$c_l = \beta z  - X$}, appearing in the second term in equation~\eqref{eq:chemistry1}. 
Also, $\alpha$ represents the supersaturation of the reactively produced melts that drive the system back towards equilibrium. Within equation~\eqref{eq:chemistry1}, $\alpha$ is an inverse reactivity, because the reactive melt rate $\Gamma$ is inversely proportional to $\alpha$. Finally, $D_X$ is the diffusivity of chemical species in the liquid phase, scaled by the porosity $\phi$. Diffusion in the liquid is much faster than that in the solid, so diffusivity of chemical species in the solid phase is neglected.

All the approximations made here are described in more detail in \citet{reesjones2018-jfm}.
For now we focus on the equilibrium concentration gradient
because this drives the reactive instability. 
In a more complete description, the equilibrium chemical concentration depends on pressure \citep{kelemen95b,longhi02}.
Thus this simple formulation with constant $\beta$ combines an assumption that the pressure is dominantly lithostatic, i.e., proportional to vertical position $z$, and an assumption that the dependence of equilibrium chemical concentration on pressure is linear. 
The latter assumption could be relaxed straightforwardly by allowing $\beta$ to vary with $z$. 
In appendix~\ref{app:c_eq}, we consider the full pressure-dependence of the equilibrium  concentration gradient and hence the reactive melting rate. The shear-driven instability creates a pressure gradient that can, in turn, feed back on the reactive instability through the  pressure-dependent reactive melt rate.
So appendix~\ref{app:c_eq} also considers a mode of coupling between the two types of instability.
We find the effects of pressure-dependence on the linear growth rate of the reactive instability are relatively small, because  $\Delta \rho/\rho_s \ll 1$, which means that the pressure is dominantly lithostatic. 
However, it is possible that the pressure-dependence may have a larger effect on the nonlinear development of channels.
\citet{spiegelman01} suggested that lateral pressure gradients that focus flow towards a mid-ocean-ridge (MOR) axis could promote the development of a diagonal solubility gradient that would further enhance convergence of flow towards the ridge axis. 
Nonlinear calculations extending those of \citet{spiegelman01} could assess the significance of this proposed mechanism at MORs. 

In the upper mantle, porosity is typically very small, so we can further simplify these equations by making the approximations \mbox{$(1-\phi) \approx 1$}, \mbox{$ \phi \boldsymbol{v}_l \approx  \boldsymbol{v}_D$}, and neglecting the terms \mbox{$ \boldsymbol{v}_s \cdot \nabla \phi \ll 1$}, \mbox{$\phi \frac{\partial X}{\partial t}\ll 1$} and $\nabla \times \phi \Delta \rho \boldsymbol{g}$.
We will revisit the role of porosity advection  \mbox{$ \boldsymbol{v}_s \cdot \nabla \phi $} later (section~\ref{sec:porosity-advection}).
We can also eliminate the undersaturation $X$ using equation~(\ref{eq:melt-rate}).
Under these approximations, we obtain a system of equations that includes the physical mechanisms that give rise to both shear and reaction driven instabilities:
\begin{align}
&\frac{\partial \phi}{\partial t} = \Gamma + \mathcal{C}, \label{eq:mass_l_summary} \\
&\mathcal{C} + \nabla \cdot \boldsymbol{v}_D,   = 0  \label{eq:mass_s_summary} \\
&0 =  \nabla \times \left[ 2 \mathrm{\textbf{D}}_s  \cdot \nabla \eta + \eta \nabla^2 \boldsymbol{u}+ \frac{4}{3} \eta \nabla \mathcal{C} \right], \:  \nabla \cdot  \boldsymbol{u} = 0, \label{eq:Momentum_summary} \\
&\boldsymbol{v}_D  =  -K\left[ 2 \mathrm{\textbf{D}}_s  \cdot \nabla \eta + \eta \nabla^2 \boldsymbol{u}+ \frac{4}{3} \eta \nabla \mathcal{C} + \nabla  (\zeta\mathcal{C}) +  \Delta \rho \boldsymbol{g}  \right]. \label{eq:Darcy_summary} \\
&  \boldsymbol{v}_D \cdot \nabla \left( \beta z  - R^{-1} \Gamma \right) = \alpha \Gamma -  \nabla \cdot( \phi D_X R^{-1} \nabla \Gamma ). \label{eq:chemistry_summary}
\end{align}

\subsection{Equations governing linear growth of instabilities} \label{sec:linear-growth}
\begin{table}
 \caption{Selected additional variables arising in linear stability analysis} \label{tab:LSA}
 \begin{tabular}{ll}
  Symbol & Property \\
    $ \sigma$ & Growth rate of normal modes    \\
    $\boldsymbol{k}$ & Wavevector of normal modes  \\
    $k$ & Amplitude of wavevector (wavenumber) \\
    $(k_x,k_y,k_z)$ & Cartesian components of wavevector \\
      $(\theta,\psi)$ & Spherical polar components of wavevector \\
          $_0$ & Background component \\ 
      $'$ & Perturbation component \\ 
    $w_{D0}$ & Background Darcy velocity \\
    $w_0$ & Background liquid velocity \\
    $\dot{\gamma}_0$ & Background stain rate  \\
    $\theta_e$ & Angle of maximum extension \\
    $\psi'$, ($ \widetilde{\psi}'$) & Scalar potential, (scaled) \\
      $\lambda^*$ & Sensitivity of shear viscosity to porosity \\
        $n$ & Exponent in porosity--permeability relationship \\
    $\delta $ & Compaction length: $ \delta = \sqrt{ K_0\left(\tfrac{4}{3}\eta_0 + \zeta_0 \right) }$ \\ 
    $\Lambda$ & Shear-driven instability scale:  $\Lambda  = 2 K_0 \eta_0 \lambda^* \dot{\gamma}_0$ \\
         $\mathrm{Da}_w$, $\mathrm{Da}_D$ & Damk{\"o}hler numbers  \\
 $G$ & Angular dependence of growth rate \\
 $\sigma_\mathrm{shear} $ & Growth rate of shear-driven instability: \\ & $\sigma_\mathrm{shear} =  {  \Lambda    }/{\delta^2 }$  \\[3pt] 

  $\sigma_\mathrm{reaction} $ & Growth rate of reaction-driven instability: \\ & $\sigma_\mathrm{reaction} = n \beta w_0/\alpha$ \\[3pt] 
  $S$ & Growth rate ratio: $S= \sigma_\mathrm{shear}/\sigma_\mathrm{reaction}$ \\
   \end{tabular} \hfill\
\end{table}

Our first goal is to determine the local growth rate of instabilities about a uniform background state with a linear incompressible shear flow $\boldsymbol{u}$ and a vertical Darcy flow $w_{D0} \boldsymbol{\tilde{z}}$, where  $\boldsymbol{\tilde{z}}$ is a unit vector in the $z$-direction.
We write all the fields as an expansion into a base state (subscript $0$) and a perturbation ($'$)
\begin{align} \label{eq:lin_decomposition}
&\phi = \phi_0 +\phi', \quad 
\mathcal{C} = \mathcal{C}_0 +\mathcal{C}', \quad 
\Gamma = \Gamma_0 +\Gamma'  \\
&\boldsymbol{u} = \boldsymbol{u}_0  +\boldsymbol{u}', \quad 
\boldsymbol{v}_D = w_{D0} \boldsymbol{\tilde{z}}  +\boldsymbol{v}_D'. \nonumber 
\end{align}
Table~\ref{tab:LSA} summarizes the notation introduced in this section.

A crucial ingredient of the reactive- and shear-driven instabilities is that the constitutive laws (material properties) depend on porosity. 
We can linearize all the constitutive laws by expanding them in a Taylor series in $\phi$ truncated at first order
\begin{align}
& K = K_0 + K',  \quad  \eta = \eta_0  +\eta',  \quad  \zeta = \zeta_0  +\zeta',
\end{align}
where all the primed variables are proportional to $\phi'$. 
For example
\begin{equation}
K(\phi_0 +\phi') = K(\phi_0) + \phi' \left. \frac{dK}{d\phi}\right|_{\phi=\phi_0} \equiv K_0 + K'.
\end{equation}
A common choice of constitutive law for permeability, and hence melt mobility assuming melt viscosity is constant, is
\mbox{$K=K^*\phi^n$}, where $K^*$ is a material property and $n$ is an exponent \citep[e.g.][]{vonbargen86}. 
Then  
\begin{equation} \label{eq:permeability_decomp}
K_0 = K^* \phi_0^n, \quad K' = n K_0 \phi' /\phi_0.
\end{equation}
Likewise for shear viscosity, a commonly used law is $\eta=\eta_0 \exp\left[-\lambda^* (\phi-\phi_0) \right]$, where $\lambda^* $ is a material property \citep[e.g.][]{kelemen97,mei02}.
Then 
\begin{equation}  \label{eq:eta_prime_def}
\eta' = -\lambda^* \eta_0 \phi' .
\end{equation}
The same methodology can be applied to any form of constitutive law that has a continuous first derivative. 
We do not state any particular constitutive law for the bulk viscosity at this stage, because $\zeta'$ does not affect the linear analysis that follows. 
We discuss this issue further in section~\ref{sec:MagmaFlow}.

\subsubsection{Background state} \label{sec:background-state}
We are interested in the development of perturbations to an initially uniform porosity field with a uniform, background upward flow of magma and a linear, incompressible mantle shear flow. 
This state is called the background or base state.

A somewhat subtle issue is that a uniform background state is only an approximate solution of equations~(\ref{eq:mass_s_summary}--\ref{eq:chemistry_summary}). 
In \citet{reesjones2018-jfm}, we showed that this approximation holds provided we are dealing with length scales much smaller than $\alpha \beta^{-1} \approx 500$~km.
Physically, the system is only weakly reactive. 
In this approximation, the background compaction rate is negligible, so $\mathcal{C}_0=0$.
This approximation is equivalent to saying that the background melting rate (which balances the background compaction rate) is negligible, an issue we discuss later in the context of mid-ocean ridges (section~\ref{sec:MagmaFlow}). 

Equation~(\ref{eq:Darcy_summary}) gives an expression for the background Darcy velocity that can also be written in terms of the liquid velocity $w_0$ using $w_{D0}= \phi_0 w_0$. 
In particular,
\begin{equation}
w_{D0} = K_0 \Delta \rho g, \quad w_{0} = \frac{K_0 \Delta \rho g}{\phi_0}, \quad g=|\boldsymbol{g}|.
\end{equation}

\begin{figure*}
  \includegraphics[width=0.7\textwidth]{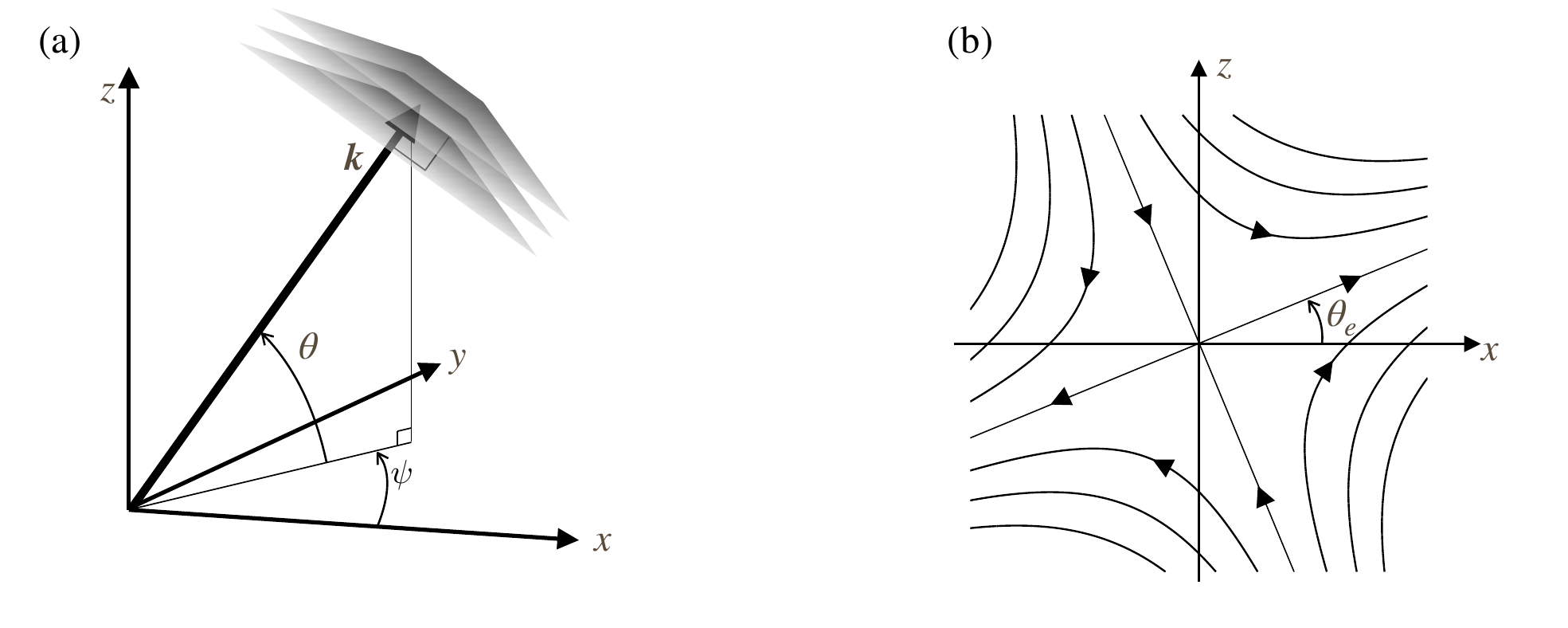}
  \caption{Sketch defining coordinates and angles. a) The three-dimensional coordinate system associated with the model of instability (\S\ref{sec:2-phase-flow}). $\boldsymbol{k}$ is the wave vector, which is normal to perturbation surfaces of constant phase. These infinite, parallel surfaces are represented by shaded patches of finite extent. b) Streamlines of the pure-shear background flow (\S\ref{sec:background-state}) with the direction of extension oriented at an angle $\theta_e$ to the $x$ axis.  \label{fig:schematic}}
\end{figure*}

Any linear incompressible shear flow satisfies equation~(\ref{eq:Momentum_summary}), because $\nabla \eta = 0$ and $\nabla \mathcal{C} =0$. 
We define 
\begin{equation}
\mathrm{\textbf{D}}_0 = \frac{1}{2}\left [\nabla \boldsymbol{u}_0 +\nabla  \boldsymbol{u}_0 ^T \right],
\end{equation}
which is the symmetric velocity gradient tensor associated with the background shear flow (note this is a constant and trace-free tensor, because the background shear flow is linear and incompressible). 
A mid-ocean ridge far from any offsets is roughly two-dimensional, so if we choose the $x$-axis to be in the direction of plate spreading, the shear is in the $x$-$z$ plane. Then a general expression for the symmetric velocity gradient tensor of this type of flow is
\begin{equation} \label{eq:D0_def}
 \mathrm{ \textbf{D}}_0  = \dot{\gamma_0}  \widetilde{ \mathrm{ \textbf{D}}}_0, \quad   \widetilde{\mathrm{ \textbf{D}}}_0 = 
\begin{bmatrix}
    \cos(2\theta_e)       &0 & \sin(2\theta_e) \\
    0       & 0 & 0  \\
   \sin(2\theta_e)       & 0 & -\cos(2\theta_e) 
\end{bmatrix},
\end{equation}
where $\dot{\gamma_0}$ is the background strain rate and $\theta_e$ is the angle of maximum extension ($\theta_e=0$ corresponds to extension in the $x$-direction and $\theta_e=\pi/2$ corresponds to extension in the $z$-direction). 
By this choice of co-ordinates, there is no extension or contraction in the $y$-direction. 
Equivalently, $\psi_e=0$ where $\psi$ is the azimuthal angle (figure~\ref{fig:schematic}).

\subsubsection{Linear equations} \label{sec:perturbation-equations}

We now substitute the linear decompositions from equations~(\ref{eq:lin_decomposition}--\ref{eq:D0_def}) into the governing equations~(\ref{eq:mass_s_summary}--\ref{eq:chemistry_summary}) and neglect any quadratic terms. 
The vorticity equation~(\ref{eq:Momentum_summary}) becomes 
\begin{equation} \label{eq:Momentum_pert1}
0 =  \nabla \times \left[ 2\mathrm{\textbf{D}}_0  \cdot \nabla \eta' + \eta_0 \nabla^2 \boldsymbol{u}'  \right], \quad \nabla \cdot  \boldsymbol{u}' = 0, 
\end{equation}
Here, we exploited the fact that the background compaction rate is negligible, which means that the effect of variations in bulk viscosity turn out to be consequently small (we discuss the effect of variation in bulk viscosity in \citet{reesjones2018-jfm} so do not elaborate further here).

Equation~(\ref{eq:Momentum_pert1}) motivates the introduction of a scalar potential $\psi'$ defined by
\begin{equation} \label{eq:psi_pert_def}
\nabla \psi' =  \left[ 2\mathrm{\textbf{D}}_0  \cdot \nabla \eta' + \eta_0 \nabla^2 \boldsymbol{u}'  \right],
\end{equation}
which (since $ \nabla \cdot  \boldsymbol{u}' = 0$) must satisfy a Poisson equation
\begin{equation} \label{eq:psi_pert_poisson}
\nabla^2 \psi' =   2\mathrm{\textbf{D}}_0  : \nabla \nabla \eta'. 
\end{equation}
Then we introduce a scaled, dimensionless potential 
\begin{equation} \label{eq:psi_poisson_scaled}
\widetilde{\psi}' = \frac{\psi'}{-2\lambda^* \eta_0 \dot{\gamma_0} }, \quad \Rightarrow \quad  \nabla^2 \widetilde{\psi}' =   \widetilde{\mathrm{\textbf{D}}}_0  : \nabla \nabla \phi', 
\end{equation}
where the implication follows using equations~\eqref{eq:eta_prime_def} and \eqref{eq:D0_def}. 
 
The remaining equations can be rewritten by substituting the constitutive laws and eliminating the Darcy velocity.
An expression for the perturbed Darcy velocity is given in appendix~\ref{app:c_eq} in equation~\eqref{eq:Darcy_pert-A-2}.
The remaining equations become
\begin{align}
&\frac{\partial \phi'}{\partial t} = \Gamma' + \mathcal{C}'. \label{eq:mass_s_pert_final} \\
&0 =(1-\delta^2\nabla^2) \mathcal{C}' + nw_0 \phi'_z +\Lambda  \widetilde{\mathrm{\textbf{D}}}_0  : \nabla \nabla \phi'  \label{eq:compaction_pert_final}  \\
& nw_0 \phi' - \delta^2   \mathcal{C}'_z +\Lambda   \widetilde{\psi}'_z = \frac{1}{\beta} \left \{  \alpha + \frac{\phi_0 w_0}{R} \partial_z - \frac{\phi_0 D_X}{R} \nabla^2 \right\}\Gamma', \label{eq:chemistry_pert_final}
\end{align}
where a subscript $_z$ is used to denote a partial derivative with respect to $z$. 
Two important parameters emerge:
\begin{align} 
& \delta = \sqrt{ K_0\left(\tfrac{4}{3}\eta_0 + \zeta_0 \right) },  \label{eq:delta_def} \\
& \Lambda  =2 K_0 \eta_0 \lambda^* \dot{\gamma}_0, \label{eq:Lambda_def}
\end{align}
where $\delta$ is the compaction length \citep{mckenzie84} and 
$\Lambda$ controls how much shear-driven compaction arises from a porosity perturbation.
The ratio $ \Lambda/\delta^2 $ determines the growth rate of the shear-driven instability (section~\ref{sec:shear_only}).
This ratio was first identified by \citet{stevenson89}, in which it is written $\sigma_m$.

\subsubsection{Normal modes} \label{sec:normal_modes}
We now look for normal mode solutions of the form
\begin{align}
&\phi'=\tilde{\phi}\exp(i \boldsymbol{k}\cdot \boldsymbol{x} +\sigma t), \nonumber \\
&\widetilde{\psi}'=\tilde{\psi}\exp(i \boldsymbol{k}\cdot \boldsymbol{x} +\sigma t), \nonumber \\
&\mathcal{C}'=\tilde{\mathcal{C}}\exp(i \boldsymbol{k}\cdot \boldsymbol{x} +\sigma t), \nonumber \\
&\Gamma'=\tilde{\Gamma}\exp(i \boldsymbol{k}\cdot \boldsymbol{x} +\sigma t),  \nonumber
\end{align}
where $\sigma$ is the growth rate of the instability, $\boldsymbol{k}=[k_x,k_y,k_z]$ is the wavevector, $\boldsymbol{x}=[x,y,z]$ is the position vector, and the prefactors are constants. We define $k$ as the magnitude of $\boldsymbol{k}$, so $k^2 = k_x^2+k_y^2+k_z^2$. The co-ordinate system is shown in figure~\ref{fig:schematic}.

We define the function 
\begin{equation} \label{eq:G_definition}
G(\boldsymbol{k};\theta_e) = k^{-2} \left[(k_x^2-k_z^2)\cos(2\theta_e) + 2k_xk_z\sin(2\theta_e)\right],
\end{equation}
such that equation~(\ref{eq:psi_poisson_scaled}) becomes $\tilde{\psi} = \tilde{\phi} G$. The use of a factor $k^{-2}$ in the definition ensures that $G$ is independent of the wavenumber (magnitude of the wavevector) and only depends on the direction. 
$G$ has a maximum value of $+1$ when $k_y=0$ and $\boldsymbol{k}$ is in the direction of maximum extension. 
$G$ has a minimum value of $-1$ when $k_y=0$ and $\boldsymbol{k}$ is perpendicular to the direction of maximum extension.
Then equations~(\ref{eq:mass_s_pert_final}--\ref{eq:chemistry_pert_final}) become
\begin{align}
&\sigma \tilde{\phi} = \tilde{\Gamma} + \tilde{\mathcal{C}}. \label{eq:mass_s_pert_final_hat} \\
&\tilde{\mathcal{C}}  = \frac{  \Lambda  G k^2 - nw_0 ik_z}{1+\delta^2 k^2} \tilde{\phi}  \label{eq:compaction_pert_final_hat}  \\
& \frac{\widetilde{\alpha} }{\beta}  \tilde{\Gamma} = \left( nw_0 +  ik_z \Lambda G \right) \tilde{\phi} - ik_z  \delta^2 \tilde{\mathcal{C}} , \label{eq:chemistry_pert_final_hat}
\end{align}
where
\begin{equation} \label{eq:alpha_tilde_def}
\widetilde{\alpha} = \alpha + \frac{\phi_0 w_0}{R} ik_z + \frac{\phi_0 D_X}{R} k^2  
\end{equation}
is an extended version of the inverse reactivity of the system, which is augmented by advection and diffusion of the undersaturated chemical species.

Finally, we substitute equation~\eqref{eq:chemistry_pert_final_hat} and then equation~\eqref{eq:compaction_pert_final_hat} into equation~\eqref{eq:mass_s_pert_final_hat} to obtain an expression for the combined growth rate of shear and reactive instabilities
\begin{align} \label{eq:dispersion_master}
&\sigma = \frac{\beta}{\widetilde{\alpha}} \left( nw_0 +  ik_z \Lambda G \right) + \left(1-\frac{\beta ik_z  \delta^2}{ \widetilde{\alpha}} \right) \frac{  \Lambda  G k^2 - nw_0 ik_z}{1+\delta^2 k^2}.
\end{align}
In subsequent sections, we explore the nature of this equation in detail and discuss the physical significance of the terms that appear in it. 
First, we relate it to previous studies of the reactive- and shear-driven instabilities in isolation.

\subsection{Shear-driven instabilities} \label{sec:shear_only}
The reactive part of the growth rate can be eliminated by setting $\beta = 0$. 
This ensures that $\tilde{\Gamma}=0$, so the reactive part of the contribution to porosity change is eliminated and we are left with the part coming from shear. 
Then equation~(\ref{eq:dispersion_master}) becomes 
\begin{align} \label{eq:dispersion_shear}
&\sigma =   \frac{  \Lambda  G k^2 - nw_0 ik_z}{1+\delta^2 k^2}.
\end{align}
The term involving $w_0$ arising from buoyancy-driven melt flow is typically neglected because buoyancy is unimportant in laboratory experiments that impose a rapid shear \citep{spiegelman03b}.
In any case, it only affects the imaginary part of the growth rate, giving rise to compaction waves.
The dependence on the compaction length $\delta$ in equation~\eqref{eq:dispersion_shear} means that when $\delta k \ll 1$ (the wavelength of the instability is much greater than the compaction length), then the growth rate approaches zero. 
Conversely, when $\delta k \gg 1$, the real part of the growth rate can be approximated
\begin{align} \label{eq:dispersion_shear_limit}
&\mathrm{real}(\sigma) \approx   \frac{  \Lambda  G  }{\delta^2 } =  \frac{2 \lambda^* \dot{\gamma}_0 }{\tfrac{4}{3} + \tfrac{\zeta_0}{\eta_0}  } G.
\end{align}
This motivates us to define
\begin{equation}
\sigma_\mathrm{shear} =  \frac{  \Lambda    }{\delta^2 } = \frac{2 \lambda^* \dot{\gamma}_0 }{\tfrac{4}{3} + \tfrac{\zeta_0}{\eta_0}  },
\end{equation}
which is the dimensional growth rate associated with shear \citep{stevenson89}.
Note that the real part of the growth rate has a distinguished direction, namely the direction of extension in the $x$-$z$ plane. 
The imaginary part comes from vertical background magma flow, which drives the perturbations upward as waves. So the vertical is an additional distinguished direction in this case.
The growth rate and angular dependence through equation~\eqref{eq:G_definition} are known from previous studies \citep[e.g.,][]{spiegelman03b, katz06}.

\subsection{Reaction-driven instabilities}  \label{sec:reaction_only}
The shear part of the flow can be eliminated by taking $\Lambda  = 0$. 
Then equation~(\ref{eq:dispersion_master}) becomes 
\begin{align} \label{eq:dispersion_reaction}
&\sigma =\frac{1}{1+\delta^2 k^2} \left[\frac{\beta nw_0}{\widetilde{\alpha}} \left( {1+\delta^2( k_x^2+k_y^2) }   \right)   - {nw_0 ik_z} \right].
\end{align}
This can be shown to be equivalent to \citet{reesjones2018-jfm} under the same assumption ($\delta k \gg 1$) mentioned previously. 
For now, note that if we additionally make the  assumption $\widetilde{\alpha} \approx \alpha$ (valid when the reaction rate $R$ is very fast), then  
\begin{align} \label{eq:dispersion_reaction_limit}
&\mathrm{real}(\sigma)  \approx \frac{\beta nw_0}{\alpha} \frac{k_x^2+k_y^2 }{k^2}      .
\end{align}
The growth rate has a cylindrical symmetry about the vertical direction.
The maximum growth rate
\begin{equation}
\sigma_\mathrm{reaction} = n \beta w_0/\alpha, 
\end{equation}
occurs when $k_z=0$, since $k_x^2 + k_y^2 = k^2 -k_z^2$. 
Thus the channels formed by the reaction-infiltration instability are vertical \citep{reesjones2018-jfm}. 
Moreover, we show in appendix~\ref{app:c_eq} that the preferred vertical orientation holds even when we consider the full pressure-dependence of the solubility gradient (at least in the context of linearised analysis).

In this section we have shown that the full dispersion relation~(\ref{eq:dispersion_master}) includes, as special cases, the results of previous studies on  both the shear-driven instability and the reaction-infiltration instability.

\section{Results: local analysis of combined instability} \label{sec:results_local}

In this section we analyse controls on the growth rate $\sigma$ that are relevant to the case of an infinite domain with a uniform base state. 

\subsection{Equilibrium dynamics at large compaction length} \label{sec:equilibrium-dynamics-large-deltak}
The simplest version of the instability involving both reaction and shear can be illustrated by considering an important limit of equation~(\ref{eq:dispersion_master}).
In particular, if the reaction rate is very fast, the system is driven to equilibrium and ${\widetilde{\alpha}} \approx {\alpha}$. 
If also we consider the short-wavelength or large-compaction-length limit discussed earlier  ($\delta k \gg 1$), then
\begin{align} \label{eq:dispersion_master_limit_1}
&\sigma = \frac{\beta}{{\alpha}} \left( nw_0 +  ik_z \Lambda G \right) + \left(1-\frac{\beta ik_z  \delta^2}{ {\alpha}} \right) \frac{  \Lambda  G k^2 - nw_0 ik_z}{\delta^2 k^2},
\end{align}
so
\begin{align} \label{eq:dispersion_master_limit_1_real}
&\mathrm{real}(\sigma) = \frac{\beta n w_0}{{\alpha}} \left(1-\frac{k_z^2}{k^2}   \right) +  \frac{\Lambda }{\delta^2} G , \nonumber \\
& \qquad \quad = \sigma_\mathrm{reaction}  \left(1-\frac{k_z^2}{k^2}   \right) +  \sigma_\mathrm{shear}  G.
\end{align}
It is important to note that all the terms involving the wavevector are independent of its magnitude; they depend only on its direction.
This is a feature of the particular limit considered that (by design) neglects the role of advection, diffusion and compaction in affecting the wavelength.  We consider these controls later. 

In this particular limit, the behaviour is controlled by the ratio of the growth rate of shear-driven to reaction-driven instabilities, which we can write
\begin{equation} \label{eq:S_def}
S = \frac{ \sigma_\mathrm{shear} }{\sigma_\mathrm{reaction}} = \frac{2 \lambda^* \alpha  }{n \beta \left(\tfrac{4}{3} + \tfrac{\zeta_0}{\eta_0} \right)  } \frac{\dot{\gamma}_0 }{w_0},
\end{equation}
where we grouped together material parameters separately to the ratio \mbox{${\dot{\gamma}_0 }/{w_0}$}. The former group might be expected to be roughly constant, provided the bulk-to-shear viscosity ratio is constant, whereas the latter will vary spatially at a mid-ocean ridge. 

\subsection{Three-dimensional effects and the orientation of porosity bands} \label{sec:orientation}
The reactive mode of instability has a cylindrical symmetry (there is no difference between the $x$ and $y$ direction; the only distinguished direction is the vertical $z$). 
Numerical calculations show that the instability leads to the formation of cylindrical, high-porosity conduits (M.~Spiegelman, unpublished work),
in accordance with laboratory experiments \citep{pec15,pec17}.
However, the shear-driven mode of instability leads to the formation of high-porosity sheets, the orientation of which depends on the direction $\theta_e$ of maximum rate of extension.
The coordinate system is aligned such that shear is in the $x$--$z$ plane, with the direction of extension having an azimuthal angle $\psi_e=0$ to that plane.  
Hence porosity sheets extend parallel to the $y$-direction. In cross-section on the $x$--$z$ plane, the sheets appear as high-porosity bands (which is how laboratory experiments are typically presented).
Here, we investigate the combined effect of the reactive and shear mechanisms in light of the fact that that this combination has two distinguished directions: the orientation of the shear flow and the direction of the solubility gradient (vertical).

We work in terms of modified spherical polar co-ordinates $(\theta,\psi)$, shown in figure~\ref{fig:schematic}a,  where \mbox{$-\pi/2 \leq  \theta \leq \pi/2$} is the inclination angle and \mbox{$0\leq \psi < 2\pi$} is the azimuthal angle. 
This definition of $\theta$ is consistent with that of $\theta_e$ (figure~\ref{fig:schematic}b). 
However, it is not the same as the angle of the porosity bands as it was defined in, for example, \citet{spiegelman03b}. 
Rather, as shown in figure~\ref{fig:schematic}a, the porosity bands are normal to the wavevector that makes an angle $\theta$ to the $x$--$y$ plane. In the particular case that we restrict attention to the $x$--$z$ plane ($\psi =0$), the angle of the porosity bands according to the definition of \citet{spiegelman03b} is $\pi/2-\theta$.
 
In the modified spherical polar co-ordinate system, 
\begin{align*} \label{eq:wavenumber_spherical}
&k_x = k\cos(\theta) \cos(\psi), \\
&k_y = k\cos(\theta) \sin(\psi), \\
&k_z = k\sin(\theta).
\end{align*}
Then equation~\eqref{eq:dispersion_master_limit_1_real} becomes
\begin{equation} \label{eq:sigma_ratio}
\frac{\mathrm{real}(\sigma)}{\sigma_\mathrm{reaction}} =  \cos^2 \theta +  S G(\theta,\psi),
\end{equation}
where, from equation~\eqref{eq:G_definition},
\begin{equation}
G=\left(\cos^2\theta\cos^2\psi -\sin^2\theta \right)\cos 2\theta_e +\cos\psi \sin2\theta \sin2\theta_e.
\end{equation}
For the particular case that $\psi =0$, 
\begin{equation}
G=\cos 2(\theta - \theta_e),
\end{equation}
since the growth rate only depends on the wavevector orientation relative to the direction of extension.
Thus $G=1$ when the wavevector is parallel to the direction of extension and $G=-1$ when they are perpendicular. 

\begin{figure*}
\centering\noindent\includegraphics[width=1.0\linewidth]{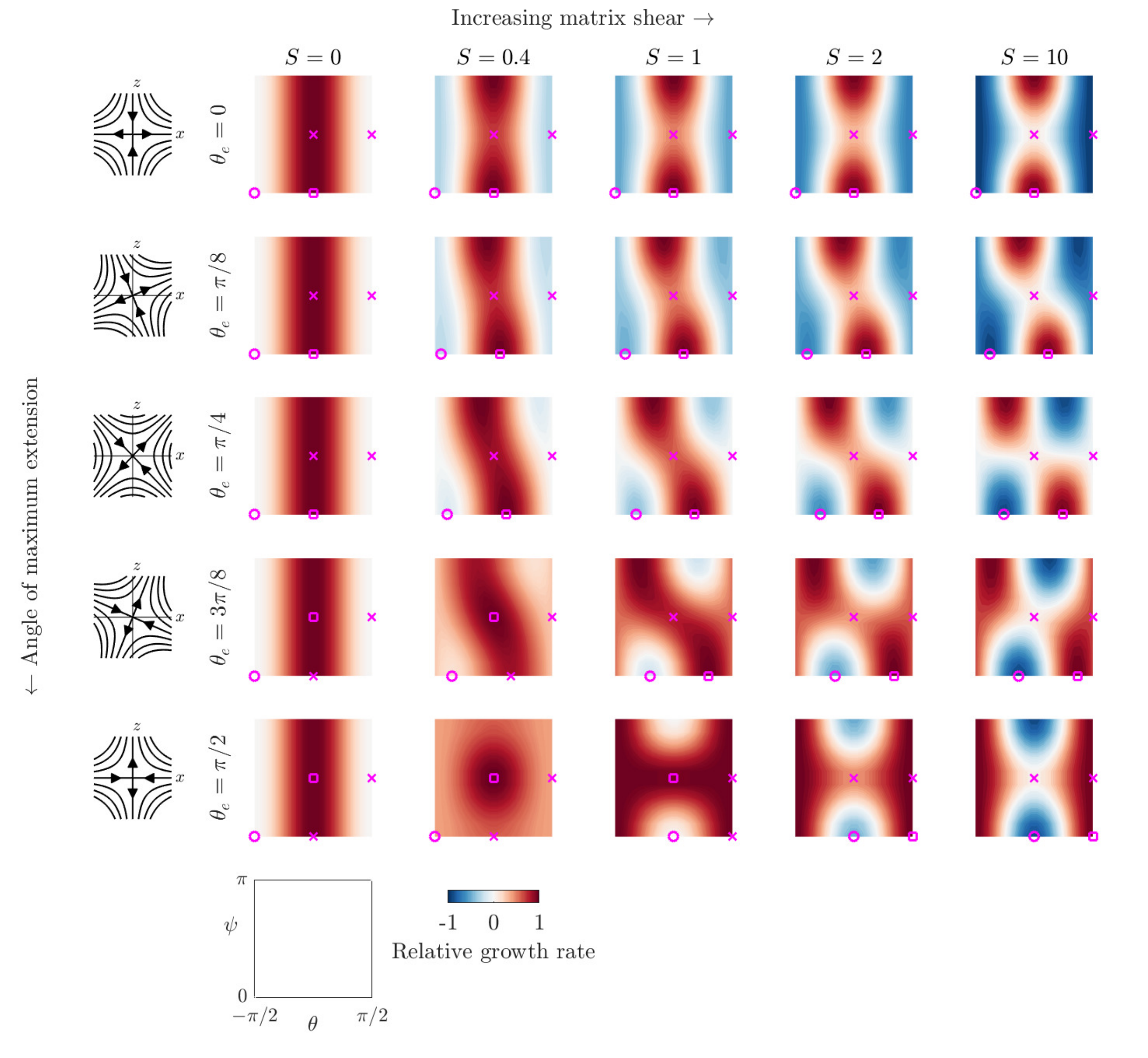}
   \caption{Growth rate from equation~\eqref{eq:sigma_ratio} in the range $-\pi/2<\theta<\pi/2$ (horizontal direction) and $0<\psi<\pi$ (vertical direction). These axis limits are shown in the labelled diagram on the bottom row. The colour scale shows the normalized growth rate relative to the maximum (growth is red, no growth is white, decay is blue). Crosses represent saddle points, squares represent maxima, circles represent minima. Icons on the left-most column indicate the orientation of the background solid shear flow.}
   \label{figure-theta-psi}
\end{figure*}

\begin{figure}
\centering\noindent\includegraphics[width=1.0\linewidth]{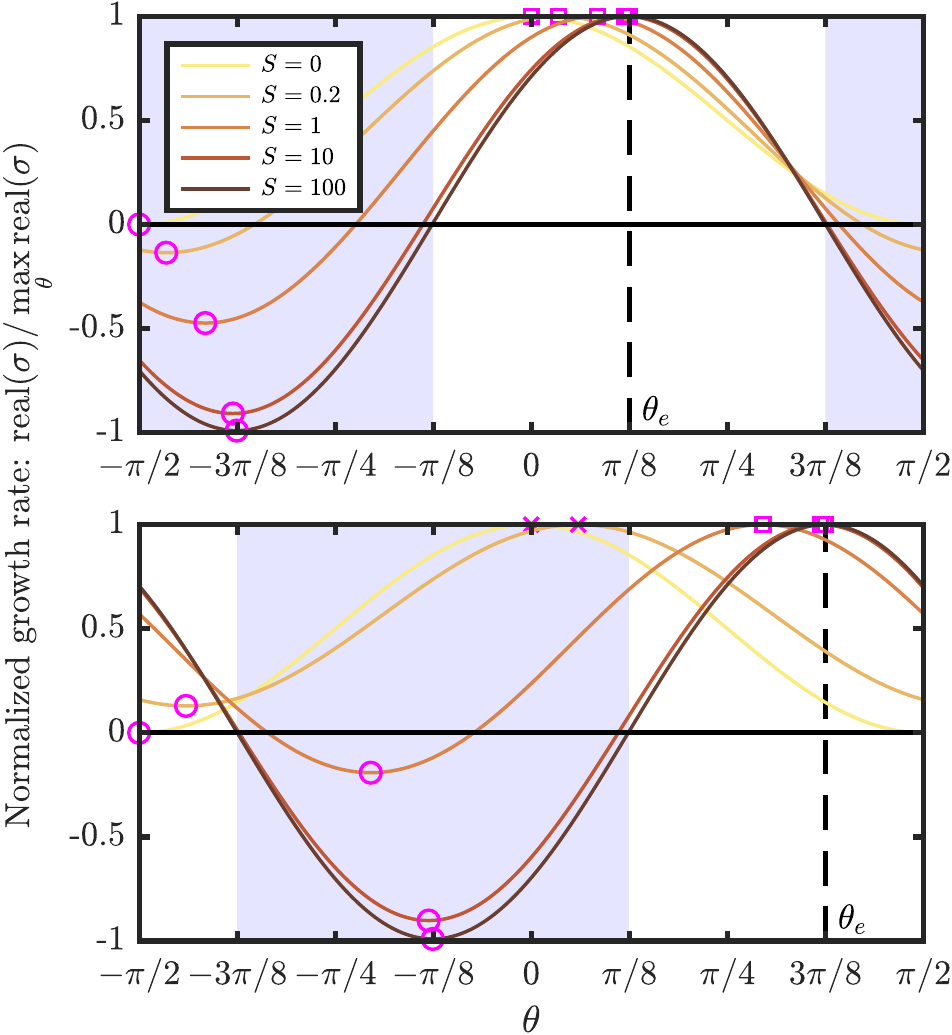}
   \caption{Cross section of the normalized growth rate from figure~\ref{figure-theta-psi} for modes restricted to the $x$--$z$ plane (i.e., $k_y=0$) for $\theta_e = \pi/8$ (upper) and $\theta_e = 3\pi/8$ (lower). Note that the normalization is relative to the maximum growth rate for this restricted set of modes, a distinction only relevant where the stationary point is a saddle point (crosses) rather than a maximum (squares).  }
   \label{figure-x-z-plane}
\end{figure}

Figure~\ref{figure-theta-psi} shows the growth rate $\text{real}(\sigma)$ from equation~(\ref{eq:sigma_ratio}) as a function of $\theta$ ($x$-axis) and $\psi$ ($y$-axis).
This function has four stationary points; we consider each in turn.
(\textit{i}) There is one stationary point at $\theta=\psi=\pi/2$, which is always a saddle point. This is a mode with a wavevector purely in the $z$-direction. Therefore it represents a tabular perturbation aligned with the $x$--$y$ plane.
(\textit{ii}) There is another stationary point when $\theta=0$ and $\psi=\pi/2$. 
This is a mode with a wavevector purely in the $y$-direction, that is, geometrically, a tabular feature in the $x$--$z$ plane.
The nature of this stationary point depends on the magnitude of the shear. 
When $S$ exceeds a critical value $S>S_c\equiv-\cos 2\theta_e$, it is a saddle.
However, when $S< S_c$, this stationary point is a maximum. 
(\textit{iii/iv}) There are two more stationary points with $\psi=0$ (so $k_y=0$) and 
\begin{equation} \label{eq:theta_ky=0}
\tan 2 \theta = \frac{2 S \sin 2\theta_e}{1+2S\cos 2\theta_e}.
\end{equation}
Figure~\ref{figure-x-z-plane} shows these modes, which are tabular features that have a wavevector in the $x$--$z$ plane at an angle $\theta$; they extend in the $y$-direction. 
Equation~(\ref{eq:theta_ky=0}) has two roots, one in the domain $-\pi/2\leq\theta<0$  and another in the domain $0\leq\theta<\pi/2$.
The former root is associated with contractional shear stress and is always a minimum.
The latter root is associated with extensional shear stress and is a maximum when $S$ exceeds the aforementioned critical value $S>S_c\equiv-\cos 2\theta_e$ and a saddle when $S<S_c$. 

\begin{figure*}
\centering\noindent\includegraphics[width=1.0\linewidth]{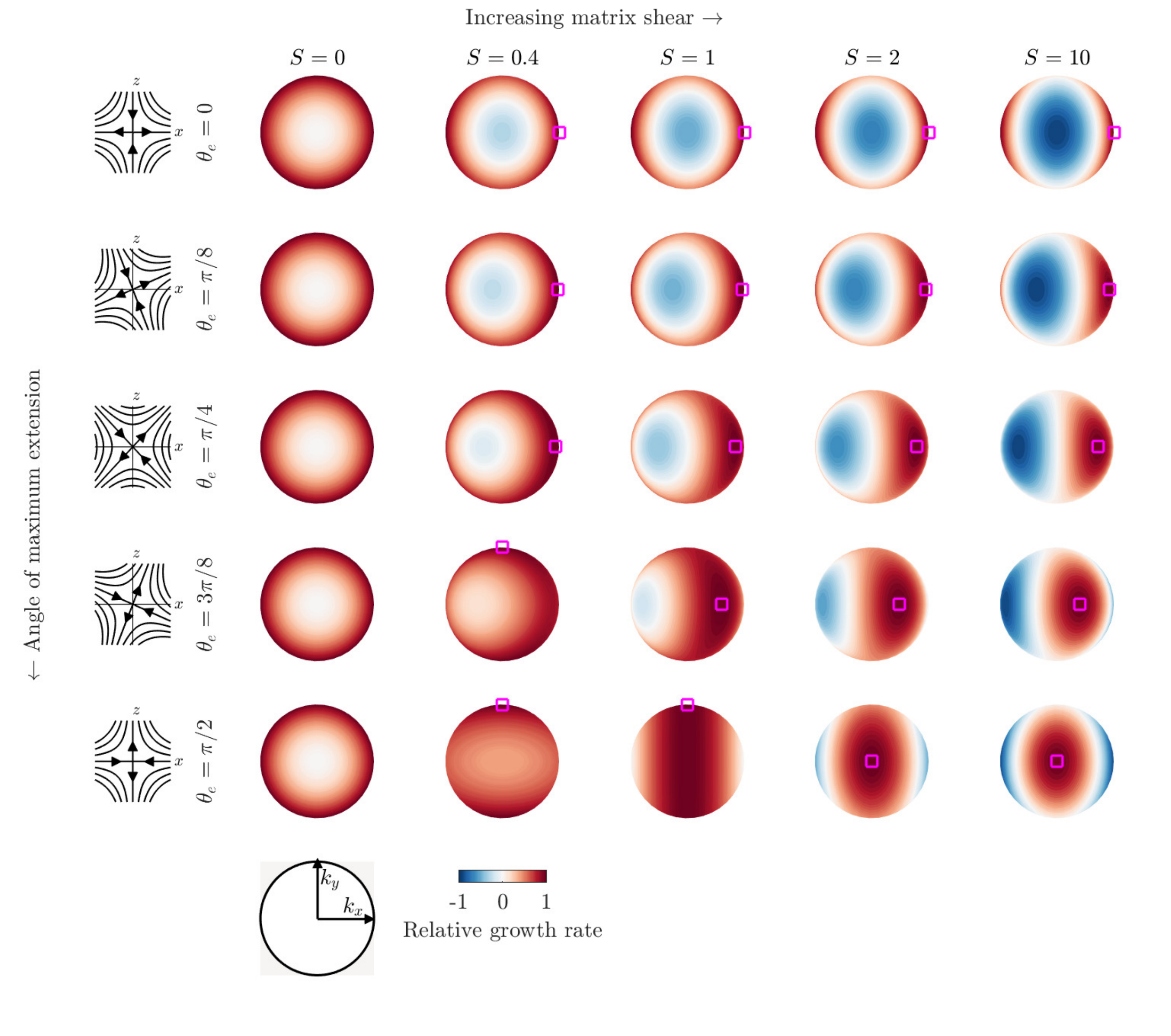}
   \caption{Growth rate from equation~\eqref{eq:sigma_ratio} projected on to the $k_x$--$k_y$ plane, as indicated on the circular icon on the bottom. The growth rate is calculated as a function of $(\theta,\psi)$ and then plotted against $[k_x(\theta,\psi),k_y(\theta,\psi)]$. The result in independent of the magnitude of the wavevector. The colour scale shows the normalized growth rate relative to the maximum (red is growth; blue is decay; white is neutral). Squares represent maxima. Squares are not shown for the column $S=0$; in this case, the maximum growth rate is achieved on the circle $k_x^2+k_y^2=k^2$. Icons on the left-most column indicate the orientation of the background solid shear flow.}
   \label{figure-kx-ky}
\end{figure*}

We expect the most unstable mode (i.e., the one that grows fastest) to be dominantly expressed in a full solution of the governing equations and hence we analyse what selects this mode.
Figure~\ref{figure-kx-ky} shows that there is a transition for a horizontally isotropic (no difference between $x$ and $y$ directions) to an anisotropic growth rate as the shear rate $S$ increases, moving from left to right in the figure.  
However, the transition depends on the angle of maximum extension in the shear flow (from top to bottom in the figure). 
If $\cos 2\theta_e \geq 0$, which corresponds to the rows with $\theta_e=0,\,\pi/8,\,\pi/4$, the transition is immediately to a state where $k_y=0$.
Geometrically, when $\cos 2\theta_e \geq 0$, the angle of maximum extension is within $\pi/4$ of the horizontal, which means that the horizontal direction is extensional and the vertical direction is contractional, as shown by the mantle flow plots in the left-most column  of figure~\ref{figure-kx-ky}.
This leads to tabular features that are orientated vertically for small $S$, because reaction dominates the instability and promotes vertical features. 
The tabular features approach an orientation perpendicular to the angle of maximum extension as $S$ increases, because this orientation is favoured by the shear-driven instability.
However, if $\cos 2\theta_e<0$, which corresponds to the rows with $\theta_e=3\pi/8,\pi/2$, 
the situation is reversed: the $x$-direction is contractional and the vertical direction is extensional, as shown by the mantle flow plots in the left-most column  of figure~\ref{figure-kx-ky}.. 
Then for small $S$, the most unstable wavevector orientation has $k_x=0$, $k_z=0$ (i.e., a  tabular feature aligned with the $x$--$z$ plane) due to the combined effect of reaction and shear. 
As in previous cases, it is reaction that promotes vertical features ($k_z = 0$).
However, in contrast to previous cases, shear-driven contraction suppresses wavevectors in the $x$--direction so $k_x=0$. 
As $S$ increases through the critical value defined above  $S_c\equiv -\cos 2\theta_e$ where the shear-driven instability dominates over the reaction-driven instability, the state switches to $k_y=0$ with orientation approaching the angle of maximum extension for large shear, as before.

Mid-ocean ridges (MORs), which we will turn to in the second half of the paper, are a horizontally extensional environment, corresponding to the upper rows of figure~\ref{figure-kx-ky}. 
So the crucial implication of this figure for MORs is that even an extremely small shear $S>0$ is sufficient to break the horizontal isotropy (symmetry under any coordinate rotation about the $z$ axis) that occurs when only the reactive instability operates.
So rather than the expecting the tube-shaped channels that arise from the pure reactive instability ($S=0$), the most linearly unstable feature with shear $S>0$ are tabular bands.  
This is important, because dunite channels (interpreted as relics of high porosity channels) have a tabular morphology, as discussed in section~\ref{sec:intro}. 

\subsection{Effect of the compaction length} \label{sec:compaction_length}
We now relax the assumption that perturbation wavelength is much shorter than the compaction length. 
Then equation~\eqref{eq:sigma_ratio} generalizes to
\begin{equation} \label{eq:sigma_ratio_compaction_length}
\frac{\mathrm{real}(\sigma)}{\sigma_\mathrm{reaction}} = \frac{1+k^2 \delta^2  \cos^2 \theta}{1+k^2 \delta^2} +   \frac{S G(\theta,\psi) k^2 \delta^2}{1+k^2 \delta^2}.
\end{equation}
This has the same angular dependence as in the small-wavelength limit $k\delta \gg 1$, so all the conclusions of section~\ref{sec:orientation} still hold, including the optimal wavevector orientation. 

\begin{figure}
\centering\noindent\includegraphics[width=1.0\linewidth]{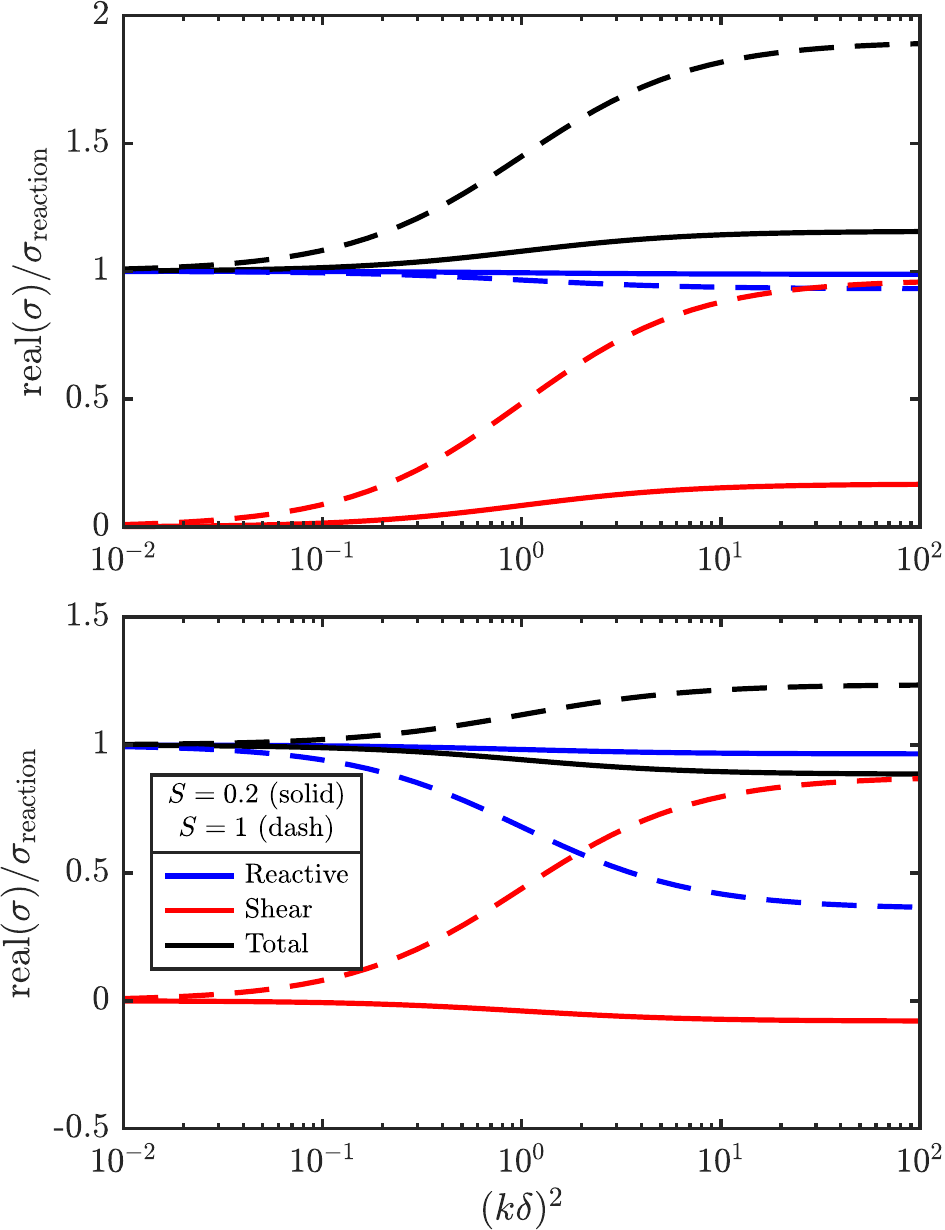}
   \caption{Dependence on normalized compaction length $k\delta$. As in figure~\ref{figure-x-z-plane}, the upper panel shows $\theta_e=\pi/8$ and the lower  panel shows $\theta_e=3\pi/8$. In both panels, we plot equation~\eqref{eq:sigma_ratio_compaction_length} for $\psi=0$ and $\theta$ chosen to maximize the growth rate according to equation~\eqref{eq:theta_ky=0}. }
   \label{figure-compaction-length}
\end{figure}

Figure~\ref{figure-compaction-length} shows how the results depend on the compaction length. It plots the normalised growth rate as a function of $k\delta$ for perturbations orientated with $\psi=0$ and $\theta$ chosen to maximize the growth rate.
As in figure~\ref{figure-x-z-plane}, there are two regimes depending on whether the $z$-direction is in extension or contraction.  
The upper panel shows the case of $\theta_e=\pi/8$, which has contraction in the $z$-direction.
Here, the contributions to the growth rate from reaction and shear vary oppositely with the compaction length. 
The reactive contribution, which is the first term on the right-hand-side of equation~\eqref{eq:sigma_ratio_compaction_length}, decreases slightly with $k\delta$. 
The amount of decrease depends indirectly on $S$, because for larger values of $S$, the angle $\theta$ that maximises the overall growth rate increases from 0 towards $\theta_e$.  This decreases $\cos^2 \theta$ in the first term of \eqref{eq:sigma_ratio_compaction_length}. By contrast, the shear-driven contribution, which is the second term on the right-hand-side of equation~\eqref{eq:sigma_ratio_compaction_length}, increases significantly with $k\delta$, starting from zero when $k\delta \ll 1$.
Thus the total growth rate is dominated by reaction when $k\delta$ is small, but there can be a cross-over as $k\delta$ increases. 
This cross-over depends on $S$. 
Indeed when $S$ is very small (e.g., the solid curve for $S=0.2$) reaction always dominates.
Physically, the shear-driven instability is suppressed when the wavelength is comparable to the compaction length (section~\ref{sec:shear_only}), while the reaction-driven instability is not directly affected by compaction in the equilibrium limit ($\widetilde{\alpha}=\alpha$). 
Disequilibrium effects introduce a direct dependence on the compaction length \citep{reesjones2018-jfm}.

The lower panel shows the case of $\theta_e=3\pi/8$, which has contraction in the $x$-direction. The behaviour is different when $S<S_c= -\cos 2\theta_e$, as illustrated by the solid curves for $S=0.2$. 
Unlike the upper panel, the shear-driven contribution decreases with $k\delta$, starting from zero when $k\delta \ll 1$ and decreasing towards a negative value when $k\delta \gg 1$.
This is because of the contraction in the direction of the optimal wavevector, consistent with figure~\ref{figure-x-z-plane}.
Thus the total growth rate decreases with compaction length in this case.
However, for $S\geq S_c$, there is extension in the direction of the optimal wavevector, and the system behaves in the same way as in the upper panel.

\subsection{Disequilibrium effects} \label{sec:disequilibrium}
So far, we made the assumption that the reaction rate was extremely fast such that $\widetilde{\alpha}\approx \alpha$. 
We now consider the role of disequilibrium effects caused by advective or diffusive chemical transport.
We re-write equation~\eqref{eq:alpha_tilde_def} 
\begin{equation} 
\widetilde{\alpha} = \alpha \left[ 1 + \frac{\phi_0 w_0}{\alpha R \delta} i k\delta \sin(\theta) + \frac{\phi_0 D_X}{ \alpha R \delta^2} (k\delta)^2 \right]  
\end{equation}
The two dimensionless parameter groups
\begin{align}
&\mathrm{Da}_w= \frac{\alpha R \delta}{\phi_0 w_0}, \quad 
 \mathrm{Da}_D =\frac{ \alpha R \delta^2}{\phi_0 D_X}, 
 \end{align}
are Damk{\"o}hler numbers that express the reaction rate relative to advective and diffusive transport, respectively. 
These are typically extremely large and hence the system is typically close to equilibrium \citep{aharonov95,spiegelman01,reesjones2018-jfm}.
Physically, advection of undersaturation is negligible when $\mathrm{Da}_w \rightarrow \infty$ and diffusion is negligible when $\mathrm{Da}_D \rightarrow \infty$.

We first restrict attention to the role of advection by taking the limit $\mathrm{Da}_D \rightarrow \infty$ so 
\begin{equation} 
\widetilde{\alpha} = \alpha \left[ 1 + i\mathrm{Da}_w^{-1}  k\delta \sin(\theta)  \right].  
\end{equation}
We substitute this expression into equation~\eqref{eq:dispersion_master} and find
\begin{align} \label{eq:dispersion_master_Da_w}
(1+\delta^2 k^2) \frac{ \mathrm{real}(\sigma)} {\sigma_\mathrm{reaction} } &= 
\frac{1+k^2 \delta^2  \cos^2 \theta}{1+\mathrm{Da}_w^{-2} (k \delta)^2 \sin^2\theta } \nonumber \\
&+\frac{ S_r  G(\theta,\psi) k^2 \delta^2  \sin^2 \theta}{1+\mathrm{Da}_w^{-2} (k \delta)^2 \sin^2\theta } \nonumber \\
&+   {S G(\theta,\psi) k^2 \delta^2},
\end{align}
where 
\begin{equation}
S_r =S  {\beta \phi_0 w_0}/{\alpha^2 R}.
\end{equation}
We make the approximation \mbox{$1+\mathrm{Da}_w^{-2} \sin^2\theta  k^2 \delta^2 \approx 1$} and also take the limit $\delta k \gg 1$.
Then equation~\eqref{eq:dispersion_master_Da_w} simplifies to
\begin{equation} \label{eq:dispersion_master_Da_w_high_delta}
 \frac{ \mathrm{real}(\sigma)} {\sigma_\mathrm{reaction} } = 
{ \cos^2 \theta}\nonumber \\
+   {S G(\theta,\psi)}\nonumber \\
+{ S_r  G(\theta,\psi)  \sin^2 \theta}.
\end{equation}
This expression can be compared to equation~\eqref{eq:sigma_ratio} and shows that the advection of undersaturated melts in the presence of shear causes an additional contribution to the growth rate. 
The contribution is proportional to the shear growth rate, but much smaller, because ${\beta \phi_0 w_0}/{\alpha^2 R} \ll 1$, so $S_r \ll S$. 
Physically, this small contribution comes from the shear-induced perturbation melt flow advecting against the background equilibrium concentration gradient combined with the background melt flow advecting the perturbed undersaturation [see left-hand-side of equation~\eqref{eq:chemistry_summary}]. 
Although this contribution is small, it is physically interesting because it arises from the coupling of the shear and reactive instability, rather than from the sum of their separate rates.

We next turn attention to the role of diffusion by taking the opposite limit $\mathrm{Da}_w \rightarrow \infty$ so 
\begin{equation} 
\widetilde{\alpha} = \alpha \left[ 1 + \mathrm{Da}_D^{-1}  (k\delta)^2  \right] = \alpha \left[ 1 + \frac{\phi_0 D_X k^2}{\alpha R}  \right].  
\end{equation}
In the second equality we used $\mathrm{Da}_D^{-1} (k \delta)^2 = {\phi_0 D_X k^2}/{ \alpha R }$, which depends on the wavelength but is independent of the compaction length.
We substitute this expression into equation~\eqref{eq:dispersion_master} and find
\begin{align} \label{eq:dispersion_master_Da_D}
(1+\delta^2 k^2) \frac{ \mathrm{real}(\sigma)} {\sigma_\mathrm{reaction} } = 
\frac{1+k^2 \delta^2  \cos^2 \theta}{1+(\phi_0 D_X /\alpha R) k ^2  }
+   {S G(\theta,\psi) k^2 \delta^2},
\end{align}
Again, in the limit $\delta k \gg 1$, equation~\eqref{eq:dispersion_master_Da_D} simplifies to
\begin{equation} \label{eq:dispersion_master_Da_D_high_delta}
 \frac{ \mathrm{real}(\sigma)} {\sigma_\mathrm{reaction} } = 
\frac{{ \cos^2 \theta}}{1+(\phi_0 D_X /\alpha R) k ^2 }
+   {S G(\theta,\psi)}.
\end{equation}
In this case there is no new contribution to the growth rate; the only change is that the reactive contribution to growth is slightly reduced, especially at high wavenumber (because diffusion acts at small length scales), as discussed in \citet{reesjones2018-jfm}. 

In the next section we embed the above considerations of perturbation growth into the background, large-scale flow beneath a mid-ocean ridge. 
We neglect both disequilibrium effects and also the role of compaction length (equivalently, the role of the wavenumber).
Figure~\ref{figure-compaction-length} indicates that the growth rate depends weakly on wavenumber provided $(k\delta)^2 \gtrsim 100$, or equivalently provided the wavelength is less than about $\delta\times 10/2\pi \approx 1.6$~km, for a compaction length $\delta\approx 1$~km.
Given that observed dunite channels are smaller than this \citep{braun02}, we make the simplifying assumption that $k\delta \gg 1$.

\section{Methods: growth of instabilities at MOR\MakeLowercase{s}} \label{sec:methods_MOR}
At mid-ocean ridges (MORs), plate spreading drives a circulation of the upper mantle.
This is a viscous shear flow and so could promote the formation of shear-driven porosity bands. 
The upwelling of the mantle beneath the ridge causes decompression melting in a roughly triangular region down to a depth of around 80~km \citep{langmuir92}.
The resulting melt upwells due to its buoyancy.
Over this depth (and hence pressure) range, there is a gradient in the equilibrium chemistry of magma that can drive reactive melting \citep{kelemen92,kelemen95,longhi02}. 
Thus mid-ocean ridges are a geological setting where both shear-driven and reaction-driven porosity localisation have the potential to occur. 
In this section, we estimate the relative and combined contributions of these mechanisms. We discuss the predicted orientation of the localized features that result from perturbation growth. 

\subsection{Background state of the mid-ocean ridge}

\begin{figure*}
  \includegraphics[width=0.9\textwidth]{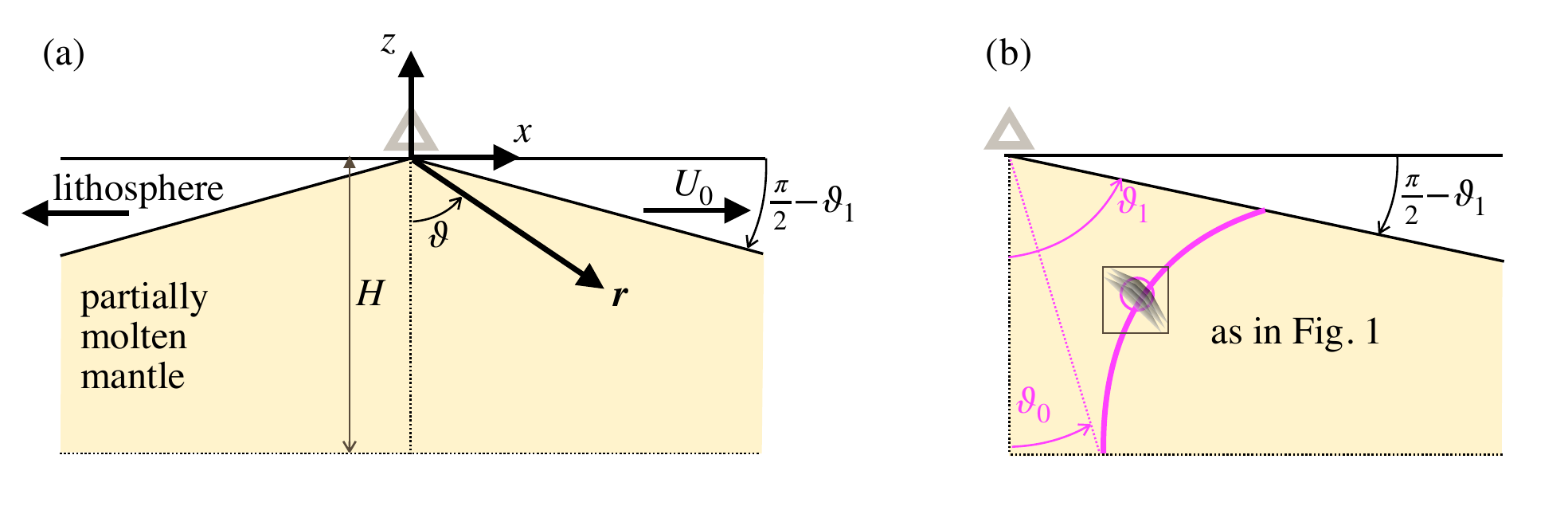}
  \caption{Sketch of mid-ocean ridge model, which is symmetric about the ridge axis. 
  (a) Coordinate system for the mid-ocean ridge model where the base of the lithosphere makes an angle $\pi/2-\vartheta_1$ from the horizontal.  The lithosphere spreads rigidly at a speed $U_0$ away from the ridge axis. The grey triangle represents the mid-ocean ridge axial volcano. This `corner-flow' geometry is taken from \citet{spiegelman87}.
  Our model is concerned with the partially-molten mantle, which is shaded yellow.
  (b) The magenta curve shows a streamline of the solid mantle corner flow, which starts from the base of the melting region at angular position $\vartheta_0$ and finishes at $\vartheta_1$. 
  We evolve a parcel of porosity bands along the streamline. 
   \label{fig:MOR_schematic}}
\end{figure*}

Our approach combines models of several aspects of mid-ocean ridges with the calculations of the growth rate of porosity bands made above.
First we estimate the background state, that is, the behaviour of the system in the absence of any small-scale localization of porosity. 
Our estimates focus on the partially molten region beneath the lithosphere, shown in figure~\ref{fig:MOR_schematic}. For present purposes, the lithosphere is defined as a rigid plate, moving uniformly away from the ridge axis at a speed $U_0$. In section~\ref{sec:CornerFlow}, we calculate the passive flow of the partially molten mantle as the response to motion of the lithosphere.
In section~\ref{sec:MagmaFlow}, we calculate the background magma flow and porosity by dividing the partially-molten region into a series of melting columns in which magma rises vertically. We assume that each of these melting columns terminates at the base of the lithosphere. 
This simple formulation for the background-state magma flow precludes capture of lateral flows associated with gradients in compaction pressure (e.g., melting-rate-pressure focusing \citep{Turner2017,Sim2020}, focusing within a decompaction channel immediately beneath the lithosphere \citep{sparks91, spiegelman93c, ghods00}). However, if melt transport throughout most of the melting region (aside from a narrow region near the base of the lithosphere) is buoyancy-driven and hence nearly vertical, our representation is valid, if approximate.
We return to this issue in section~\ref{sec:implications}.

\subsubsection{Mantle flow and strain rate}  \label{sec:CornerFlow}

We estimate the background flow by treating the mantle as a uniform, isoviscous material that flows in response to prescribed plate motion. 
This is sometimes called a passive or kinematic model. 
In treating the mantle as isoviscous, we are neglecting the variation of shear viscosity with the background porosity. 
Based on the parameters estimated subsequently ($\lambda^* =26$, $\phi_0\lesssim 2\times 10^{-3}$), this is a reasonable approximation since $e^{-26\times2\times10^{-3}}\approx 0.95$, i.e., only about a 5\% reduction in viscosity. 
Equations governing the flow of an isoviscous material in a triangular region are given by \citet{batchelor67} and applied to a mid-ocean ridge melting region by \citet{spiegelman87}. 
Here, we summarize these results. 

Figure~\ref{fig:MOR_schematic}(a) shows the co-ordinate system, centred on the ridge axis, with $x$ the horizontal distance from the ridge and $z$ the vertical distance from the surface, measured upwards, such that $z=-H$ is the bottom of the melting region. 
The corner-flow solution is most naturally expressed in polar co-ordinates $(r,\vartheta)$, where $r$ is the distance to the origin and $\vartheta$ is the angle to the downward vertical (so $\vartheta = 0$ is straight down). 
The partially molten region is triangular and extends to $\vartheta = \pm \vartheta_1 $, where \mbox{$\pi/2-\vartheta_1 $} is the dip of the bottom of the lithosphere. 

There is a separable solution in this geometry where the radial flow $u_r$ and tangential flow $u_\vartheta$ can be written 
\begin{equation}
    u_r =  - U_0  \Theta'(\vartheta) , \quad u_\vartheta = U_0  \Theta(\vartheta).
\end{equation}
The function $\Theta(\vartheta)$ that satisfies the relevant boundary conditions (that the flow is symmetric about the ridge axis and uniformly translating at speed $U_0$ in the lithosphere) is 
\begin{equation}
     \Theta(\vartheta) = \frac{\vartheta \cos \vartheta -  \sin \vartheta \cos^2 \vartheta_1  }{C},
\end{equation}
where $C=\vartheta_1  - \sin \vartheta_1  \cos \vartheta_1 $ is a constant that depends only on the geometry of the melting region.

\begin{figure*}
    \begin{center}
        \includegraphics[width=0.95\textwidth]{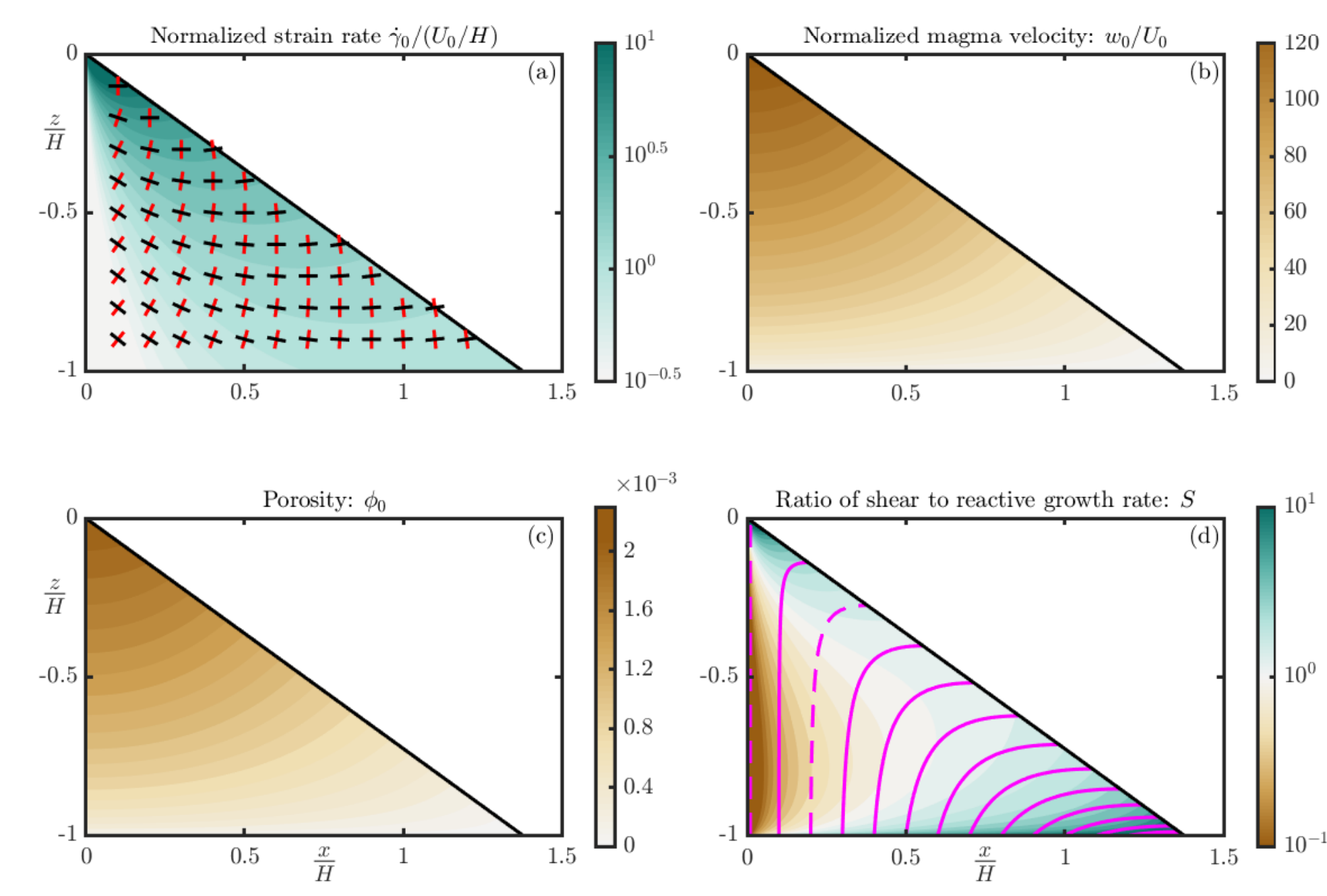}
    \caption{Background state of a mid-ocean ridge. (a) strain rate with black/red markers showing the direction of extension/contraction, respectively; (b) magma velocity; (c) porosity. Note that the maximum porosity is about $2\times 10^{-3}$, which is equivalent to 0.2\%. Panel (d) ratio of local growth rate due to shear vs reaction. Brown colours indicate where reactive instability is favoured. Green colours indicate where shear is favoured. Magenta curves show streamlines of the solid flow, along which the growth of the instability is integrated (section~\ref{sec:streamlines}). Two streamlines are dashed -- these are used as examples in figure~\ref{fig:MOR_growth_spectrum}. The parameters used are discussed in section~\ref{sec:parameters}.}
    \label{fig:MOR_base}
    \end{center}
\end{figure*}

We can convert from polar to scaled Cartesian co-ordinates 
$\tilde{z}=-\tilde{r} \cos \vartheta$, $\tilde{x} =\tilde{r}  \sin \vartheta$, where all distances are scaled by $H$, e.g., \mbox{$\tilde{r}=r/H$}.
Then the scaled solid velocity $\tilde{\boldsymbol{u}} ={  \boldsymbol{u} }/{U_0 }$ has components
\begin{equation} 
\tilde{u}_x = \frac{\tilde{x}\tilde{z}/\tilde{r}^2 - \arctan(\tilde{x}/\tilde{z})}{ C},  \quad \tilde{u}_z =  \frac{\tilde{z}^2/\tilde{r}^2 - \cos^2 \vartheta_1 }{C}.
\end{equation}
The scaled  velocity gradient tensor is
\begin{equation} \label{eq:scaled_nabla_u}
 \tilde{\nabla} \tilde{\boldsymbol{u}}  = \frac{1}{C \tilde{r}^4} 
\begin{bmatrix}
   -2 \tilde{z} \tilde{x}^2      &0 & 2\tilde{x}^3 \\
    0       & 0 & 0  \\
   -2\tilde{x} \tilde{z}^2     & 0 & 2 \tilde{z} \tilde{x}^2  
\end{bmatrix},
\end{equation}
which has a symmetric part
\begin{equation}
 \tilde{ \mathrm{ \textbf{D}} }  = \frac{1}{C \tilde{r}^4} 
\begin{bmatrix}
   -2 \tilde{z} \tilde{x}^2      &0 & \tilde{x}(\tilde{x}^2-\tilde{z}^2) \\
    0       & 0 & 0  \\
   \tilde{x}(\tilde{x}^2-\tilde{z}^2)      & 0 & 2 \tilde{z} \tilde{x}^2  
\end{bmatrix}.
\end{equation}
Then, by comparison with equation~\eqref{eq:D0_def}, we find that the dimensional strain rate is
\begin{align}
\dot{\gamma} &= \frac{U_0}{H }  \frac{1}{C  }\frac{|\tilde{x}|}{\tilde{r}^2}
= \frac{U_0}{H} \frac{1}{C  } \frac{|\sin \vartheta|}{  \tilde{r}}, \label{eq:gamma-dot-polars} 
\end{align}
and the direction of maximum extension is 
\begin{equation} \label{eq:theta_e}
\theta_e  = \vartheta - \frac{\pi}{4}\mathrm{sign}(\vartheta).
\end{equation}
Thus when $x<-z$, the angle of maximum extension is below the horizontal ($\theta_e<0$) and vice-versa when $x>-z$. 
The strain rate is zero beneath the ridge axis ($x=0$) and increases in approach to the base of the lithosphere (as $x$ increases or $r$ decreases). 
The term involving $\mathrm{sign}(\vartheta)$ in equation~\eqref{eq:theta_e} arises because $\dot{\gamma}>0$ by definition, leading to the the modulus operator in equation~\eqref{eq:gamma-dot-polars}. In cylindrical polar co-ordinates, the only non-zero components of the symmetric rate-of-strain tensor are $D_{r \vartheta}=D_{\vartheta r} = \dot{\gamma}$ (which can be seen from the separable solution in terms of $\Theta(\vartheta)$).
Figure~\ref{fig:MOR_base}(a) shows this background strain rate. 

The other information that we need for the magma flow calculations in the following section is the mantle upwelling speed $W_b$ at the base of the melting column (where $\tilde{z}=-1$ and $\tilde{r}^2=1+\tilde{x}^2$). 
Here, 
\begin{equation}  \label{eq:W_b}
W_b = U_0 \frac{ (1+\tilde{x}^2)^{-1}-\cos^2\vartheta_1  }{C}  .
\end{equation}
As expected, this equation says that the upwelling speed decreases with distance from the ridge axis.

\subsubsection{Magma flow and porosity} \label{sec:MagmaFlow}

We estimate the background magma flow and porosity by dividing the melting region into a series of one-dimensional melting columns.  The methodology for calculating the flow in a melting column was originally devised by \citet{Ribe1985}.
The presentation and notation is based on \citet{ReesJonesRudge2020}, which develops a revised estimate of magma velocities based on the magmatic response to the deglaciation of Iceland \citep{maclennan02}.

We start from the equations of two-phase flow presented in section~\ref{sec:2-phase-flow}.
Making the same simplifications and taking the special case of a one-dimensional flow at steady state, equations~(\ref{eq:mass_l_summary}, \ref{eq:mass_s_summary}) imply that
\begin{equation} \label{eq:1D-flux}
\frac{d}{dz} (\phi w)  = \Gamma,
\end{equation}
where $w$ is the vertical component of the magma velocity.
Then we take $\Gamma = \Gamma_0$, where $\Gamma_0$ is the rate of decompression melting, which we assume to be uniform within the melting column. 
This is a reasonable approximation in the present context, but it neglects the drop in productivity when a mineral phase is exhausted from the residue \citep{hirschmann99}. Moreover, it does not apply for volatile-driven melting at depths beneath the dry (i.e., volatile-free) solidus, or at the transition to dry melting \citep{keller16}. 

As an aside, note that in our earlier development of the method in section~\ref{sec:methods} we could have used an alternative melting-rate parameterization.
In place of equation~\eqref{eq:chemistry_summary}, we could have used
\begin{equation} \label{eq:melting_rate_alternative}
\Gamma= \Gamma_0 + \frac{\beta}{\alpha} \phi \boldsymbol{v}_l\cdot \tilde{\boldsymbol{z}},
\end{equation}
in which case $\Gamma_0$ would have entered the base state calculations but would not have affected the linear equations governing the perturbations.
The most important difference is that the background compaction rate would no longer have been negligible, which would have complicated the analysis. 
Physically, the background compaction rate is a stabilizing influence on the reaction-infiltration instability, the magnitude of which is very sensitive to the dependence of bulk viscosity on porosity \citep{hewitt10,reesjones2018-jfm}.
The parameterization of equation~\eqref{eq:melting_rate_alternative} also neglects the disequilibrium effects (section~\ref{sec:disequilibrium}).
Thus we use equation~\eqref{eq:chemistry_summary} to calculate the melting rate throughout this study, which maintains consistency with previous studies of the reaction-infiltration instability \citep{aharonov95,reesjones2018-jfm}. 
 
Equation~\eqref{eq:1D-flux} can be integrated, with the constant of integration chosen such that $\phi w = 0$ at the base of the melting column $z=-H$,
\begin{equation} \label{eq:1D-flux-integrated}
\phi w  = \Gamma_0 (z+H) = F_\mathrm{max} W_b (1+\tilde{z}),
\end{equation}
in which we made use of the fact that $ \Gamma_0 H = F_\mathrm{max} W_b$,
where $F_\mathrm{max}$ is the maximum degree of melting 
and $W_b$ is the mantle upwelling velocity at the base of the column.

Then we take equation~\eqref{eq:Darcy_summary} and assume that background magmatic segregation is entirely driven by buoyancy to obtain
\begin{equation} \label{eq:1D-flux-Darcy}
\phi w  = Q_0 \phi^n, \quad Q_0 = K^* \Delta \rho g, 
\end{equation}
where $K^*$ is the prefactor in the permeability--porosity relationship $K=K^* \phi^n$, and $\Delta \rho g$ is the buoyancy associated with the density difference between solid and liquid phases (see section~\ref{sec:linear-growth}).
We combine equations~\eqref{eq:1D-flux-integrated} and \eqref{eq:1D-flux-Darcy} to obtain
\begin{align}
&\phi = \left[ F_\mathrm{max} W_b (1+\tilde{z}) \right]^{\tfrac{1}{n}} Q_0^{-\tfrac{1}{n}}, \label{eq:phi_0} \\
&w =  \left[ F_\mathrm{max} W_b (1+\tilde{z}) \right]^{\tfrac{n-1}{n}} Q_0^{\tfrac{1}{n}}.
\end{align}
We combine this last expression with equation~\eqref{eq:W_b} to obtain an expression for the background rate of magmatic upwelling relative to the plate half-spreading rate,
\begin{equation} 
\frac{w_0}{U_0} = \left[\frac{F_\mathrm{max}}{ C } \left((1+\tilde{x}^2)^{-1} -\cos^2\vartheta_1\right)\right]^{\tfrac{n-1}{n}}  
  \times [1 + \tilde{z}]^{\tfrac{n-1}{n}} \left[{ Q_0}/{U_0} \right]^{\tfrac{1}{n}}. \label{eq:w_0-relative}
\end{equation}
Here, we relabelled $w$ as $w_0$, using a subscript to refer to a background property, consistent with the convention introduced in section~\ref{sec:linear-growth}. 
Figure~\ref{fig:MOR_base}(b) plots the ratio $w_0/U_0$, and figure~\ref{fig:MOR_base}(c) shows the corresponding background porosity, denoted $\phi_0$. 
These plots are sensitive to the efficiency of melt extraction $Q_0$. 
If melt extraction is more efficient ($Q_0$ is higher), then the relative melt velocity  $w_0/U_0$ is increased, reflecting the proportionality  to $Q_0^{1/n}$ in equation~\eqref{eq:w_0-relative}.
The steady-state porosity $\phi_0$ is a balance between melt production and melt extraction.
Thus if melt extraction is more efficient ($Q_0$ is higher), then $\phi_0$ is decreased, reflecting the proportionality to $Q_0^{-1/n}$ in equation~\eqref{eq:phi_0}.

\subsubsection{Choice of parameters} \label{sec:parameters}
Our calculations use a reference set of parameters to illustrate the typical behaviour of the model.  This includes a prescribed dip of the base of the lithosphere of $\pi/5$ (36$^\circ$), as an intermediate case between a very steep boundary ($\pi/4$, or 45$^\circ$) and a much shallower case. 
Later, in section~\ref{sec:sensitivity}, we consider the behaviour when $\pi/12$  (15$^\circ$), which might be appropriate if the lithosphere is interpreted to be the cold thermal boundary, which thickens gradually at intermediate to fast spreading rates.

Reference material parameters are estimated from laboratory experiments and micromechanical models. 
We use a porosity--permeability exponent $n=2$, appropriate for small porosity as found in figure~\ref{fig:MOR_base}(c), from the micromechanical model of \citet{rudge18}.
At larger porosity, the exponent may be higher according to laboratory experiments \citep[e.g.][]{wark98,connolly09,miller14}.
The sensitivity of shear viscosity to porosity was estimated experimentally to be $\lambda^* =26$ by \citet{mei02}.
Micromechanical models \citep{takei09a,rudge18b} indicate that this factor decreases with increasing porosity, but is similar to the estimate $\lambda^* =26$ when porosity is fairly small (see, e.g., Fig.~13 of \citet{rudge18b}). 
The bulk-to-shear viscosity ratio has been extensively debated and has not been directly measured experimentally. 
Micromechanical models offer different predictions depending on assumptions about the microphysics of creep. 
There are two main categories. 
Models that assume viscous deformation at the microscale have bulk viscosity proportional to $\eta/\phi$, so the bulk-to-shear viscosity ratio is very large \citep[$O(10^2)$, e.g.,][]{simpson10b}. 
Models that assume diffusion at the microscale (either volumetric or grain boundary diffusion) have a much weaker sensitivity to porosity and the bulk-to-shear viscosity ratio is moderate \citep[$O(1)$, e.g.,][]{takei98,rudge18b}.
We take the estimate  $\zeta_0/\eta_0 = 5/3$ as our reference case and consider the possibility that the ratio is much higher in section~\ref{sec:sensitivity}. 

The melt velocity depends on the maximum degree of melting, for which we take $F_\mathrm{max}=0.2$ as a typical value. 
It also depends on the ratio $Q_0/U_0$.
As a reference value we take $Q_0/U_0 = 6.3\times10^4$, which for $U_0 = 3$~cm/yr corresponds to a maximum melt velocity of $4$~m/yr. $Q_0$ is sensitive to the reference permeability of the mantle, which is poorly constrained.
\citet{ReesJonesRudge2020} argue that the maximum melt velocity is faster than this. Here we make a relatively conservative choice such that in our reference case, both reaction and shear have the potential to make similar contributions to the growth of porosity bands. We consider faster and slower melt segregation in section~\ref{sec:sensitivity}. 
The overall amount of reactive melts generated depends on the parameter group $\beta H/\alpha\approx 0.2$ (probably within a range 0.15--0.3), based on previous studies of the reaction-infiltration instability \citep{aharonov95,reesjones2018-jfm}

\subsection{Combined instability at MORs}
We now use reference parameters to combine this MOR background state with the calculation of the linear growth rate.  We estimate the growth of instabilities caused by reaction and shear.
Figure~\ref{fig:MOR_schematic}(b) illustrates our approach, which is based on evolving a local parcel of porosity bands along each streamline of the solid mantle flow, as we now describe in detail.

\subsubsection{The role of porosity advection} \label{sec:porosity-advection}
Our previous calculations in section~\ref{sec:results_local} concerned the growth rate of instability in an infinite, uniform medium. 
For a mid-ocean ridge however, we must consider how to treat the advection of porosity in equation~(\ref{eq:mass_s_2}). We write that equation as
\begin{equation}
\frac{\partial \phi}{\partial t} +  \boldsymbol{v}_s \cdot \nabla \phi = f, \label{eq:porosity-advection}
\end{equation}  
where $f$ is a general source term representing the effect of compaction and reaction.
This equation contains the only time derivative and the only solid-velocity advection term in the overall set of equations and its treatment has been considered by previous studies.
\citet{spiegelman03b} showed that the rotational part of a linear flow causes an evolution in the angle of porosity bands. 
For simple shear, $\boldsymbol{v}_s \propto [z,0,0]$, \citet{spiegelman03b} used a generalized linear analysis where the wavevector depends on time such that $\phi' \propto \exp(i \boldsymbol{k}(t) \cdot \boldsymbol{x} )$. 
\citet{butler10} showed that both a pure and also a simple shear flow cause the wavelength of porosity bands to increase.
\citet{Gebhardt2016} extended the methodology of \citet{spiegelman03b} to a general flow with translation and shear, and applied it to the solid velocity field of a mid-ocean ridge setting. 

In this section, we use the same approach as \citet{Gebhardt2016}. 
The only new aspect of our calculation involves the growth rate $\sigma$, where we account for growth of instabilities by reactive infiltration as well as shear.
We now give a slightly expanded justification for the methodology. We introduce a local co-ordinate system about some arbitrary point $\boldsymbol{x}_0$ and let  $\boldsymbol{x}$ be the (small) displacement from that point.
We then Taylor expand the solid velocity to first order about the point using the velocity gradient tensor 
\begin{equation} \label{eq:Taylor}
\boldsymbol{v}_s(\boldsymbol{x}_0+ \boldsymbol{x})\approx \boldsymbol{v}_0 +  \nabla  \boldsymbol{v}_s \cdot \boldsymbol{x} , \quad \boldsymbol{v}_0= \boldsymbol{v}_s(\boldsymbol{x}_0),
\end{equation} 
where the velocity gradient tensor $ \nabla  \boldsymbol{v}_s$ is evaluated at $\boldsymbol{x}_0$.
Thus the velocity gradient tensor in the approximation is locally a constant, so the corresponding term \mbox{$ \nabla \boldsymbol{v}_s \cdot \boldsymbol{x} $} is a linear shear flow of the type discussed in section~\ref{sec:methods}.
We express the porosity as a generalized normal mode:
\begin{equation} \label{eq:generalized_normal-modes}
\phi = \phi_0 +  \exp[ i \boldsymbol{k}(t)\cdot \boldsymbol{{x}} +s(t)],
\end{equation} 
where $\phi_0$ is the background state and $\boldsymbol{k}(t)$ is the wavevector of a disturbance and $s(t)$ determines its amplitude.
The wavevector and amplitude evolve according to
\begin{align} 
&\frac{D \boldsymbol{k}}{Dt} +   \nabla  \boldsymbol{v}_s ^T \cdot \boldsymbol{k} = 0,  \label{eq:wavenumber_evolution} \\
&\frac{Ds}{Dt}    = \sigma, \label{eq:amplitude_evolution} 
\end{align} 
where $ \frac{D}{Dt} \equiv \frac{\partial }{\partial t} +  \boldsymbol{v}_0 \cdot \nabla$ is the Lagrangian derivative (or equivalently the derivative along a streamline of the solid flow).
These equations are sufficient to ensure that the porosity advection equation~(\ref{eq:porosity-advection}) is satisfied, provided  $f\approx f_0 +  \sigma (\phi - \phi_0)$, where $f_0$ is the background part of the source term. 
This can be verified by substituting equation~\eqref{eq:generalized_normal-modes} into  equation~\eqref{eq:porosity-advection} with the velocity expanded using equation~\eqref{eq:Taylor}.
The terms involving the uniform translation $ \boldsymbol{v}_0 $ are accounted for by the switch to the Lagrangian derivative.
The terms involving the wavevector can be shown to cancel by taking the scalar (dot) product of equation~\eqref{eq:wavenumber_evolution} with the vector $\boldsymbol{{x}}$.

This approach to evolution of porosity perturbations is local, and is only valid when the solid velocity varies on some scale very much larger than the wavelength of the perturbation.
It can be made precise in certain situations, including that of \citet{spiegelman03b}, which considers a uniform background state and a linear shear flow.
However, as \citet{Gebhardt2016} discuss, for mid-ocean ridges it should be thought of as a reasonable, if \textit{ad hoc}, estimate of the behaviour of porosity bands.
This is because the calculation of $\sigma$ was for an infinite domain with a uniform background porosity, uniform magma flow field and linear mantle shear flow. 
These will all be reasonable approximations provided the wavelength of the porosity bands is small, in which case the bands vary on scale much shorter than that over which the background porosity evolves, and so the growth rate calculation should be reasonably accurate. 

\subsubsection{Integration along streamlines} \label{sec:streamlines}
We now integrate equations~(\ref{eq:wavenumber_evolution},\ref{eq:amplitude_evolution}) along streamlines of the solid flow, as shown in figures~\ref{fig:MOR_schematic}(b) and \ref{fig:MOR_base}(d). 
It is possible to do this in Cartesian co-ordinates, but preferrable to work in polar co-ordinates because the equation of a streamline is particularly simple. 
Consider a streamline $(\tilde{r}(t),\vartheta(t))$ that starts at $(\tilde{r}_0,\vartheta_0)$ when $t=0$. 
By choosing $\tilde{r}_0 = 1/\cos \vartheta_0$, all the streamlines start at the base of the melting region.
Then a streamline is defined by
\begin{equation} \label{eq:r-def}
\tilde{r} \Theta(\vartheta) = \mathrm{constant} \Rightarrow \tilde{r}(\vartheta) = \frac{ \Theta(\vartheta_0)}{\cos \vartheta_0 }\frac{1}{\Theta(\vartheta)}.
\end{equation}
Thus given $\vartheta$, we have an explicit expression for $\tilde{r}$ and hence ($\tilde{x},\tilde{z}$) along the streamline, if required. 
Then we have an evolution equation for $\vartheta$, namely
\begin{equation} \label{eq:vartheta_evolution}
\frac{d \vartheta}{d t} = \frac{u_\vartheta}{r}  = \frac{U_0}{H} \frac{ \Theta(\vartheta)}{  \tilde{r}(\vartheta)}.
\end{equation}
Since we are integrating along streamlines, the material derivatives $D/Dt$ can be replaced by full derivatives $d/dt$. 
Finally, we can change variables to make $\vartheta$ the independent variable instead of $t$ using equation~\eqref{eq:vartheta_evolution}. 
Thus 
\begin{align} 
&\frac{d \boldsymbol{k}}{d\vartheta} = -\frac{  \tilde{r}(\vartheta)}{ \Theta(\vartheta)}   \tilde{\nabla} \tilde{\boldsymbol{u}} ^T \cdot \boldsymbol{k}  ,  \label{eq:wavenumber_evolution-A2} \\
&\frac{ds}{d\vartheta}  =\frac{  \tilde{r}(\vartheta)}{ \Theta(\vartheta)} \frac{H \sigma}{U_0} . \label{eq:amplitude_evolution-A2} 
\end{align}
This system of equations can be integrated over the range of $\vartheta_0 \leq \vartheta \leq \vartheta_1$ using any standard ODE solver
(we use the MATLAB routine ODE45). 
This procedure is repeated for a range of initial positions specified by $\vartheta_0$.
The system is linear in the initial amplitude so we take $s=0$ as an initial condition. 
We also need to specify an initial condition on the wavevector (see section~\ref{sec:ic_wavevector}).
Note that the change of variables to $\vartheta$ breaks down exactly beneath the ridge axis (by construction, since $\vartheta$ is not varying along that streamline). 
For that special case, it is easiest to use $z$ as the independent variable. 

\subsubsection{Overall growth rate}
The amplitude of the instability evolves along a streamline according to equation~\eqref{eq:amplitude_evolution-A2}. 
We neglect the imaginary part of the growth rate (which gives rise to transient waves) and use the simplest expression for the real part of the growth rate~\eqref{eq:sigma_ratio}, in which the contributions from reaction and shear can be computed separately. 
Thus we split the amplitude $s$ from equation~\eqref{eq:generalized_normal-modes} into a part arising from reaction $s_\mathrm{reaction}$ and a part arising from shear $s_\mathrm{shear}$.

For reaction-driven instabilities, we find
\begin{equation} \label{eq:s_reaction}
\frac{d}{d\vartheta} s_\mathrm{reaction}  = n \frac{\beta H}{\alpha} \frac{w_0}{U_0} \frac{  \tilde{r}(\vartheta)}{ \Theta(\vartheta)} \cos^2 \theta. 
\end{equation}
Thus the magnitude of the reactive growth rate depends on the following dimensionless parameters: the permeability exponent $n=2$, the overall amount of reactive melts generated $\beta H/\alpha = 0.2$, and the relative magma velocity ${w_0}/{U_0}$ given by equation~\eqref{eq:w_0-relative}. 
This latter ratio can be significantly greater than 1, allowing the contribution from the reactive instability to be significant.

For shear-driven instabilities, we find 
\begin{equation}  \label{eq:s_shear}
\frac{d}{d\vartheta} s_\mathrm{shear}   = \frac{2 \lambda^*}{\tfrac{4}{3}+\tfrac{\zeta_0}{\eta_0} } \frac{|\sin \vartheta|}{C \Theta(\vartheta)}  G(\theta,\psi), 
\end{equation}
in which we used equation~\eqref{eq:gamma-dot-polars}.
Crucially, this is independent of the plate half-spreading rate.
Although a faster spreading rate increases the strain rate, which increases $\sigma_\mathrm{shear}$, it also increases the solid mantle velocity, thereby reducing the time spent to move through the partially molten region (proportional to $H/U_0$). 
Instead, the the shear-driven instability depends only on a combination of rheological properties $ {2 \lambda^*}/\left(\tfrac{4}{3}+\tfrac{\zeta_0}{\eta_0} \right)$, relative position in space (via $\vartheta$ and $\vartheta_1$), and orientation of the wavevector. 

\subsubsection{Initial conditions and evolution of the wavevector} \label{sec:ic_wavevector}
We specify the wavevector at the start of a streamline as an initial condition.
The wavevector then evolves according to  equation~\eqref{eq:wavenumber_evolution-A2}, 
where 
$ \tilde{\nabla} \tilde{\boldsymbol{u}}$, obtained from the corner flow, is given by equation~\eqref{eq:scaled_nabla_u}.
We choose an initial condition $k_y=0$. 
Then by equation~\eqref{eq:wavenumber_evolution-A2}, $k_y$ remains zero.
This choice of initial condition is motivated by the fact that a mid-ocean ridge has extension in the horizontal, which favours a wavevector orientation with $k_y=0$, as discussed in section~\ref{sec:orientation}. 

We take two approaches to specify the initial condition. 
First, we prescribe a single initial orientation of the wavevector.  
Second, we prescribe a uniform distribution of wavevector orientation, an approach also taken  by \citet{Gebhardt2016}. 
 
\begin{figure}
\centering\noindent\includegraphics[width=1.0\linewidth]{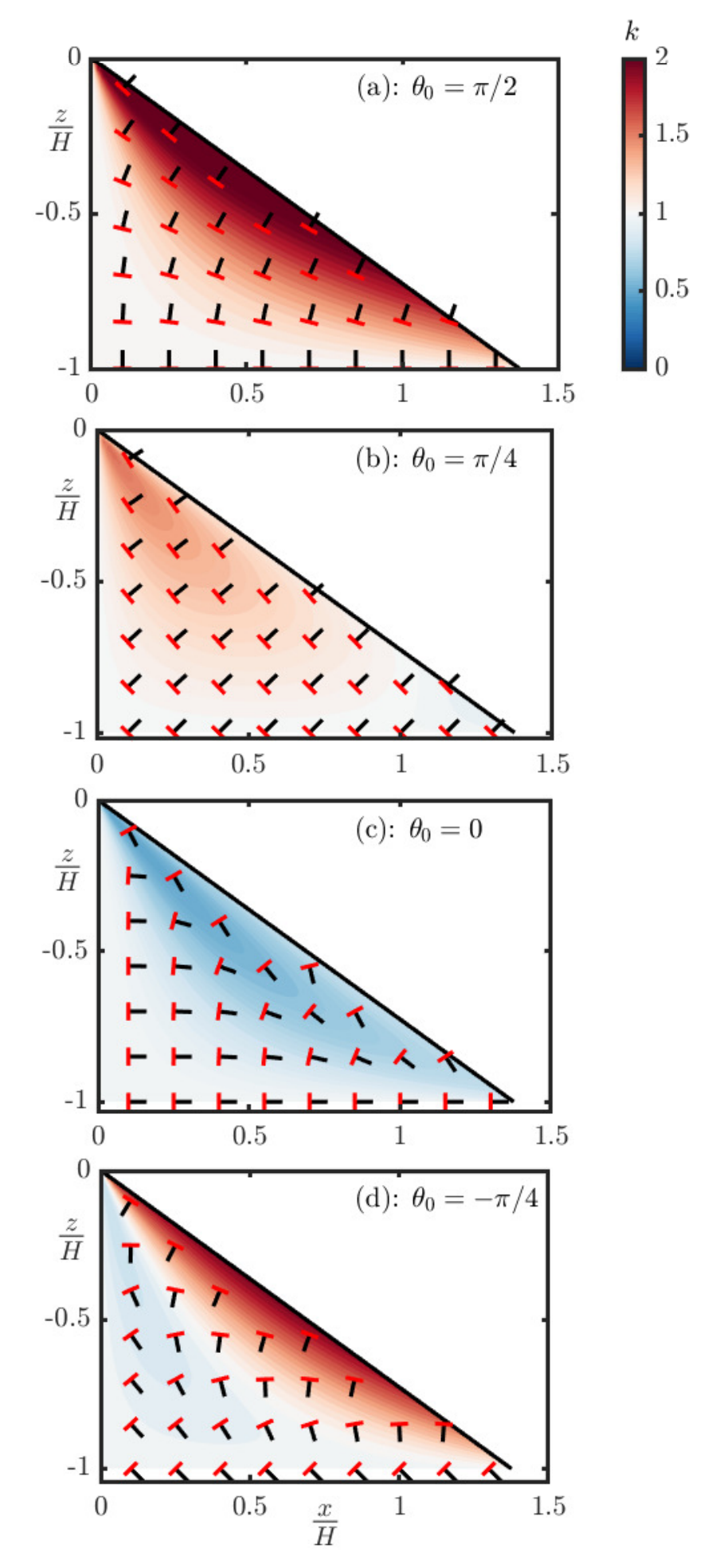}
   \caption{Effect of shear on orientation and magnitude of wavevector. In each panel, the initial magnitude of the wavevector is 1 and the initial orientation is: (a) $\theta_0=\pi/2$, (b) $\theta_0=\pi/4$, (c) $\theta_0=0$, (d) $\theta_0=-\pi/4$. Black line segments show the orientation of the wavevector and red line segments show the corresponding porosity bands, which are perpendicular to the wavevector. The colourscale shows the wavenumber, with red colours corresponding to increased wavenumber (reduced wavelength) and blue colours corresponding to reduced wavenumber (increased wavelength). }
   \label{fig:MOR_wavenumber}
\end{figure}

Figure~\ref{fig:MOR_wavenumber} shows the effect of the corner flow on the wavevector, starting from a single initial orientation. 
The flow acts to rotate the wavevector in a clockwise sense to the right of the ridge axis (decreasing $\theta$).
The wavenumber variation depends on whether the flow resolved along it is extensional or contractional. However, the maximum change of wavenumber (and hence of wavelength) is a factor of about 2 of its initial value. 
Given that the growth rate is only weakly sensitive to wavelength (assuming it is smaller than the compaction length), we can infer that the effect of solid flow on the wavenumber is mainly through rotation.

\section{Results: growth of instabilities at MOR\MakeLowercase{s}} \label{sec:results_MOR}

Results are described in the following order.
First, in section~\ref{sec:max_growth}, we calculate the maximum possible growth of the instability by assuming that the wavevector is always instantaneously in the optimal orientation.
Second, in section~\ref{sec:orientation_growth}, we calculate the growth for a prescribed initial wavevector orientation that evolves along streamlines.
Third, in section~\ref{sec:dist_growth}, we  calculate the growth for a distribution of initial wavevector orientations.
Finally, in section~\ref{sec:sensitivity}, we discuss how the results depend on the choice of parameters.

\subsection{Orientation-independent, maximal growth} \label{sec:max_growth}

To facilitate an understanding of instability growth in the context of our mid-ocean ridge background state, we first present results where growth rates are maximised over all possible perturbation orientations $\theta$. In particular, for reaction we take $\cos^2\theta=1$ in equation~\eqref{eq:s_reaction}, and for shear we take $G(\theta,\psi)=1$ in equation~\eqref{eq:s_shear}. This approach gives an upper bound on growth, because the factors involving the wavevector orientation are always less than or equal to 1 in magnitude. 

Figure~\ref{fig:MOR_base}(d) compares the ratio of  local maximum growth rate due to shear versus that due to reaction.  
Reaction is dominant beneath the ridge axis and shear is dominant off-axis along the base of the lithosphere. This reflects the pattern of background strain rate (which is zero on the axis) and magma segregation speed (which decreases away from the axis), as shown in other panels of figure~\ref{fig:MOR_base}.

\begin{figure}
\centering\noindent\includegraphics[width=1.0\linewidth]{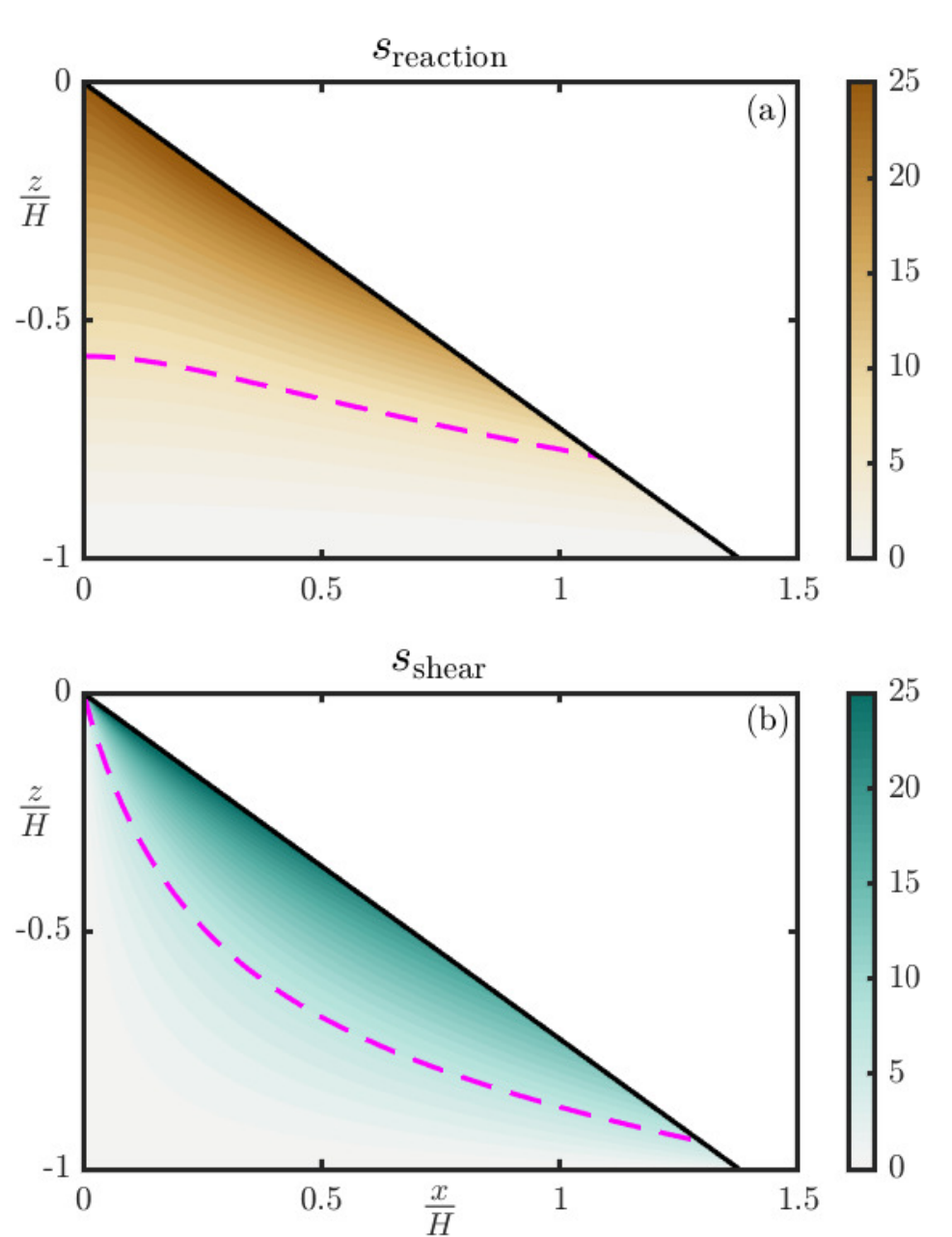}
   \caption{Total growth of (a) reactive and (b) shear-driven instabilities. The dashed magenta curve highlights the contour $s=7$, which corresponds to an increase in amplitude of about $10^3$. }
   \label{fig:MOR_growth_upper}
\end{figure}

Figure~\ref{fig:MOR_growth_upper} shows the maximum possible accumulated growth of perturbations. 
The maximum possible growth is comparable between the two mechanisms, but this is sensitive to the choice of parameters. 
If the magma velocity were assumed faster or the ratio of bulk to shear viscosity were higher, the reactive growth would be much larger than the shear-driven growth (see section~\ref{sec:sensitivity} for a fuller sensitivity analysis).
Even the reference parameter choices (that allow both reaction and shear to contribute) emphasize an important result: beneath the ridge axis, reaction dominates in the formation of channels. The dashed magenta contours in figure~\ref{fig:MOR_growth_upper} highlight where the amplitude factor $s=7$, which corresponds to a growth in amplitude of $\exp(s) \approx 10^3$ according to the definition in equation~\eqref{eq:generalized_normal-modes}.
This suggests that deep channels were probably formed through reactive, rather than shear-driven instability. This would only be reinforced by inclusion of volatile-driven reactive melting \citep{keller16}. 

\begin{figure*}
\centering\noindent\includegraphics[width=1.0\linewidth]{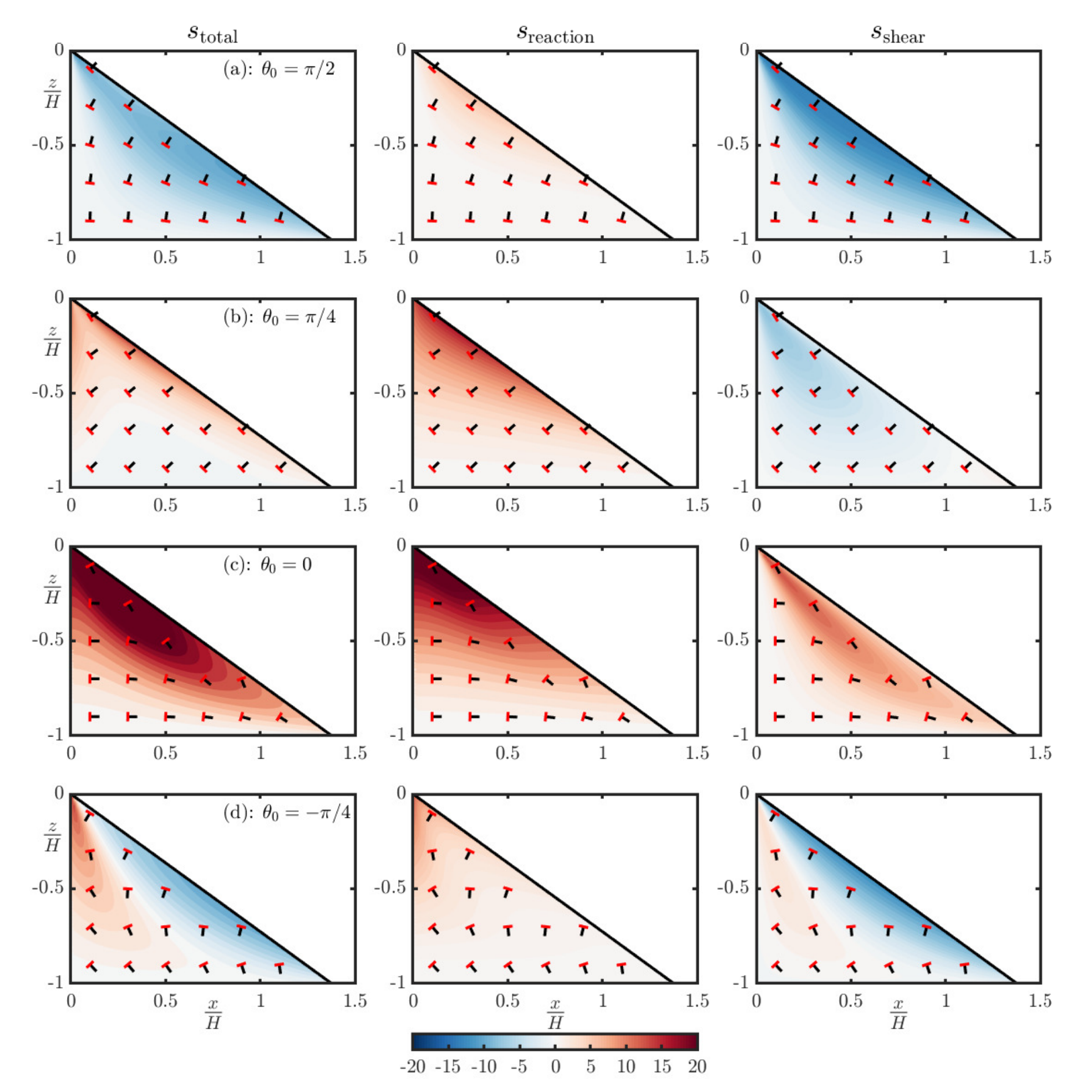}
   \caption{The growth of instabilities initialized with a single wavevector orientated with row: (a) $\theta_0 =\pi/2$; (b) $\theta_0 =\pi/4$; (c) $\theta_0 =0$; (d) $\theta_0 =-\pi/4$. The left column shows the total growth $s_\mathrm{total}$ which is the sum of the growth from reaction $s_\mathrm{reaction}$ (middle column) and the growth from shear $s_\mathrm{shear}$ (right column). The colour-scale is the same for all panels (and the range is clipped). Black line segments show the orientation of the wavevector and red line segments show the corresponding porosity bands, which are perpendicular to the wavevector.  }
   \label{fig:MOR_growth_combined}
\end{figure*}

\subsection{Orientation dependence of growth} \label{sec:orientation_growth}

We now turn our attention to the consequences of wavevector orientation. Figure~\ref{fig:MOR_growth_combined} shows a set of four models, plotted in terms of accumulated growth of the perturbation.  In each case, we set the perturbation wavevector to have orientation $\theta_0$ as it enters the melting region from below.  The wavevector then evolves along streamlines of the corner flow. The left column shows the total growth $s_\mathrm{total}$, which is the sum of the growth from reaction $s_\mathrm{reaction}$ (middle column) and the growth from shear $s_\mathrm{shear}$ (right column). Overall it is evident that accumulated growth is sensitive to the initial orientation.
 
Row (a) shows that when the initial wavevector is vertical (\mbox{$\theta_0=\pi/2$}; the porosity bands are initially horizontal), the instability is suppressed. There is contraction across the bands, so the shear-driven instability has a negative growth rate. The bands remain close enough to horizontal that the reactive mode of instability only grows slowly. The net effect is that the overall growth is negative. 

Row (b) shows that when the  initial wavevector has an angle \mbox{$\theta_0=\pi/4$}, there is some growth of the instability.
Again, there is contraction across the bands, so shear-driven instability has a negative growth rate. 
The bands also have a larger component in the vertical direction, so reactively-driven instability is now more significant.
In total, the reaction-driven instability is large enough to offset the shear-driven suppression of the instability and give net positive growth.

Row (c) shows that when the initial wavevector is horizontal \mbox{($\theta_0=0$)} and the porosity bands are initially vertical, the instability grows rapidly. 
These bands start in the orientation most favourable to reactively driven instability.  However, they are rotated into an orientation with relatively large extension across bands, so the shear-driven instability is also significant.
Rotation along streamlines slightly reduces the reactive growth, but this reduction is insignificant. 
In total, reaction and shear cooperate to drive strong growth of porosity bands.

Row (d) shows that when the  initial wavevector has an angle \mbox{$\theta_0=-\pi/4$}, the instability is partially suppressed.
There is some initial growth of both reactive and shear-driven instability.
However, following the streamlines, the wavevector is rotated into an orientation where there is very little reactive growth and there is decay caused by contraction across the bands.
Thus, in total, by the time the lithosphere is reached, the porosity bands are suppressed.

\begin{figure*}
\centering\noindent\includegraphics[width=1.0\linewidth]{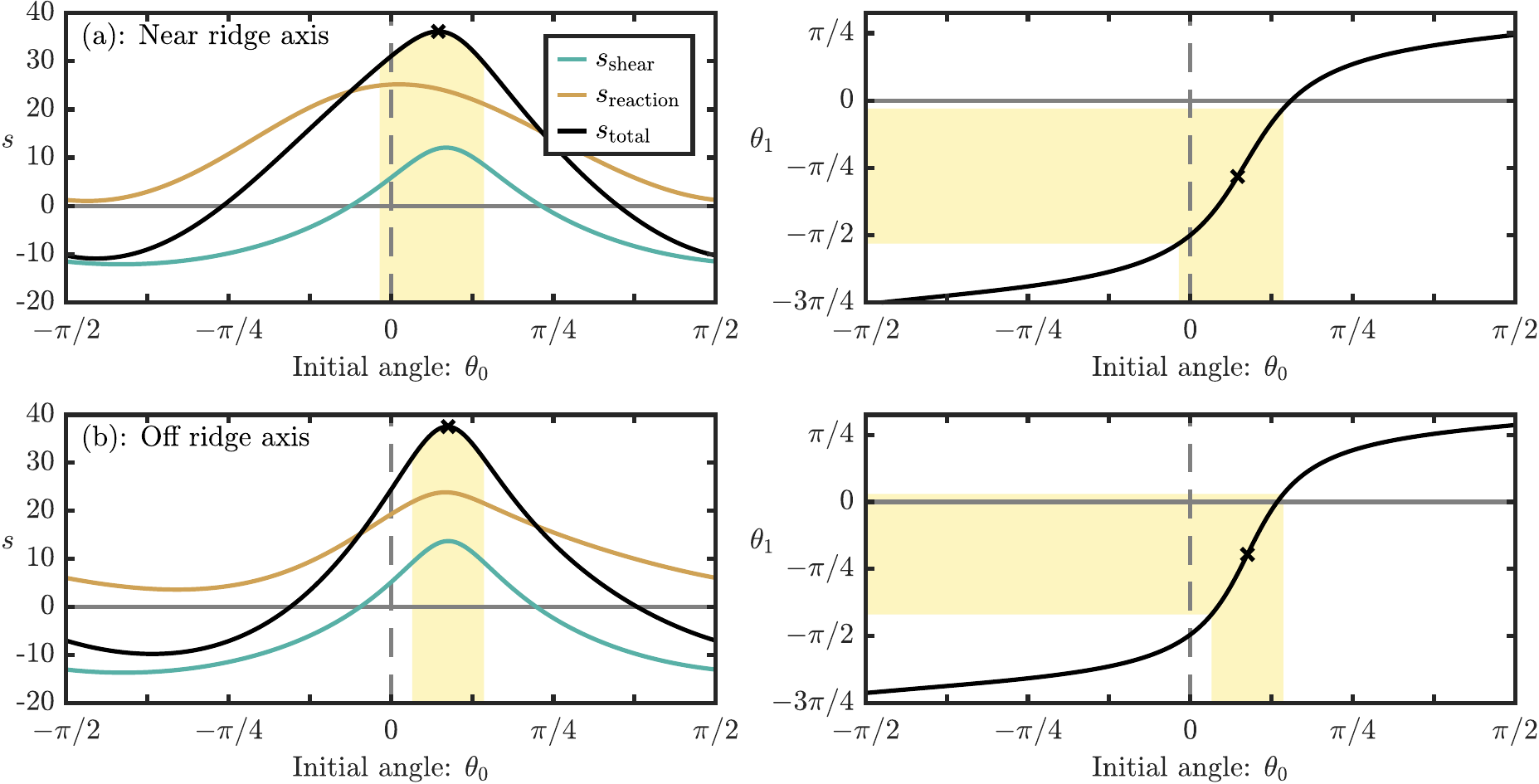}
   \caption{The effect of the orientation of the wavevector $\theta_0$ on the total accumulated growth $s$ at the end of a streamline (left column) and orientation of the final wavevector $\theta_1$  (right column). A black cross marks the most unstable initial wavevector orientation. The pale yellow shaded region highlights wavevector orientations with a total growth within 80\% of the maximum. The top row (a) is an example for a streamline near the ridge axis $\tilde{x}_0 = 0.01$ and the bottom row (b) is an example for a streamline further from the ridge axis $\tilde{x}=0.2$. These streamlines are shown as dashed magenta curves in figure~\ref{fig:MOR_base}(d).}
   \label{fig:MOR_growth_spectrum}
\end{figure*}

\subsection{Growth with a spectrum of initial wavevector orientations} \label{sec:dist_growth}
Figure~\ref{fig:MOR_growth_spectrum} considers the evolution of a spectrum of initial wavevector orientations.  This is evaluated along two particular streamlines, one near the ridge axis and one further off-axis, shown by the dashed curves in figure~\ref{fig:MOR_base}(d). 
At the base of the melting region, independent of position $\vartheta_0$, we assume that the initial magnitude of the wavevector is uniformly distributed, i.e., independent of $\theta_0$.
The spectrum is then evolved along streamlines. We plot the final accumulated growth and final wavevector angle for two particular streamlines, attained when these streamlines terminate at the lithosphere. 

Row (a) shows results for a streamline near the ridge axis. 
Here, the reactive growth favours wavevectors orientated very close to horizontal (vertical porosity bands). 
The shear-driven instability favours wavevectors orientated with slightly positive initial angle for reasons discussed above.
The overall growth favours a wavevector orientation that is intermediate between these angles.
The final orientation $\theta_1$ associated with the greatest accumulated growth is rotated by the shear flow.   
So high-porosity bands might be expected to correspond to a final wavevector orientation  in the  pale yellow shaded region range, roughly $-\pi/2 < \theta_1 < 0 $. 

Row (b) shows results for a streamline further from the ridge axis. 
The results are similar to row (a).
The only important difference is that reactive growth favours wavevectors orientated with a slightly positive initial angle.
This is because such bands accumulate more reactive growth as they are rotated clockwise into the vertical orientation by the shear.
For this example, the overall growth favours an initial wavevector angle that is very similar to that favoured by both reaction-driven growth and shear-driven growth separately, with the peak $s$ occurring at a very similar angle for each mechanism. 
Further calculations (not shown) find similar patterns of behaviour even further off-axis, so the results in row (b) are illustrative of the general pattern away from the immediate vicinity of the ridge axis (row a). 

\begin{figure*}
\centering\noindent\includegraphics[width=1.0\linewidth]{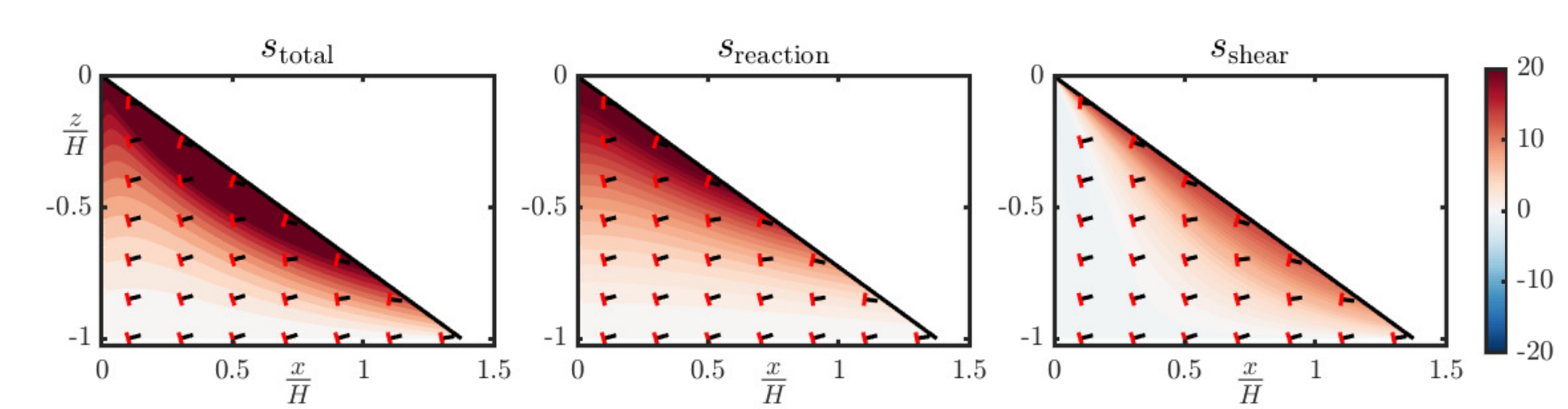}
   \caption{The evolution of the most unstable initial orientation of the wavevector from the bottom row of Figure~\ref{fig:MOR_growth_spectrum}. Other figure details are as in Figure~\ref{fig:MOR_growth_combined}. }
   \label{fig:MOR_growth_favoured}
\end{figure*}

Figure~\ref{fig:MOR_growth_favoured} shows the evolution of perturbations that have maximal ultimate growth over all initial orientations $\theta_0$.  This maximum is evaluated for each streamline separately, with the aim of highlighting the perturbations that would be most likely expressed in a full, non-linear solution (albeit the full non-linear solution may behave differently, as discussed in section~\ref{sec:limitations}).
An alternative approach is to optimize over the initial wavevector orientation independently at each point in the interior of the melting region, rather than just at the end of each streamline. 
We present calculations using this alternative approach in appendix~\ref{sec:alternative}.

Figure~\ref{fig:MOR_growth_favoured} demonstrates that the most unstable orientation varies only slightly with distance off the axis.
At every location, the most unstable wavevector is close to horizontal and the corresponding porosity bands are close to vertical. 
The calculations reinforce the point that was  discussed when considering the upper bound on growth shown in figure~\ref{fig:MOR_growth_upper} --- that all channels are initially formed by the reactive mode of instability, and that axial channels are dominated by reactive growth.
In this case, shear actually reduces the growth for the deeper part of the streamline. For off-axis streamlines, shear contributes to the growth of instability as the streamlines approach the lithosphere. 
However, the roots of channels are always associated with reactive rather than shear-driven instability. 

\subsection{Parametric sensitivity} \label{sec:sensitivity}

\begin{figure*}
\centering\noindent\includegraphics[width=1.0\linewidth]{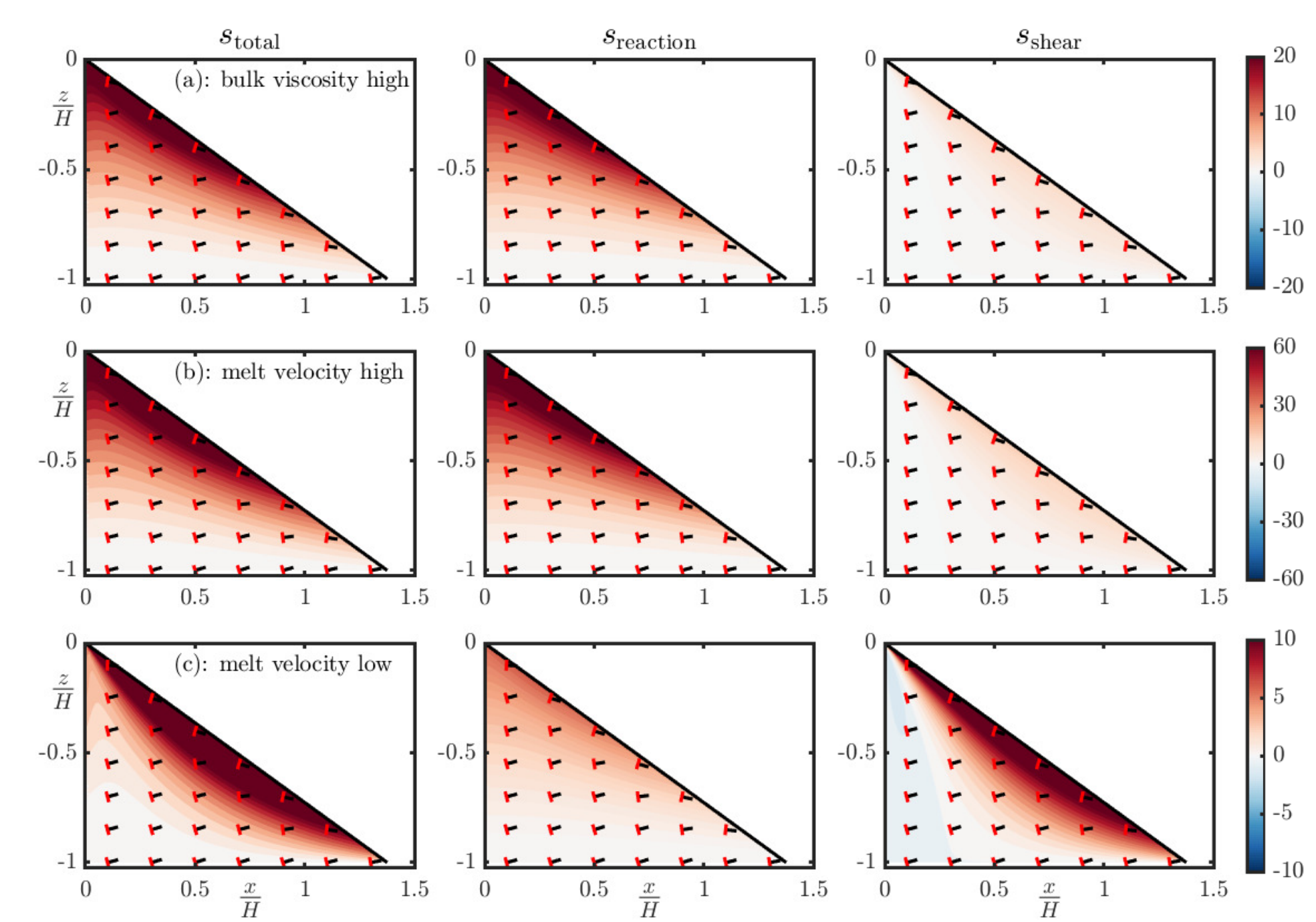}
   \caption{Sensitivity experiments showing the results with the optimal initial wavevector orientation. Row (a) has a higher bulk-to-shear viscosity ratio $\zeta_0/\eta_0 =  10$, which is 6 times larger than the reference case. Row (b) has a higher melt velocity ratio $Q_0/U_0=6.3\times 10^6$, which is 10 times greater than the reference case. Row (c) has a lower melt velocity ratio $Q_0/U_0=4 \times 10^3$, which is 16 times smaller than the reference case.   Note the different colour scales. Other figure details are as in Figure~\ref{fig:MOR_growth_favoured}. }
   \label{fig:MOR_growth_combined_sensitivity}
\end{figure*}

\begin{figure*}
\centering\noindent\includegraphics[width=1.0\linewidth]{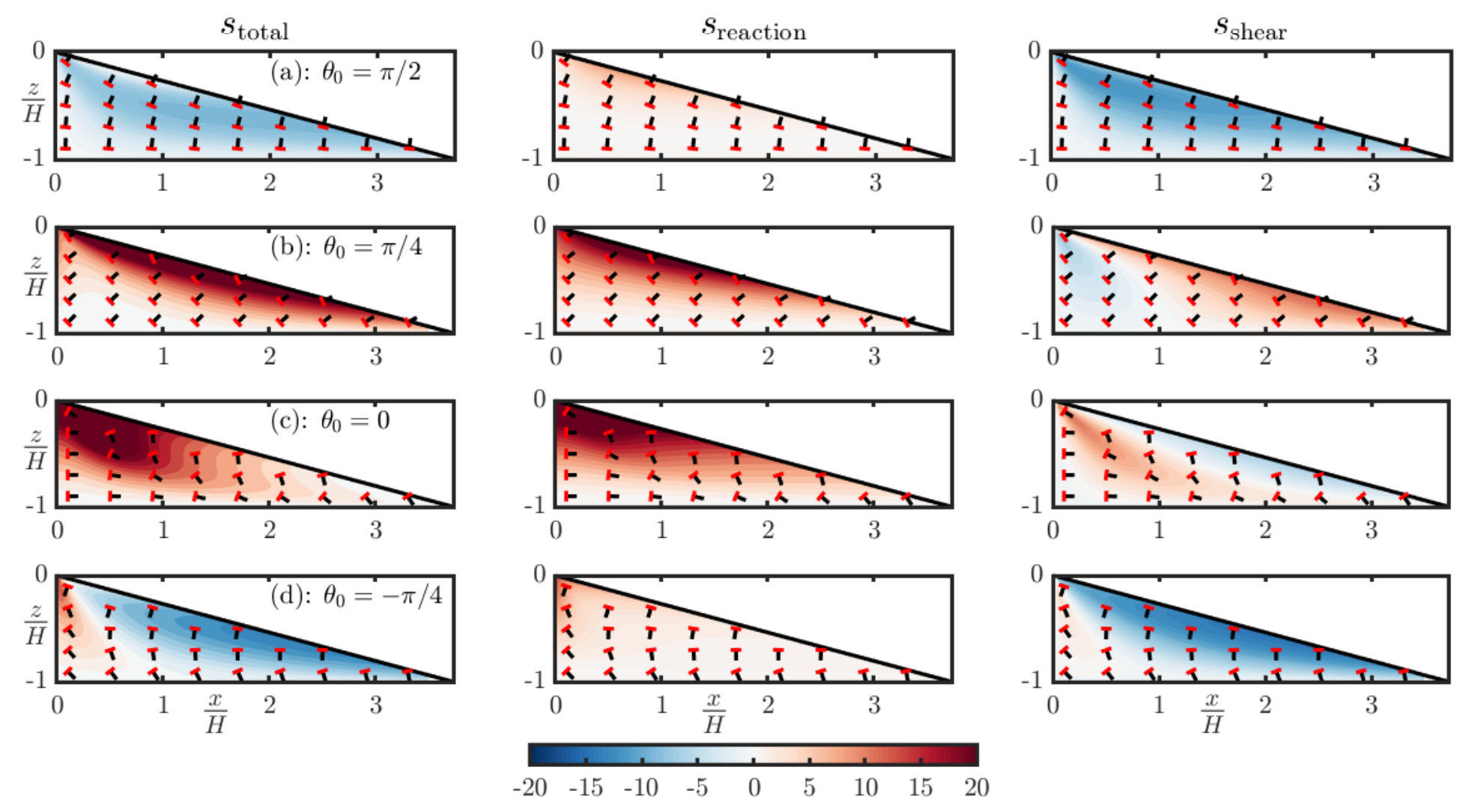}
   \caption{Sensitivity experiment with a shallower dip $\pi/12$ of the lithospheric base showing a range of initial wavevector orientations. Other figure details are as in Figure~\ref{fig:MOR_growth_combined}. }
   \label{fig:MOR_growth_combined_shallow_dip}
\end{figure*}

Figure~\ref{fig:MOR_growth_combined_sensitivity} shows that this emphasis on the importance of reactive instability is robust to changes in the parameters considered. 
In row~(a), the ratio of bulk to shear viscosity is increased to $\zeta_0/\eta_0 = 10$, which is six times higher than the reference case. 
This has the effect of suppressing the shear-driven mode of instability, such that the total accumulated growth is dominated by reaction throughout the melting region. 
Consequently, the most unstable orientation of the wavevector is close to horizontal, since this produces the vertical, high porosity channels favoured by reaction.
The solid flow still plays a role in rotating the wavevector.
In row (b), we show that a similar pattern is obtained by increasing the relative magma flow speed $Q_0/U_0$ by a factor of 10, which corresponds to a maximum melt speed of $13$~m/yr.  This increases the reactive growth rate rather than decreasing the shear-driven growth rate, but the relative effect is the same as for variations in the viscosity ratio (note the different colour scale). 
The only situation that allows shear to play a greater role is when the bulk-to-shear viscosity ratio is low (as in the reference case) and the melt velocity is also low.  For this case, taking $Q_0/U_0$ as 16 times smaller than the reference case, which corresponds to a maximum melt speed of $1$~m/yr, row~(c) shows that in most of the domain, shear is more important than reaction. Very close to the ridge axis, reaction remains dominant. 
However, as we discussed in section~\ref{sec:parameters}, this segregation rate is probably too slow to satisfy observational constraints.
Thus, under more realistic choices of parameters, our model predicts that reaction-driven instability plays the dominant role. 

In figure~\ref{fig:MOR_growth_combined_shallow_dip} we consider a reduced lithospheric slope, such as might correspond to a faster spreading rate. 
The results are more complex than in the reference geometry, so we plot them for a set of different initial wavevector orientation to explore this complexity (as in fig.~\ref{fig:MOR_growth_combined}). 
The pattern of reaction-driven growth is similar to calculations with a more steeply dipping lithospheric base. This is as expected, given that it is not directly sensitive to the strain-rate field.
However, the pattern of shear-driven growth differs, especially in rows (b) and (c). 
In row (b), we find a positive contribution to the instability from shear as the streamlines approach the lithosphere, whereas the equivalent contribution in figure~\ref{fig:MOR_growth_combined}(b) was slightly negative. 
Conversely, in row (c), we find a slightly negative contribution whereas the equivalent contribution in figure~\ref{fig:MOR_growth_combined}(c) was positive. 
These differences reflect the different orientation of strain rate that results when the base of the lithosphere has a shallower dip. 
Nevertheless, the overall picture of a reactive instability that dominates the dynamics and favours 
sub-vertical high-porosity channels persists under these conditions.

\section{Discussion} \label{sec:discussion}
\subsection{Summary of results}
The results above investigate channelized melt extraction from the mantle and the combined role of two known mechanisms of flow localisation, reaction- and shear-driven instability. 
The theoretical framework developed in section~\ref{sec:methods} allows us to simultaneously describe these two mechanisms in a manner consistent with published results obtained separately for the reaction-infiltration instability and the shear-driven instability. 
In section~\ref{sec:results_local}, we showed that the relative importance of shear-driven versus reaction-driven instability is governed by the dimensionless ratio given in equation~\eqref{eq:S_def}, which we rewrite here:
\begin{equation} 
    S = \frac{ \sigma_\mathrm{shear} }{\sigma_\mathrm{reaction}} = \frac{2 \lambda^*   }{n  \left(\tfrac{4}{3} + \tfrac{\zeta_0}{\eta_0} \right)  }\frac{\alpha}{\beta} \frac{\dot{\gamma}_0 }{w_0}.
\end{equation}
The ratio $S$ represents the ratio of growth rates due to shear and due to reaction. 
$S$ is controlled by a particular combination of mechanical material properties 
\begin{equation*}
    2\lambda^*    n^{-1} \left(\tfrac{4}{3} + \tfrac{\zeta_0}{\eta_0} \right)^{-1},
\end{equation*} 
the reactivity of the system $\beta/\alpha$ (which has units of~m$^{-1}$), the background rate of melt flow $w_0$, and the background solid strain rate $\dot{\gamma}_0$.
The dimensionless parameter $S$ is distinct from both the ``Fiji'' number 
\begin{equation}
    \Phi g = \frac{\Delta \rho g}{\frac{\eta_0 \dot{\gamma}_0}{0.3\delta}}
\end{equation} 
and the Damk{\"o}hler number discussed in the review of \citet{kohlstedt09}. 
The parameter $\Phi g$ is related to the ratio of shear-driven melt velocity to buoyancy-driven melt velocity and would come into the imaginary part of the growth rate. The imaginary part is associated with porosity waves; here we chose to focus on the real part of the growth rate, which controls the amplitude of porosity bands. 
The Damk{\"o}hler number controls the degree of disequilibrium. 
We showed that this plays only a modest role in affecting the overall growth rate; instead, the crucial parameter is the reactivity $\beta/\alpha$, which appears in $S$. 

The definition of $S$ highlights the crucial role played by the material properties of partially molten rocks, some of which are not well constrained.
Indeed, there remain important differences between micromechanical models.  Perhaps the greatest uncertainty is the case of the bulk viscosity, as demonstrated by the contrast between model predictions  \citep{takei09a,takei09b,simpson10b,rudge18b}.
As a consequence of these uncertainties, robust, leading-order features of laboratory experiments \citep[e.g.,][]{holtzman07} such as the orientation of high-porosity bands, their size, spacing and rate of emergence remain challenging to predict quantitatively \citep{alisic16}.
Because of this gap in our knowledge, we face significant uncertainty in extrapolating between the laboratory scale and the mantle scale. 
The laboratory experiments are performed in closed capsules with a fixed melt fraction deforming at very high strain rates, whereas the MOR system is open to melt flow and deforms a million times more slowly.
In this context, we have opted for the simplest form of model that captures the shear-driven formation of porosity bands.  In particular, we have not considered a power-law shear viscosity \citep{katz06}, anisotropic viscosity \citep{takei15, qi15}, or the stabilizing influence of surface tension \citep{parsons08,king11,bercovici16}.
Our methodology could be extended to include these effects, but such extensions are most worthwhile once there is a more settled and complete understanding of the  shear-driven formation of porosity bands in isolation. 

Our results also clarify the geometric controls on the combined instability. 
The reaction-infiltration instability has only one preferred direction in the context of our model --- the vertical. 
This is because the background magma flow direction and the solubility gradient are aligned with gravity. 
In the absence of shear, there is rotational symmetry about the vertical, so the instability leads to the formation of tube-shaped regions of elevated porosity. 
The presence of a large-scale, solid shear flow breaks this symmetry.
Provided the deviatoric stress in the horizontal direction is extensional rather than contractional, even a small amount of shear favours the formation of tabular, high-porosity bands (section~\ref{sec:orientation}). 
This situation applies at a mid-ocean ridge, for example.
The orientation of the resulting bands is intermediate between that favoured by reaction (vertical) and that favoured by shear, and is controlled by the parameter $S$. 
Thus the tabular geometry of dunite bodies in ophiolites is consistent with a reactive origin combined with extension in the horizontal direction.
The favoured orientation is only weakly affected by considering the full pressure-dependence of the solubility gradient that drives the reaction infiltration instability (see appendix~\ref{app:c_eq}, figure~\ref{figure-x-z-plane_A}).
Nonetheless, the pressure dependence of the solubility gradient may be important in the nonlinear development of channels, particularly in the presence of strong lateral pressure gradients. 
This suggests an important role for numerical models, as discussed in section~\ref{sec:implications}.

We briefly discussed the dependence of the growth rate on wavelength, which is potentially significant in setting the length scale of dunites.
Growth is suppressed at length scales that exceed the compaction length, so channels have a smaller scale in the direction of the wavevector than the compaction length, which is thought to be about a kilometre in the mantle.
In previous work, we showed that the reaction-infiltration instability can lead to channels growing with a scale consistent with geological observations, within the considerable parametric uncertainty \citep{braun02,reesjones2018-jfm}.
Given this parametric uncertainty and the incomplete understanding of the length scales of shear-driven porosity bands, it is premature to draw definitive conclusions. 

\begin{figure*}
  \includegraphics[width=1.0\textwidth]{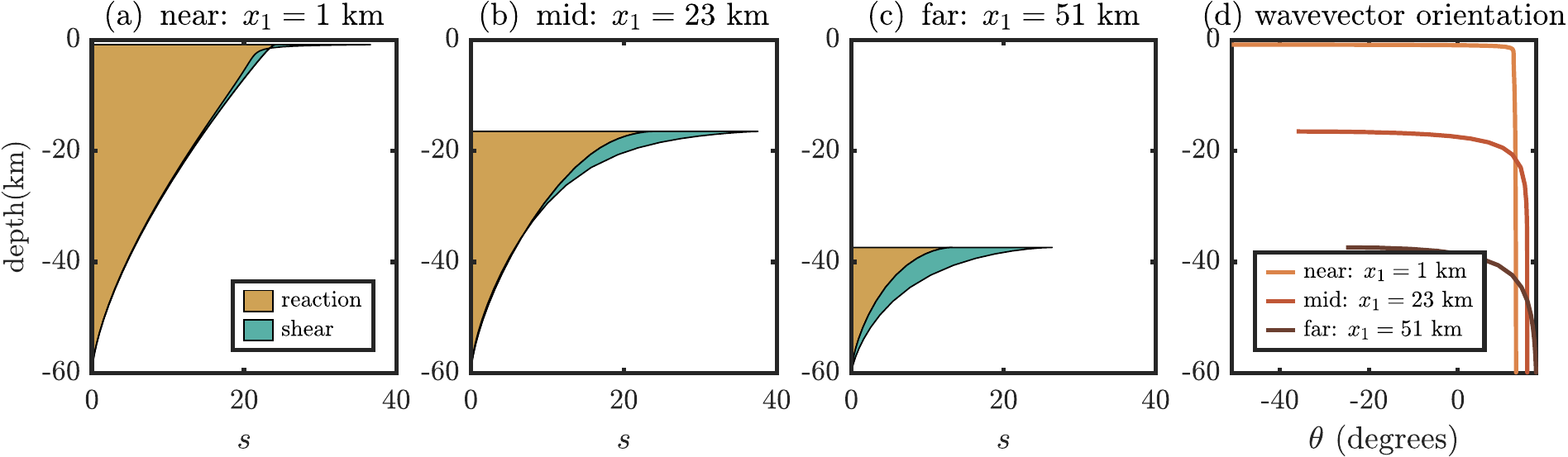}
  \caption{(a--c) The relative and combined importance of reaction (brown) and shear (green) to the amplitude of the porosity bands $s$. Plots show the evolution along three streamlines that finish at positions $x_1$ increasingly far from the ridge axis. (d) The rotation of the wavevector along the same three streamlines.   Calculations use the reference set of parameters. Cases (a) and (b) correspond to the streamlines plotted in figure~\ref{fig:MOR_growth_spectrum}. The conversion to dimensional units is based on $H=60$~km for illustration.
  \label{fig:MOR_summary}}
\end{figure*}

We applied our model of the combined instability to a mid-ocean ridge using the method proposed by \citet{Gebhardt2016}, as described in section~\ref{sec:methods_MOR}.  
Our results (section~\ref{sec:results_MOR}) are summarised in figure~\ref{fig:MOR_summary}, where panels a--c show the accumulated growth $s$ as a function of depth for three different corner-flow streamlines that ascend from the bottom of the melting region.  The contribution of shear (green) is significant only at shallow depths and/or far from the ridge axis.  
Geochemical evidence requires channels to form at least 15~km beneath the Moho \citep{kelemen97} and U-series disequilibrium and reactive-flow models suggest it is possibly much deeper \citep{jull02, keller16,liu19}.  Together this indicates that instantaneous growth rates are dominated by reaction and affected by shear only along the base of the lithosphere.

Nonetheless, the shear associated with tectonic-scale flow is important in that it promotes the tabular geometry of channels and affects the orientation of channels by rotating their tops away from the ridge axis, which corresponds to a decrease in wavevector angle along the streamline, as shown in figure~\ref{fig:MOR_summary}d.
Most of this rotation happens very close to the upper end of a streamline, where the streamline turns the corner and terminates at the base of the lithosphere.
The most unstable initial orientation of bands is close to vertical, with their tops tilted slightly towards the ridge axis.
Bands with this orientation grow rapidly, mainly due to the reactive instability.
Near the end of the streamline, the bands are rotated further and can end up lying between the vertical and the horizontal.
We explored whether these results are robust to the choice of parameters used, including the angle from the horizontal to the base of the lithosphere.
The only combination of parameters that favoured shear over reaction is a very slow melt velocity ($\lesssim 1$~m/yr) with a small bulk-to-shear viscosity ratio ($O(1)$).   
There is evidence for much more rapid melt extraction ($\gtrsim 10$~m/yr) based on the Icelandic deglaciation \citep[e.g.][]{jull96,maclennan02,Eksinchol2019,ReesJonesRudge2020} and U-series data \citep[e.g.][]{iwamori94,kelemen97,jull02,stracke06,Elliot2014}.
Given much more rapid melt extraction, it is most likely that reaction is dominant over shear in general. 

\subsection{Limitations of analysis} \label{sec:limitations}

Our calculations, like those of \citet{Gebhardt2016}, rely on a separation of scales between the sub-compaction-length scale of porosity localisation ($\lesssim 1$~km) and the tectonic scale of the mid-ocean ridge ($\sim 100$~km).
This separation arguably enables us to embed our idealised calculation of the local growth rate from section~\ref{sec:results_local} in tectonic models of mid-ocean ridge magmatism. 
However, we have not shown that these scales are truly separate and, indeed, there are reasons to question this. For example, local perturbation growth could create a pattern of anisotropic material properties (e.g., permeability, viscosity) that would feedback on the large-scale dynamics.
It is also worth noting that the same scale separation presents a severe challenge for numerical models that discretise a two- or three-dimensional space. To capture the interaction between the large and small scales requires either extremely high grid resolution over a very large domain or a more sophisticated, multi-scale approach \citep[e.g.,][]{kevrekidis2003}. 

The linearity of the current approach is another important limitation. The linearized equations apply rigorously only to the initial stages of channel formation; their formal validity breaks down as the perturbation to the background state grows. 
Exponential growth cannot continue indefinitely and, instead, the channel amplitude saturates \citep{spiegelman03a, liang10,king10}. At the same time, the lithological imprint of channelised flow (replacive dunites) is advected and rotated by the mantle flow.  Lithological structure may serve to lock in the pattern of reactive channels and limit overprinting by shear-driven growth in contrasting directions. 
This indicates that two-dimensional numerical models have an important role to play in understanding the nonlinear evolution of channels, channel coalescence and the overall arrangement of magmatic localisation.
Unfortunately, such numerical models sometimes fail as the localisation becomes more pronounced \citep[e.g.][]{katz13, Vestrum2020}, perhaps reflecting our incomplete understanding of the physics of partially molten rocks.  Numerical diffusion may also mask behaviour that is expected based on the results obtained here. In one possible example of this, \citet{katz10b} found that shear-driven porosity bands did not emerge in two-dimensional simulations of melt transport beneath a mid-ocean ridge, even with exaggerated viscous weakening by porosity. However, \citet{katz12} demonstrated that inherited lithological heterogeneity in the mantle can lead to sharp localisation of melting and melt transport. And while the chemical heterogeneity imposed in that case may be extreme, models by \citet{keller2017} predicted the emergence of channelized flow due to volatile-rich flux melting beneath a mid-ocean ridge.  Our approach based on linearization of the governing equations complements these numerical studies. 

Other potentially significant issues relate to the background state about which we linearize. 
It is probably a good approximation to consider that the solid flow is only minimally affected by the magma flow provided the porosity remains small. 
However, we used a simple Newtonian viscosity for the solid flow; this could be extended to consider a non-Newtonian or anisotropic viscosity. 
These effects alter the behaviour of the shear-driven instability, reducing the angle of porosity bands. 
Nevertheless, the principle remains that the bands grow most rapidly at an angle between that favoured by reaction and that favoured by shear. 
We also assumed that the magma flow was purely vertical, driven by buoyancy. 
While this can be the dominant contribution to magma transport, there are other mechanisms (discussed in the following section) that give rise to a more complex, not purely vertical, pattern of magma flow at mid-ocean ridges.

Another related issue that could modify our predictions for mid-ocean ridges is the background compaction rate associated with melt segregation. 
If the compaction viscosity is a decreasing function of porosity, background compaction acts to stabilize the system against exponential perturbation growth. 
\citet{hewitt10} showed, in the context of a melting-column model, that this effect is significant if the bulk viscosity has an inverse dependence on porosity.
The stabilising effect is much weaker if the compaction viscosity varies only logarithmically with porosity \citep{rudge18, reesjones2018-jfm}.
This could be reassessed in the mid-ocean ridge geometry. 

\subsection{Implications} \label{sec:implications}
A leading-order observation about mid-ocean ridges is that the volcanic zone is much narrower ($\sim$10~km) than the lateral extent of the partially molten region ($\sim$100~km), which means that melt must be focused laterally towards the ridge axis. \citet{spiegelman87} and \citet{phippsmorgan87} hypothesised that dynamic pressure gradients suck melt towards the ridge. \citet{sparks91} and \citet{spiegelman93c} proposed that melts migrate to the ridge through a sub-lithospheric decompaction channel.
More recently, \citet{Turner2017} and \citet{Sim2020} have argued for `melting-pressure focusing' associated with gradients in compaction pressure. 
In addition to these mechanisms, \citet{katz06} suggested that shear-driven porosity bands create an effectively anisotropic permeability, and that they have an orientation such this anisotropy focuses melt toward the ridge \citep[see also][]{phippsmorgan87,daines97,kohlstedt09,liu19}.
In contrast, the present calculations indicate that high-porosity channels typically have a sub-vertical orientation in the region moderately close to the ridge (see figure~\ref{fig:MOR_growth_combined_sensitivity} and \ref{fig:MOR_summary}d). Beneath the lithosphere, they remain sub-vertical but tend to rotate away from the ridge axis due to the corner flow. The anisotropic permeability structure that might arise from this pattern would not contribute much to melt focusing.  On the other hand, melt focusing by pressure gradients might lead to reaction-induced channels that point toward the ridge axis, aligned with the magmatic flow direction \citep{rabinowicz2005}.
Numerical models could be used to test the hypothesis that lateral pressure gradients additionally drive the formation of diagonal reactive dissolution channels through pressure-dependence of the solubility gradient (appendix~\ref{app:c_eq}), thereby enhancing focusing \citep{spiegelman01}. 

Coherent alignment of melt within the mantle could give rise to anisotropy of seismic wavespeeds.  Measured anisotropy might therefore be related to the predictions above if the influence of melt can be disentangled from other causes of anisotropy. \citet{kendall94} and \citet{blackman97} recognised the potential for grain-scale alignment of melt to shape the pattern of anisotropy beneath mid-ocean ridges. Later, \citet{Holtzman2010} argued that localised melt-fraction perturbations (channels or bands) would also induce seismic anisotropy. To explain observations of seismic anisotropy beneath some mid-ocean ridges, \citet{Holtzman2010} and \citet{nowacki12} invoke sheets of higher melt fraction sub-parallel to the lithosphere-asthenosphere boundary (LAB) at some $50$~km from the ridge axis to create a tilted transverse isotropy (TTI). While such alignment was predicted by \citet{katz06} and \citet{Holtzman2010} on the basis of the shear-driven instability, results presented here cast doubt on it and hence on the TTI hypothesis. However, plate spreading at the Main Ethiopian Rift is associated with a fast seismic direction that is ridge-parallel \citep{kendall05, hammond14}. The sub-vertical, tabular magmatic structures that we predict to arise from the shear-modified reaction-infiltration instability are consistent with such anisotropy.  Alignment of magmatic features sub-parallel to the shallowly dipping LAB may instead arise by the dynamic response to magmatic flow toward an impermeable boundary \citep{sparks91, hewitt08}. Alternatively, seismic anisotropy may arise from a grain-scale texture (e.g., lattice- or melt-preferred orientation \citep{holtzman03b, qi18}) rather than accumulated growth of macroscopic localisation patterns, as initially proposed by \citet{kendall94}.

\subsection{Conclusions}

\begin{itemize}
    \item We developed a consistent framework to model the combined growth of reaction-driven and shear-driven instabilities in the partially molten mantle and calculated their linear growth rate.  We applied that framework to the melting region beneath mid-ocean ridges.
    \item The reactive-infiltration instability is dominant over most of the melting region and, in particular, along its base and close to the ridge axis, where mantle flow is vertical.  This gives rise to sub-vertical, high-porosity channels that form sufficiently deep within the melting region to explain the observed disequilibrium between erupted lavas and the harzburgitic upper mantle.
    \item The presence of even a small amount of horizontal extension favours tabular channels over tube-shaped channels, consistent with the morphology of observed dunites.
    \item The shear-driven instability may contribute to further growth of channels along the base of the lithosphere. However, the orientation of these channels is set by prior growth in the reactive-flow regime and hence they remain sub-vertical, despite some rotation by the corner flow.
    \item Within the limitations of our study, shear-driven melt bands do not function as a mechanism for melt focusing at mid-ocean ridges.  Moreover, we predict that bands are not aligned consistently with the shallow-dipping seismic anisotropy that has been obtained by inversions.
\end{itemize}

\begin{acknowledgments}
Insightful comments by P.~Kelemen and an anonymous reviewer helped us to improve the manuscript. D.W.R.J.~acknowledges research funding through the NERC Consortium grant NE/M000427/1. The research of R.F.K.~leading to these results received funding under the European Union's Horizon 2020 research and innovation programme, grant agreement number 772255.
We thank the Deep Carbon Observatory of the Alfred P.~Sloan Foundation.
\end{acknowledgments}

\section*{Data availability statement}
Software that implements the theoretical methods presented this article is archived by Zenodo and available at https://doi.org/10.5281/zenodo.4618182 \citep{davidreesjones_2021_CODE}. 

\bibliographystyle{gji}

\bibliography{references}

\appendix
\section{Pressure-dependant reactive melting}\label{app:c_eq}
The reaction-infiltration instability is driven by a chemical solubility gradient. 
In the main text, we assumed that this gradient was vertical.
However, chemical solubility depends not on depth directly but rather on pressure. 
If the pressure is dominantly lithostatic, then the solubility gradient will be vertical, motivating our assumption in the main text. 
The purpose of this appendix is to investigate the potential role of pressure-dependent melting by relaxing the assumption that the thermodynamic pressure is dominantly lithostatic.
The shear-driven instability creates a pressure gradient that can, in turn, feedback on the reactive instability through the  pressure-dependent  reactive melt rate.
So this appendix allows us to consider another potential mode of coupling between the two types of instability.

We redo the linear stability from section~\ref{sec:methods} to account for the full pressure-dependence of the reactive melt rate. 
To focus on this effect, we strip out other parts of our earlier analysis from the outset.
These assumptions led to the simplified growth rate reported in section~\ref{sec:equilibrium-dynamics-large-deltak}, so we compare with results reported in that section. 

\subsection{Models for pressure-dependent reactive melt rate} \label{app:c_eq_models}
We start with the model for the reactive melt rate originally given in equation~\eqref{eq:chemistry1}, which we replace by
\begin{equation}
    \boldsymbol{v}_D \cdot \nabla c_\mathrm{eq}   = \alpha \Gamma, \label{eq:chemistry1-A}
\end{equation}
where we repeated our earlier simplification ($\phi \boldsymbol{v}_l \approx \boldsymbol{v}_D$). 
The solubility gradient $\nabla c_\mathrm{eq}$ plays the role of $\nabla (\beta z)$ earlier. 
We next assume that the solubility gradient is linearly related to the pressure gradient. 
Over the full-depth of the melting region this is a simplification, but will be valid locally, consistent with the approach taken in section~\ref{sec:methods}. 
Thus
 \begin{equation} 
      \nabla c_\mathrm{eq}   = - m \nabla P_\mathrm{therm}, \label{eq:c_eq(P)}
\end{equation}
where $m>0$ is a proportionality constant and $\nabla P_\mathrm{therm}$ denotes the thermodynamic pressure gradient. 

The concept of thermodynamic pressure needs to be carefully considered for systems out of equilibrium.
\citet{jull96} consider two main possibilities, $\nabla P_\mathrm{therm} = \nabla P_l$ and  $\nabla P_\mathrm{therm} = \nabla ( P_l -  \zeta\mathcal{C})$, in which $\nabla P_l$ is given by equation~\eqref{eq:Momentum2}:
\begin{equation}
    \nabla P_l =  2\mathrm{\textbf{D}}_s  \cdot \nabla \eta + \eta \nabla^2 \boldsymbol{u}+ \frac{4}{3} \eta \nabla \mathcal{C} + \nabla  (\zeta\mathcal{C}) + \overline{\rho}\boldsymbol{g}.\label{eq:Momentum2-A}
\end{equation}
\citet{jull96} argue in favour of the second possibility, which corresponds to \mbox{$-1/3$}~times the trace of the solid stress tensor. 
For the purposes of this appendix, the distinction between these definitions has only a marginal effect on the results, so we consider both possibilities for completeness. 

Next, we define the lithostatic pressure gradient
\begin{equation} \label{eq:P_lith}
\nabla P_\mathrm{lith}  = \overline{\rho}\boldsymbol{g} \approx \rho_s \boldsymbol{g} = -\rho_s g \hat{\boldsymbol{z}},
\end{equation}
where the approximation $\overline{\rho}\approx \rho_s$ is consistent with the assumption $\phi \ll 1$ we made in section~\ref{sec:methods}. 
Then we define the non-lithostatic part of the liquid pressure gradient as
\begin{equation}
\nabla P_m =  2\mathrm{\textbf{D}}_s  \cdot \nabla \eta + \eta \nabla^2 \boldsymbol{u}+ \frac{4}{3} \eta \nabla \mathcal{C} + \nabla  (\zeta\mathcal{C}), 
\end{equation}
so 
\begin{equation}
    \nabla P_l = \nabla P_m + \nabla P_\mathrm{lith}. 
    \label{eq:Momentum3-A}
\end{equation} 

If the thermodynamic pressure is dominated by the lithostatic contribution, then we can combine equations~\eqref{eq:c_eq(P)} and \eqref{eq:P_lith} to obtain
$
      \nabla c_\mathrm{eq}   = m \rho_s g \hat{\boldsymbol{z}},
$
which is consistent with the simplified approach in the main text provided 
\begin{equation}
\beta = m \rho_s g \quad \Leftrightarrow \quad m = \frac{ \beta }{\rho_s g}. 
\end{equation}
Finally, we consider the solubility gradient under the two potential definitions of thermodynamic pressure above. 

If $\nabla P_\mathrm{therm} = \nabla P_l$, then 
 \begin{equation} \label{eq:c_eq_v1}
      \nabla c_\mathrm{eq}   = \beta \left[ \hat{\boldsymbol{z}} - \frac{\nabla P_m}{\rho_s g} \right] .
\end{equation}
The dimensionless ratio ${\nabla P_m}/{\rho_s g}$ is the additional melting factor arising from consideration of the pressure-dependence of the solubility gradient. 

If $\nabla P_\mathrm{therm} = \nabla (P_l-\zeta \mathcal{C})$, then 
 \begin{equation}   \label{eq:c_eq_v2}
      \nabla c_\mathrm{eq}   = \beta \left[ \hat{\boldsymbol{z}} - \frac{\nabla P_m}{\rho_s g} + \frac{\nabla (\zeta \mathcal{C}) }{\rho_s g} \right] .
\end{equation}

\subsection{Revised linear stability analysis}
We now redo the linear stability analysis with the generalized reactive melting model. 
We first rewrite equation~\eqref{eq:Darcy_summary} for the Darcy velocity in terms of $P_m$,
\begin{equation}
\boldsymbol{v}_D  = K  \Delta \rho g \boldsymbol{\hat{z}}   -K \nabla P_m  . \label{eq:Darcy_summary-A}
\end{equation}
As before, this can be decomposed into a base state and perturbation: $\boldsymbol{v}_D = K_0\Delta\rho g \boldsymbol{\hat{z}}  +\boldsymbol{v}_D'$, where 
\begin{equation}
\boldsymbol{v}_D'  =n w_0  \boldsymbol{\hat{z}}  \phi' -K_0 \nabla P_m'  . \label{eq:Darcy_pert-A}
\end{equation}
We next expand out the pressure gradient term, making the same simplification that $\mathcal{C}_0=0$ as in the main text,
\begin{align}
 \nabla P_m' & =  \nabla P_l', \nonumber \\
 & =  2\mathrm{\textbf{D}}_0  \cdot \nabla \eta' + \eta_0 \nabla^2 \boldsymbol{u}'  +\left(\tfrac{4}{3} \eta_0 + \zeta_0 \right)\nabla \mathcal{C}'  , \nonumber \\
  & =   \nabla \psi' +\left(\tfrac{4}{3} \eta_0 + \zeta_0  \right) \nabla \mathcal{C}'  ,  \nonumber \\
    & =   -2\lambda^* \eta_0 \dot{\gamma_0} \nabla \widetilde{\psi}' +\left(\tfrac{4}{3} \eta_0 + \zeta_0  \right) \nabla \mathcal{C}'  .
 \end{align}
 Therefore 
 \begin{equation} \label{eq:K_0-gradP_pert}
 -K_0\nabla P_m' = \Lambda  \nabla \widetilde{\psi}' - \delta^2 \nabla \mathcal{C}',
 \end{equation}
 using the definition of $\delta$ in equation~\eqref{eq:delta_def} and $\Lambda$ in equation~\eqref{eq:Lambda_def}. So
 \begin{equation}
\boldsymbol{v}_D'  =n w_0  \boldsymbol{\hat{z}}  \phi' + \Lambda  \nabla \widetilde{\psi}' - \delta^2 \nabla \mathcal{C}'  . \label{eq:Darcy_pert-A-2}
\end{equation}
As an aside, this gives another interpretation of $\Lambda$ as a factor that controls the perturbed melt flow arising from the shear-driven instability. 

We next linearize the generalized reactive melt rate equation~\eqref{eq:chemistry1-A}, 
Note that the base-state solubility gradient is $ \beta \hat{\boldsymbol{z}}$, independent of the additional terms arising from the pressure-dependence of the solubility gradient. 
The perturbation to the melt rate is given by
\begin{equation}
    \boldsymbol{v}_D' \cdot  \beta \hat{\boldsymbol{z}} + K_0 \Delta \rho g \hat{\boldsymbol{z}} \cdot \nabla c_\mathrm{eq}'   = \alpha \Gamma'. \label{eq:chemistry1-A-pert}
\end{equation}
The extra terms from the pressure-dependence of the solubility gradient depend on the definition of thermodynamic pressure as discussed above. 
If $\nabla P_\mathrm{therm} = \nabla P_l$, we linearize equation~\eqref{eq:c_eq_v1} as follows
 \begin{equation}   \label{eq:c_eq_v1_pert}
      \nabla c_\mathrm{eq}'   = \frac{\beta}{\rho_s g} \left[  - {\nabla P_m'} \right] .
\end{equation}
If $\nabla P_\mathrm{therm} = \nabla (P_l-\zeta \mathcal{C})$, we linearize equation~\eqref{eq:c_eq_v2} as follows
 \begin{equation}   \label{eq:c_eq_v2_pert}
      \nabla c_\mathrm{eq}'   = \frac{\beta}{\rho_s g} \left[  - {\nabla P_m'} + {\zeta_0 \nabla \mathcal{C}' } \right] .
\end{equation}
From now on, we stick with the latter of these models, seeing that the former can be obtained from it by removing the term coming from $\zeta_0 \nabla \mathcal{C}'$. 

We next take equation~\eqref{eq:chemistry1-A-pert} and combine it with equations~\eqref{eq:Darcy_pert-A}, \eqref{eq:K_0-gradP_pert} and \eqref{eq:c_eq_v2_pert} to obtain
\begin{align}
  \frac{\alpha}{\beta} \Gamma' &=   n w_0   \phi' -K_0 \partial_z P_m' +  \frac{\Delta \rho}{\rho_s}   \left[  - K_0 \partial_z P_m' + K_0 \zeta_0 \partial_z \mathcal{C}'  \right]    , \nonumber \\ 
  &=   n w_0   \phi' -\left(1+\frac{\Delta \rho}{\rho_s}\right) K_0 \partial_z P_m' +  \frac{\Delta \rho}{\rho_s}       K_0 \zeta_0 \partial_z \mathcal{C}'     , \nonumber \\ 
   &=   n w_0   \phi' + \left(1+\frac{\Delta \rho}{\rho_s}\right) \left(\Lambda  \partial_z  \widetilde{\psi}' - \delta^2  \partial_z  \mathcal{C}' \right)+  \frac{\Delta \rho}{\rho_s}         \zeta_\mathrm{rel} \delta^2 \partial_z \mathcal{C}'     , \label{eq:Gamma_pert_A}
\end{align}
where we define 
\begin{equation}
 \zeta_\mathrm{rel}\equiv \frac{1}{1+\tfrac{4}{3} \tfrac{\eta_0}{\zeta_0}}<1,
\end{equation}
such that $\zeta_\mathrm{rel} \delta^2 = \zeta_0 K_0$. 
The difference between the two definitions of thermodynamic pressure amounts to inclusion or exclusion of the term involving  $\zeta_\mathrm{rel} $.

Following the approach of section~\ref{sec:normal_modes} (in which we showed that $\tilde{\psi}=G\tilde{\phi}$), we take normal modes of equation~\eqref{eq:Gamma_pert_A} as follows:
\begin{align}
 \frac{\alpha}{\beta} \tilde{\Gamma} &=      n w_0   \tilde{\phi} + \left(1+\frac{\Delta \rho}{\rho_s}\right) \left(\Lambda G  ik_z   \tilde{\phi} - \delta^2  ik_z  \tilde{\mathcal{C}} \right)\nonumber \\&\quad+  \frac{\Delta \rho}{\rho_s}         \zeta_\mathrm{rel} \delta^2 ik_z \tilde{\mathcal{C}}   , \nonumber \\ 
&=      n w_0   \tilde{\phi} + \left(1+\frac{\Delta \rho}{\rho_s}\right)  \Lambda G  ik_z   \tilde{\phi} \nonumber\\&\quad -   ik_z \delta^2 \left[1+\frac{\Delta \rho}{\rho_s} \left(1- \zeta_\mathrm{rel}\right)\right]  \tilde{\mathcal{C}} .
\end{align}
This expression is a generalized version of equation~\eqref{eq:chemistry_pert_final_hat}. 
Next we combine with equations~\eqref{eq:mass_s_pert_final_hat} and \eqref{eq:compaction_pert_final_hat}  in the same way as before to obtain a generalized expression for the growth rate
\begin{align} \label{eq:dispersion_master_A}
&\sigma = \frac{\beta}{\alpha}  nw_0 + \frac{\beta}{\alpha}  \left(1+\frac{\Delta \rho}{\rho_s}\right)   ik_z \Lambda G  \nonumber \\&\quad +\left(1-\frac{\beta ik_z  \delta^2}{ {\alpha}} \left[1+\frac{\Delta \rho}{\rho_s} \left(1- \zeta_\mathrm{rel}\right)\right] \right) \frac{  \Lambda  G k^2 - nw_0 ik_z}{1+\delta^2 k^2}.
\end{align}
The additional terms involving $\Delta\rho/\rho_s$ appear in the imaginary part of $\sigma$ so affect the propagation speed of porosity waves. 

We now make the same approximations as in section~\ref{sec:equilibrium-dynamics-large-deltak}. 
Throughout this study, we focus on the real part of the growth rate. 
We also take the large-compaction-length limit  ($\delta k \gg 1$).
Then    
\begin{align} \label{eq:dispersion_master_real_A}
&\mathrm{real}(\sigma) = \frac{\beta}{\alpha}  nw_0 -\frac{\beta}{\alpha}nw_0  \left[1+\frac{\Delta \rho}{\rho_s} \left(1- \zeta_\mathrm{rel}\right)\right]\frac{k_z^2}{ k^2}   +   \frac{  \Lambda  G    }{ \delta^2}, 
\end{align}
so
\begin{align} \label{eq:dispersion_master_limit_1_real-A}
&\mathrm{real}(\sigma)  = \sigma_\mathrm{reaction}  \left(1-\left[1+\frac{\Delta \rho}{\rho_s} \left(1- \zeta_\mathrm{rel}\right)\right] \frac{k_z^2}{k^2}   \right) +  \sigma_\mathrm{shear}  G.
\end{align}
We let $\mathcal{R}= \left[1+\tfrac{\Delta \rho}{\rho_s} \left(1- \zeta_\mathrm{rel}\right)\right]$ and note that $k_z = k\sin\theta$. Then 
\begin{align} \label{eq:dispersion_master_limit_1_real-A-R}
&\mathrm{real}(\sigma)  = \sigma_\mathrm{reaction}  \left(1-\mathcal{R} \sin^2\theta   \right) +  \sigma_\mathrm{shear}  G.
\end{align}
This is identical to equation~\eqref{eq:dispersion_master_limit_1_real} when $\mathcal{R}=1$.
Indeed all the various thermodynamic models of reactive melting can be treated according to the following cases:
\begin{equation}
\mathcal{R} = \begin{cases}
1, &\text{lithostatic only},\\ 
1+\tfrac{\Delta \rho}{\rho_s},   &\text{$\nabla P_\mathrm{therm} = \nabla P_l$,}\\ 
1+\tfrac{\Delta \rho}{\rho_s} \left(1- \zeta_\mathrm{rel}\right),&\text{$\nabla P_\mathrm{therm} = \nabla (P_l-\zeta \mathcal{C})$.}
\end{cases}
\end{equation}

Accounting for the pressure-dependence of the solubility gradient slightly reduces the reactive part of the growth of the instability. 
However, for partially molten rocks, $\Delta \rho/\rho_s \approx 1.1$. 
So even for the most different thermodynamic model $\nabla P_\mathrm{therm} = \nabla P_l$, $\mathcal{R}$ is only 10\% different from the purely lithostatic case. 
Using $\nabla P_\mathrm{therm} = \nabla (P_l-\zeta \mathcal{C})$, reduces this already small difference further, since $\zeta_\mathrm{rel}<1$ and so $\mathcal{R}$ is closer to 1. 

\begin{figure}
\centering\noindent\includegraphics[width=1.0\linewidth]{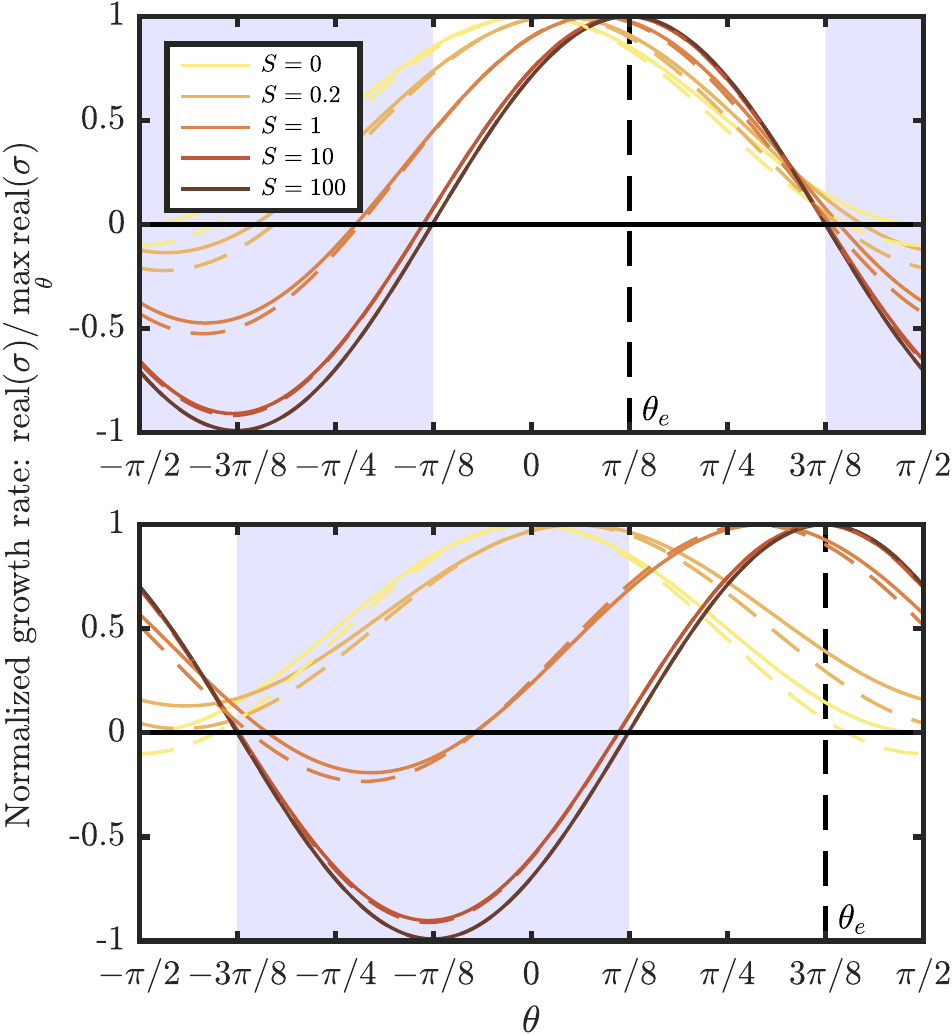}
   \caption{Generalized version of figure~\ref{figure-x-z-plane} (see that figure caption for further details). The solid curves correspond to $\mathcal{R}=1$. The dashed curves correspond to $\mathcal{R}=1.1$ and hence account for the dependence of the solubility gradient on the dynamic pressure. For the reactive instability only ($S=0$, yellow curves), the preferred wavevector angle is always $\theta=0$ corresponding to vertical channels,  independent of $\mathcal{R}$.}
   \label{figure-x-z-plane_A}
\end{figure}

Figure~\ref{figure-x-z-plane_A} shows that the angular dependence of the growth rate from equation~\eqref{eq:dispersion_master_limit_1_real-A-R} is only minimally affected by the choice of $\mathcal{R}$. 
In particular, the most unstable orientation satisfies 
\begin{equation} \label{eq:theta_ky=0_R}
\tan 2 \theta = \frac{2 S \sin 2\theta_e}{\mathcal{R}+2S\cos 2\theta_e},
\end{equation}
which is very weakly affected by $\mathcal{R}$. 
At small $S$, the favoured orientation is close to $k_z=0$ ($\theta=0$) so the extra stabilization due to $\mathcal{R}>1$ is unimportant, as shown by equation~\eqref{eq:dispersion_master_limit_1_real-A-R}. 
At large $S$, the growth rate is dominated by shear, so the extra stabilizing of the reactive part of the growth rate is again unimportant, as shown by equation~\eqref{eq:theta_ky=0_R}.

Physically, the pressure-perturbations associated with the shear instability are coupled to the reactive instability through the pressure-dependence of the solubility gradient. 
However, this coupling only affects the imaginary part of the growth rate.
The real part of the growth rate is also affected by pressure perturbations associated with the reactive instability (not the shear-driven instability). 
Both of these effects are relatively small, because  $\Delta \rho/\rho_s \ll 1$. 

This analysis is only preliminary for two reasons. 
First, the consideration of terms $\Delta \rho/\rho_s$ goes beyond the Boussinesq approximation made throughout this study (see section~\ref{sec:2-phase-flow}). 
Second, perhaps more importantly, our study is restricted to a linear stability analysis with a vertical base upwelling of magma.
So reactive melting is driven by the vertical part of the solubility gradient, which is dominantly lithostatic. 
In a fully developed nonlinear state, lateral pressure gradients could drive reactive melting and this might be important in understanding the coalescence of channels \citep{spiegelman01}.

\subsection{Further considerations for MORs}
The importance of the non-lithostatic solubility gradient depends on its magnitude relative to the lithostatic pressure gradient. The latter scales like $\rho_s g$. 

Pressure gradients arising from buoyancy-driven flow scale like $\Delta \rho g$, so their relative contribution scales like $\Delta \rho/\rho_s$ --- the same scaling we identified in the previous section. 

The viscous corner flow (section~\ref{sec:CornerFlow}) gives rise to a dynamic pressure gradient. 
This can be estimated from Stokes equation for an incompressible fluid of constant viscosity 
\begin{equation} \label{eq:stokes}
\nabla P = \eta \nabla^2 \boldsymbol{u}.
\end{equation} 
The pressure gradient is singular at the corner ($x=z=0$ on the diagram in figure~\ref{fig:MOR_schematic}a) and decreases rapidly away from this point. 
To estimate the scale of this dynamic pressure gradient $\nabla P_\mathrm{dyn}$, we scale equation~\eqref{eq:stokes} as follows. 
We estimate $\eta \sim \eta_0$, $\boldsymbol{u} \sim U_0$, $\nabla \sim 1/L_\mathrm{dyn}$, where $L_\mathrm{dyn}$ is the distance from the corner. 
Then 
\begin{equation} \label{eq:stokes_scaled}
\nabla P_\mathrm{dyn} \sim \frac{\eta_0 U_0}{L_\mathrm{dyn}^2}. 
\end{equation} 
We then compare this estimate to the lithostatic pressure gradient $\rho_s g$ and find they are comparable when 
\begin{equation} \label{eq:L_dyn}
L_\mathrm{dyn} \sim \left(\frac{\eta_0 U_0}{\rho_s g}\right)^{1/2} \approx 1\,\mathrm{km},
\end{equation} 
where we used the rough estimates \mbox{$\eta_0 = 10^{19}$~Pa\,s,} \mbox{$U_0 = 10$~cm/yr,} \mbox{$\rho_s=3\times 10^3$~kg/m$^3$} and \mbox{$g=10$~m/s$^2$}.
Note that this estimate is factor of $(\Delta \rho / \rho_s)^{1/2}$ smaller than the melt focussing length scale estimated by \citet{spiegelman87}.
In conclusion, within a distance of about a kilometre from the corner, the dynamic pressure gradient is larger than the lithostatic pressure gradient. 
However, outside this region, the lithostatic pressure gradient is much larger than the dynamic pressure gradient (the latter decreases with the inverse square of distance from the corner). 
Therefore, we can neglect the contribution of dynamic pressure gradients almost everywhere beneath MORs.

Finally, we estimate the scale of compaction pressure gradients. 
\citet{Turner2017} and \citet{Sim2020} have argued for `melting-pressure focusing' associated with gradients in compaction pressure and we base our estimates on these studies. 
The base state compaction rate $\mathcal{C}_0$ scales like the base state melting rate $\Gamma_0$ (ignoring a minus sign), which we estimated in section~\ref{sec:MagmaFlow}. 
By combining estimates in that section, we find
\begin{equation}
\mathcal{C}_0 \sim \Gamma_0 \sim \frac{F_\mathrm{max} U_0}{H}.
\end{equation}
Then the compaction pressure $P_c$ can be estimated
\begin{equation}
P_c \sim \zeta_0 \mathcal{C}_0 \sim  \frac{F_\mathrm{max}  \zeta_0  U_0}{H}.
\end{equation}
To estimate the compaction pressure gradient, we must consider the length scale over which the compaction pressure gradient varies. 
For a triangular melting region an appropriate scale is the depth $H$, since the width also scales like this. 
In practice, the length scale is probably somewhat smaller than $H$ \citep{Sim2020}, leading to a somewhat higher gradient. 
Thus 
\begin{equation}
\nabla P_c \sim \ \frac{F_\mathrm{max}  \zeta_0  U_0}{H^2}.
\end{equation}
Finally, we compare this with the lithostatic pressure gradient 
\begin{equation}
\frac{\nabla P_c}{\rho_s g}  \sim \frac{F_\mathrm{max}  \zeta_0  U_0}{H^2 \rho_s g} \approx 10^{-3},
\end{equation}
where we used the rough estimates  \mbox{$F_\mathrm{max} = 0.2$,} \mbox{$\zeta_0 = 10^{20}$~Pa\,s,}  and \mbox{$H=60$~km} in addition to the estimates used in equation~\eqref{eq:L_dyn}.
Thus the compaction pressure gradient is a very small fraction of the lithostatic pressure gradient. 
The same conclusion can be reached by comparing figure~7 of \citep{Sim2020} with the lithostatic pressure $\rho_s g H \approx 2 \times 10^3$~MPa.\

These arguments do not mean that dynamic and compaction pressure gradients can be neglected when considering melt flow (after all the lithostatic pressure gradient does not drive any flow), only that they can generally be neglected when considering thermodynamic pressure. 

\section{Alternative approach to wavevector optimization for MOR\MakeLowercase{s}} \label{sec:alternative}

For the main MOR results (section~\ref{sec:results_MOR}), we optimized the initial wavevector orientation to maximize the total growth accumulated at the end of each streamline. 
This is relevant if we focus at the location where geological observations of dunite channels are most readily made. 
However, if we focus on the fabric at depth, it might make more sense to ask what initial wavevector orientation maximizes the total growth at each point in space separately. 
The results are a little harder to interpret, because the corresponding initial wavevector will not be consistent along each streamline.

\begin{figure*}
\centering\noindent\includegraphics[width=0.9\linewidth]{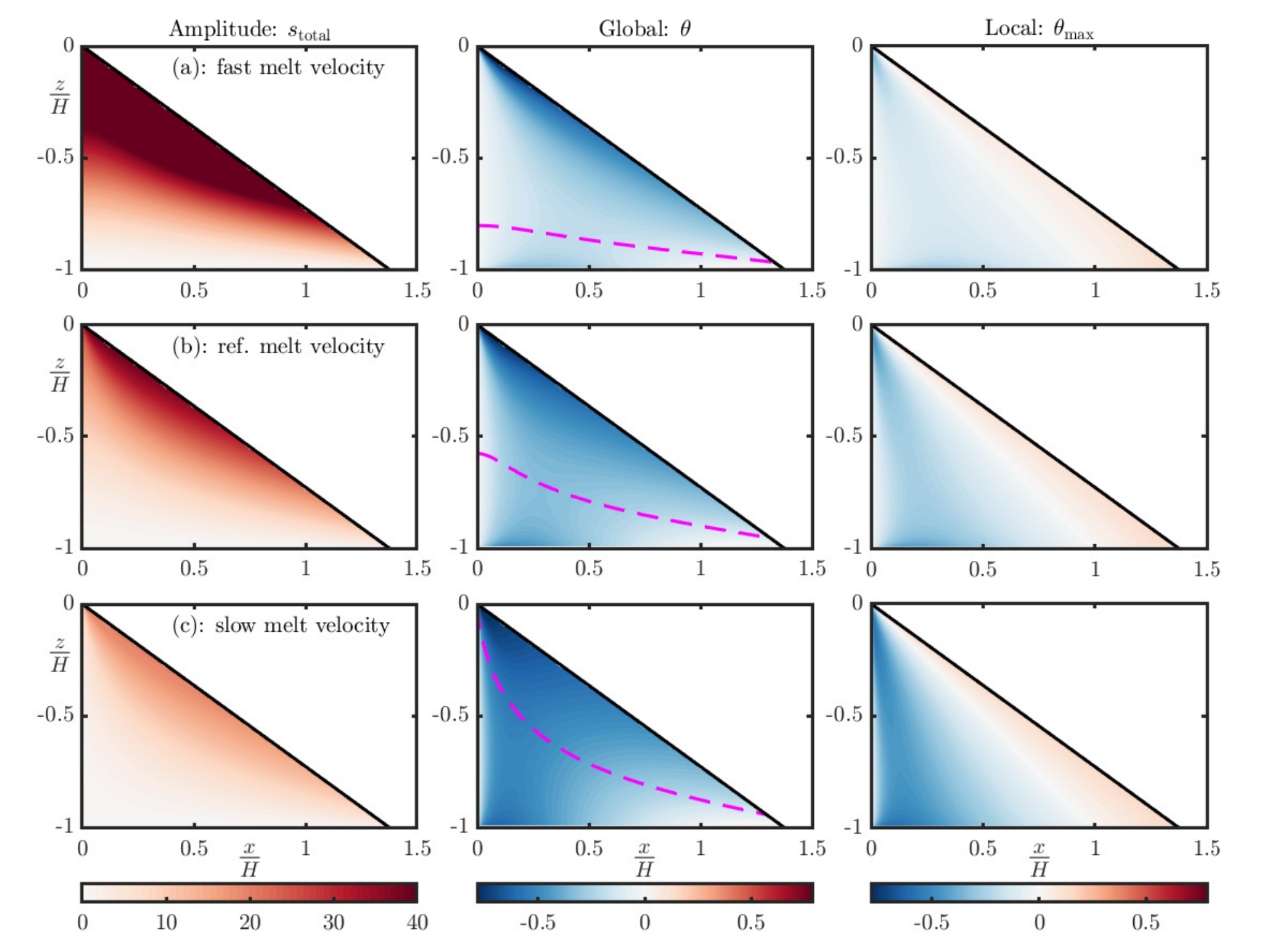}
   \caption{Alternative approach based on optimizing the total growth at each interior point separately, not just at the end of a streamline. The left column shows the amplitude $s$, the middle column shows the wavevector orientation, the right column shows the optimal orientation in terms of the local growth rate based on equation~\eqref{eq:theta_ky=0}. We report results at a range of different melt segregation rates. Row (a) has a higher melt velocity ratio $Q_0/U_0=6.3\times 10^6$, which is 10 times greater than the reference case. Row (b) is the reference case. Row (c) has a lower melt velocity ratio $Q_0/U_0=4 \times 10^3$, which is 16 times smaller than the reference case.   Note the colour scale in (a) is clipped. For the middle column, we added dashed magenta contours of $s_\mathrm{total}=7$. This gives an indication of the level below which the perturbations are very small.  }
   \label{fig:MOR_theta_max}
\end{figure*}

Figure~\ref{fig:MOR_theta_max} shows the results of this alternative optimization procedure at three different melt segregation speeds. 
We also plot in the right column the optimal wavevector orientation based on equation~\eqref{eq:theta_ky=0}, which reflects the local growth \textit{rate}.
The preferred wavevector orientation is always tilted in a slightly negative orientation (meaning that porosity bands are tilted slightly away from the ridge axis. 
Near the ridge axis (and across a wide part of the domain in the case of fast melt segregation), the wavevector angle is close to zero and porosity bands are close to vertical.
This is because reaction is the dominant mode of instability here. 
Elsewhere, the wavevector is tilted over at angles up to about $-\pi/4$, with more negative values at slower melt segregation rates.
This reflects the relative importance of shear in this case. 
However, for slower melt segregation, the amplitude of the porosity bands is not very high, especially at depth.
The locally most unstable orientation has a similar but distinct pattern (comparing the middle and right columns).
It is important, therefore, to track the accumulation of growth and the rotation of the wavevector along streamlines because the most unstable orientation at a particular point cannot be inferred purely from the local behaviour there.

\end{document}